\documentclass{aa}

\usepackage{graphicx}

\usepackage{psfig,latexsym,longtable,lscape,rotating}
\newcommand{\Xisol}{X_{i,\odot}}

\newcommand{\afebase}{$[\alpha /Fe]_b$}
\newcommand{\afenew}{$[\alpha /Fe]_n$}
\newcommand{\zbase}{$Z_{b}$}
\newcommand{\znew}{$Z_{n}$}

\newcommand{\mybold}{\bf }

\begin{document}

\title{Nearby early--type galaxies with ionized gas. III}

\subtitle{Analysis of line-strength indices with new stellar population models\thanks{Tables 5-7 and 9-12 are only available in electronic form
at the CDS via anonymous ftp to cdsarc.u-strasbg.fr (130.79.128.5)
or via http://cdsweb.u-strasbg.fr/cgi-bin/qcat?J/A+A/}}

\author{F. Annibali \inst{1,2}, A. Bressan \inst{2,3,4}, R. Rampazzo \inst{3}, W. W. Zeilinger \inst{5} and L. Danese \inst{2}.}


\offprints{F.~Annibali}

 \institute{ 
STSCI, 3700 San Martin Drive, Baltimore, MD 21218, USA \\
\email{annibali@stsci.edu} 
\and 
SISSA, Via Beirut 4, 34014 Trieste, Italy 
\and
INAF - Osservatorio Astronomico di Padova, Vicolo dell'Osservatorio 5, 35122 Padova, Italy
\and
Instituto Nacional de Astrof\'{\i}sica, Optica y Electr\'onica, Apdos. Postales 51 y 216, C.P. 72000 Puebla,  M\'exico
\and
Institut f\" ur Astronomie der Universit\" at  Wien, T\"
urkenschanzstra$\ss$e 17, A-1180 Wien, Austria 
}

\date{Received date; accepted date}


\authorrunning{Annibali et al.}
\titlerunning{Nearby early--type galaxies with ionized gas.III.}

\abstract
{}
{The paper is devoted to the study  of the underlying
stellar population of a sample of 65 nearby early-type galaxies, 
predominantly located in low density environments,
a large fraction of which show emission lines.
}
{
Ages, metallicities, and [$\alpha$/Fe] ratios
have been derived 
through the comparison of Lick indices measured at different galacto-centric
distances (7 apertures and 4 gradients)
with new simple stellar population (SSP) models that account for the
presence of $\alpha$/Fe--enhancement. 
The  SSPs cover a wide range
of ages ($10^{9}-16 \times 10^{9}$ yr), metallicities 
(0.0004 $\le$ Z $\le$0.05), and [$\alpha$/Fe] ratios (0-0.8). 
To derive the stellar population parameters, we use an algorithm 
that provides,  together
with the most likely solution in the (age, Z, [$\alpha$/Fe]) space,
the probability density function 
along the age-metallicity degeneracy.}
{We derive a large spread in age, with SSP-equivalent ages
ranging from a few to 15 Gyrs.
Age does not show any significant trend with central
velocity dispersion $\sigma_c$, but 
E galaxies appear on average older than lenticulars. 
On the contrary, a significant increasing trend of metallicity and
[$\alpha$/Fe] with $\sigma_c$ is observed, testifying that the chemical
enrichment was more efficient and the duration of the star formation
shorter in more massive galaxies. 
These latter two relations do not depend on galaxy morphological type.

We have also sought possible correlations 
with the local galaxy density $\rho_{xyz}$, but 
neither metallicity nor $\alpha$--enhancement show clear trends.
However, we find that while
low density environments (LDE) 
($\rho_{xyz} \le 0.4$) contain 
very young objects (from 1 Gyr to 4 Gyr),
none of the galaxies in the higher density environments (HDE) 
(40 \% of galaxies with a measured density) 
is younger than 5 Gyrs.
Considering the lack of environmental effect on the [$\alpha$/Fe] ratio
and the high value of [$\alpha$/Fe] in some young massive objects,
we argue that 
young galaxies in LDE are more likely due to recent rejuvenation 
episodes.
By comparing the number
of ``rejuvenated'' objects with the total number of galaxies in our sample,
and by means of simple two-SSP component models, we estimate that, on average,
the rejuvenation episodes  
do not involve more than 25 \% of the total galaxy mass.

The good quality of the data also allow us to analyze the
gradients of stellar populations.
An average negative metallicity gradient 
$\Delta \log Z/\Delta \log (r/r_e) \sim -0.21$ is firmly detected,
while the age and $\alpha$--enhancement spatial distributions
within r$_e$/2 appear quite flat.
These observations suggest that, in a given galaxy,
the star formation proceeded
on similar timescales all across the central r$_e$/2 region, but with an
increasing efficiency toward the center.}
{}


 \maketitle

\section{Introduction}

This is the third paper of a series dedicated to the study of early-type
galaxies with emission lines (Rampazzo et al. \cite{Ram05}, hereafter 
Paper~I; Annibali et al. \cite{Annibal05}, hereafter Paper~II). A recent
work  by Falc\' on-Barroso et al. (\cite{Falcon06}) (see also Sarzi et al. \cite{Sarzi06}) 
based on SAURON observations shows that the incidence of ionized-gas
emission in early-type galaxies is about 75\%. By morphological type,
lenticular galaxies display a slightly higher content of ionized gas
than elliptical galaxies (83\% versus 66\%), in good agreement with the
pioneering survey performed by Macchetto et al. (\cite{Mac96}), who found
emission in 85\% of the lenticulars and 68\% of the ellipticals in their
sample.  Falc\' on-Barroso et al. (\cite{Falcon06}) found similar
percentages when their sample was divided according to environment (83\%
field, 66\% cluster). Furthermore, the incidence of ionized gas shows no
correlation with either luminosity or the presence of a bar in the galaxy.
The above percentages suggest that the phenomenon of emission lines is
quite common among early-type galaxies: early-type galaxies without emission
lines are more exceptions rather than the rule in this class of
galaxies. The first step to understanding the evolution of these galaxies
and the nature of their ionized gas consists in studying the galaxy underlying stellar population.

For this purpose we make use of the of absorption-line indices that have been
proven to be a powerful tool to disentangle age and
metallicity  effects (e.g.,  
Buzzoni et al.~\cite{Buz92};
Worthey~\cite{Wor92};
Gonz\' alez \cite{G93} (G93); 
Worthey~\cite{Wo94};
Buzzoni et al.~\cite{Buz94}; 
Worthey et al.~\cite{W94}; 
Bressan et al. \cite{Br96};
Leonardi \& Rose~\cite {LR96}; 
Worthey \& Ottaviani~\cite{WO97}; 
Trager et al.~\cite{Tra98}; 
Longhetti et al.~\cite{L98a}; 
Vazdekis~\cite{Vaz99};
Longhetti et al.~\cite{L99}; 
Longhetti et al.~\cite{L00}; 
Trager et al.~\cite{TFWG00}; 
Kuntschner ~\cite{Ku00}; 
Beuing et al.~\cite{Beu02}; 
Kuntschner et al.~\cite{Ku02}; 
Mehlert et al.~\cite{Mehl03}). 
More recently the interpretation has been faced with the
more difficult task of also accounting for
the presence of non--solar scaled abundance patterns in
early-type galaxies, and of deriving age, metallicity, and 
[$\alpha$/Fe] ratio at the same time (e.g., Tantalo et al. \cite{Tan96};
Tantalo \& Chiosi \cite{TC04a}; 
Thomas et al. \cite{Thom03}; Thomas et al. \cite{Thom05}).  
Indeed, it has been shown that the $\alpha$/Fe element ratio 
plays a fundamental role in deriving the formation epochs. 
This ratio, in fact,
quantifying the relative importance of Type II and Type Ia supernovae in
the enrichment of the ISM,  carries information about the timescale over
which the star formation occurred (Greggio \& Renzini \cite{GR83};
Matteucci \& Greggio \cite{MG86}; Pagel \& Tautvaisiene \cite{PT95};
Thomas et al. \cite{Thom98}).   

In this paper we try to constrain the epoch of the baryon assembly
of our sample of 65 nearby early-type galaxies 
predominantly located in low density environments,
a large fraction of which show emission lines
 (Paper I and II),
and seek possible correlations with
their morphological type 
(elliptical and lenticular), velocity dispersion, and environmental density. 
A large part of this paper is dedicated to the development of
our own $\alpha$--enhanced models and to the comparison with the models 
already presented in the literature. 
The stellar population parameters are derived by means of a new 
algorithm based on the probability density function. 
The algorithm
provides,  together with the most probable solution in the (age, Z,
[$\alpha$/Fe]) space, the solutions along the age-metallicity
degeneracy that are within 1~$\sigma$  error from the observed index
values. Our analysis is based on the use of three different
line--strength index diagnostics, including the Mgb index, the $<$Fe$>$
index, and a Balmer line (H$\beta$, H$\gamma$, or H$\delta$). The use of
higher order Balmer lines allows us to minimize spurious effects that may
arise from uncertainties in the emission correction of the H$\beta$
absorption line.
The paper is organized as follows. Section~2 
presents an overview of the sample, summarizing its main properties 
and focusing on the delicate step of the correction for emission
infilling of the most commonly used age indicator, the H$\beta$ absorption line. 
In Sect. 3 a complete description of the procedure adopted to compute the
new $\alpha$--enhanced SSPs is provided. Section~4 describes how the new
SSP models have been implemented within an algorithm that allows us to
derive the stellar population parameters, i.e., age, metallicity, and
$\alpha$/Fe ratio. Section~5 presents the results.  
In particular we analyze both the
stellar population parameters derived for the apertures and gradients,
and their relations with velocity dispersion and local galaxy density. 
In Sect. 6 we discuss our results and compare them with the
literature. Finally, a summary and our conclusions are presented in 
Sect. 7.

\section{Further analysis of the line--strength indices data set}

\subsection{A sample overview}

The procedure of data reduction and analysis
of the 65 nearby early-type galaxies studied here 
has been already discussed in Papers I and II.
Table~1 summarizes the main characteristics of the sample. 
Column (1) gives the galaxy identification name; 
Col. (2) and (3) provide the galaxy morphological classification
according to RSA (Sandage \& Tammann \cite{RSA}) and 
RC3 (de Vaucouleurs et al. \cite{RC3}), respectively: only in few cases 
do the two catalogues disagree in the distinction between E and S0 classes;
Col. (4) gives the galaxy systemic velocity, V$_{hel}$, which is lower
than $\sim$5000 km~s$^{-1}$; Col. (5) provides the richness 
parameter $\rho_{xyz}$ (Tully \cite{Tu88}): it represents the density of
galaxies brighter than -16 apparent B-mag in the vicinity of the entry,
in galaxies $\times$Mpc$^{-3}$. The local density was determined on a
three--dimensional grid with 0.5 Mpc spacings. A Gaussian smoothing
function was used to calculate the contribution of each member of the
sample brighter than -16 B-mag to the density of the specific location.
In the Tully catalogue, the galaxies (considered in the local
environment) within 40 Mpc (3000  km~s$^{-1}$) with an apparent
magnitude B-mag $\leq$-16 are 2189. 
If one entry is at a distance larger than 40 Mpc, the local density 
is not recorded. The galaxies of our sample are mainly located in low
density environments. The local density of our galaxies varies from
$\rho_{xyz}$ $\approx$ 0.1, characteristic of very isolated galaxies, to
$\rho_{xyz}$ $\approx$ 4, which is characteristic of denser galaxy
regions in the Virgo cluster. 
For comparison, in the Tully~\cite{Tu88} catalogue, objects like NGC~1399
and NGC~1389, Fornax cluster members, have values of $\rho_{xyz}$=1.59
and 1.50, respectively.  Thus, our sample, even though biased towards low
density environments, contains a fraction  of galaxies in relatively
dense environments.

The sample spans a large range in central velocity dispersion, $\approx$
from 115 to 340 km~s$^{-1}$ (see Papers~I and II for detailed 
descriptions). Our long--slit spectra cover the (3700 - 7250) \AA\
wavelength range with a spectral resolution of $\approx$7.6~\AA\ at
5550~\AA. 
For the detailed description of the data reduction procedure and of the 
derivation of line--strength indices in the Lick system, we refer to
Papers~I and II. We only recall here that for each galaxy, 
25 Lick indices (21 belonging to the original Lick set (see 
Trager et al. (\cite{Tra98}) for the pass-band definitions) 
plus 4 higher order Balmer lines introduced by 
Worthey \& Ottaviani  (\cite{WO97})) have been measured for 7 luminosity
weighted apertures (with radii:  1.5\arcsec, 2.5\arcsec, 
10\arcsec,  r$_e$/10, r$_e$/8, r$_e$/4, and r$_e$/2), 
corrected for the galaxy ellipticity, and  4 gradients 
(0 $\leq$ r $\leq$r$_e$/16, r$_{e}$/16 $\leq$ r $\leq$r$_e$/8, r$_{e}$/8
$\leq$ r $\leq$r$_e$/4, and r$_{e}$/4 $\leq$ r $\leq$r$_e$/2).

\subsection{H$\beta$ correction for emission infilling}

Because of the presence of emission in a significant fraction of the
sample, the problem of possible contamination of the H$\beta$ absorption
line deserves particular attention. In Papers~I and II we have shown two
different methods to correct the H$\beta$ index for possible emission
infilling, based respectively on the measure of the  [OIII]$\lambda$5007
\AA \ and of the H$\alpha$ emission lines.
The first method is based on the correlation found by G93 between the 
H$\beta$ and the [OIII] emission in his sample of early-type galaxies,
such that ${\rm EW(H\beta_{em})/EW([OIII]\lambda5007)=0.7}$. 
The second method rests on the possibility of measuring the H$\alpha$
emission at 6563 \AA, from which the emitted flux in the 
H$\beta$ line results in ${\rm F_{H\beta}=1/2.86 F_{H\alpha}}$ (Osterbrock
\cite{Osterb89}). In Paper~I and II, we have shown that the corrections
derived from the two different methods are statistically similar. We
then provided the  final H$\beta$ indices, applying the correction from
the  [OIII]$\lambda$5007. 
We noticed, however, that the scatter of the
relation derived by G93 is large, and that small variations in the
H$\beta$ index correspond to large differences in the derived 
stellar population parameters when compared with models. 

A comparison of the emission corrections derived from the 
[OIII]$\lambda$5007 and H$\alpha$ lines, respectively, is also performed 
by Denicol{\'o} et al. (\cite{Den05a}) for their sample of early-type galaxies.
The authors show that the differences in the emission corrections derived 
through the two methods are small for the majority of galaxies, but argue that 
it is very dangerous to rely on the use of the [OIII]$\lambda$5007  
alone because the ${\rm [OIII]\lambda5007/H\beta}$ ratio may be 
very different from galaxy to galaxy. They use the H$\alpha$ to correct the 
${\rm H\beta}$ index. 

The uncertainties concerning the emission corrections derived from the 
H$\alpha$ line
mainly rest on the assumption of a value for the  H$\alpha$ in absorption.
Denicol{\'o} et al. (\cite{Den05a}) adopt an absorption value  
on the basis of the 
H$\alpha$  EW measured for their sample of stars (mainly K).
Differently, as described in Papers~I and II, we computed 
the emitted flux in the H$\alpha$ line using,  as a reference, the spectrum of 
an elliptical galaxy (NGC1426) lacking emission lines or dust.
This galaxy is located in the low tail of the 
Mg2-$\sigma$ relation. However, we verified that the use of a giant elliptical 
reflects in a difference of 0.03 $\AA$ in the value of the computed H$\beta$
emission. Denicol{\'o} et al. (\cite{Den05a}) allowed for a reasonable range of
H$\alpha$ absorption strengths and determined that the difference in the
emission correction is smaller than 0.1 $\AA$.
In this paper we adopt the direct estimate derived from
the H$\alpha$ emission line for the correction of the H$\beta$ index. 
The adopted H$\beta$ values can be easily recovered from
Papers~I and II, where we provided, together with the final indices in
the Lick system, the H$\beta$ emission estimates derived both from the 
[OIII]$\lambda$5007 and H$\alpha$ lines.

\begin{table}
\tiny{
\begin{tabular}{llccc}
\multicolumn{5}{c}{\bf Table 1 Sample overview} \\
\hline\hline
\multicolumn{1}{c}{Ident}
& \multicolumn{1}{l}{RSA}
& \multicolumn{1}{c}{RC3}
& \multicolumn{1}{c}{V$_{hel}$}
& \multicolumn{1}{c}{$\rho_{xyz}$}  \\
\multicolumn{1}{c}{}
& \multicolumn{1}{c}{}
& \multicolumn{1}{c}{}
& \multicolumn{1}{c}{km~s$^{-1}$}
& \multicolumn{1}{c}{Gal. Mpc$^{-3}$}  \\
 \hline
 & & & & \\
 NGC~128  & S02(8) pec    & S0 pec sp   & 4227 &     \\
 NGC~777  & E1             & E1           & 5040 &      \\
 NGC~1052 & E3/S0        & E4            & 1475  & 0.49\\
 NGC~1209 & E6            & E6:           & 2619 & 0.13 \\
 NGC~1297 & S02/3(0)     & SAB0 pec:   & 1550  & 0.71 \\
 NGC~1366 & E7/S01(7)     & S0 sp       & 1310 & 0.16  \\
 NGC~1380 & S03(7)/Sa     & SA0         & 1844 & 1.54 \\
 NGC~1389 & S01(5)/SB01  & SAB(s)0-:    &  986 & 1.50 \\
 NGC~1407 & E0/S01(0)     & E0           & 1766 & 0.42 \\
 NGC~1426 & E4            & E4            & 1443 & 0.66 \\
 & & & & \\
 NGC~1453 & E0            &    E2         & 3906 &  \\
 NGC~1521 & E3            &    E3         & 4165 &   \\
 NGC~1533 & SB02(2)/SBa & SB0-          &  773 & 0.89     \\
 NGC~1553 & S01/2(5)pec & SA(r)0        & 1280 & 0.97  \\
 NGC~1947 & S03(0) pec    & S0- pec     & 1100  & 0.24 \\
 NGC~2749 & E3            & E3           & 4180 &   \\
 NGC~2911 & S0p or S03(2)& SA(s)0: pec  & 3131  \\
 NGC~2962 & RSB02/Sa      & RSAB(rs)0+ & 2117 & 0.15  \\
 NGC~2974 & E4            & E4           & 1890  & 0.26 \\
 NGC~3136 & E4            & E:           & 1731  & 0.11 \\
 & & & & \\
 NGC~3258 & E1            & E1          & 2778 & 0.72 \\
 NGC~3268 & E2            & E2          & 2818 & 0.69 \\
 NGC~3489 & S03/Sa       & SAB(rs)+   &  693  & 0.39    \\
 NGC~3557 & E3            & E3          & 3038 & 0.28 \\
 NGC~3607 & S03(3)        & SA(s)0:     &  934 & 0.34  \\
 NGC~3818 & E5      & E5      &1701 & 0.20\\
 NGC~3962 & E1            & E1          & 1822 & 0.32 \\
 NGC~4374 & E1      & E1      &1060 & 3.99 \\
 NGC~4552 & S01(0)        & E           &  322 & 2.97 \\
 NGC~4636 & E0/S01(6)    & E0-1        &  937 & 1.33 \\
 & & & & \\   
 NGC~4696 &(E3)     & E+1 pec  &2958 & 0.00 \\
 NGC~4697 & E6      & E6      &1241 & 0.60\\
 NGC~5011 & E2      & E1-2    &3104 & 0.27 \\
 NGC~5044 & E0      & E0      &2704 & 0.38 \\
 NGC~5077 & S01/2(4)      & E3+        & 2764 & 0.23 \\
 NGC~5090 & E2      & E2      &3421 &  \\
 NGC~5193 & S01(0)  & E pec  &3711 &  \\
 NGC~5266 & S03(5) pec & SA0-: & 3074 & 0.35 \\
 NGC~5328 & E4            & E1:          & 4671 &  \\
 NGC~5363 & [S03(5)]       & I0:          & 1138 & 0.28 \\
  & & & & \\  
 NGC~5638 & E1         & E1   & 1676 & 0.79 \\
 NGC~5812 & E0         & E0   & 1930 & 0.19 \\
 NGC~5813 & E1         & E1-2 & 1972 & 0.88 \\
 NGC~5831 & E4         & E3   & 1656 & 0.83 \\
 NGC~5846 & S01(0)        & E0+         & 1709 & 0.84 \\
 NGC~5898 & S02/3(0)     & E0           & 2267& 0.23 \\
 NGC~6721 & E1            & E+:          & 4416 & \\
 NGC~6758 & E2 (merger)& E+:  & 3404 &     \\
 NGC~6776 & E1 pec     & E+pec & 5480 & \\
 NGC~6868 & E3/S02/3(3) & E2           & 2854 & 0.47 \\
  & & & & \\ 
NGC~6875 & S0/a(merger) & SAB(s)0- pec: & 3121 &      \\
NGC~6876 & E3             & E3          & 3836 &  \\
NGC~6958 & R?S01(3)       & E+         & 2652 & 0.12 \\
NGC~7007 & S02/3/a      & SA0-:      & 2954 & 0.14  \\
NGC~7079  & SBa            & SB(s)0     & 2670 & 0.19 \\
NGC~7097 & E4             & E5          & 2404 & 0.26 \\
NGC~7135 & S01 pec       & SA0- pec   & 2718 &  0.32 \\
NGC~7192 & S02(0)         & E+:         & 2904 &  0.28 \\
NGC~7332 & S02/3(8)       & S0 pec sp & 1207 &  0.12  \\
NGC~7377 & S02/3/Sa pec & SA(s)0+  & 3291 &   \\
& & & & \\
IC~1459    & E4             & E          & 1659 & 0.28  \\
IC~2006    & E1             & E          & 1350 & 0.12 \\
IC~3370    & E2 pec         & E2+       & 2934 & 0.20 \\
IC~4296    & E0             & E          & 3762  \\
IC~5063    & S03(3)pec/Sa   & SA(s)0+: & 3402 &   \\
& & & & \\\hline\hline
\end{tabular}}
\label{tab1}
\end{table}

\section{Modelling line-strength indices for simple stellar populations}
 
\subsection{Standard models}

Following the procedure described in Bressan et al. (\cite{Br96}), 
to which we refer for details, we have derived line--strength indices 
for SSPs. Indices are constructed by means of a central band-pass 
and two pseudo-continuum band-passes on either side of the central band.
Molecular bands are expressed in magnitude, while atomic features are
expressed in equivalent width (EW). The definition in EW is:

\begin{equation}
\label{eq1}
I_{EW}= \left(1- F_R/F_C \right) \Delta \lambda,
\end{equation}

\noindent while the definition in magnitude is

\begin{equation}
\label{eq2}
I_{mag}= -2.5 \log \left(F_R/F_C \right), 
\end{equation}

\noindent where $F_R$ and $F_C$ are the 
fluxes in the line and in the pseudo-continuum,
respectively, and $\Delta \lambda$ is the width of the central band. 
The flux $F_C$ is obtained 
by interpolating the fluxes in the blue and red pseudo-continua 
bracketing the line of interest to the central wavelength of the 
absorption band.
We used the passband definitions provided in  Trager et al. 
(\cite{Tra98}) for the original 21 Lick indices, and 
in Worthey \& Ottaviani (\cite{WO97}) for the 
higher-order Balmer lines. The integrated indices 
for SSPs are based on the
Padova library of stellar models  (Bressan et al. \cite{Bre94}) and 
accompanying isochrones  (Bertelli et al. \cite{Ber94}).
The SSPs indices are calculated using the following method.
We derive both 
the pseudo-continuum flux $F_C$ from the library of stellar spectra
used by Bressan et al. (\cite{Bre94}) and implemented by the revision 
of Tantalo et al. (\cite{Tan96}), and the line strength index
$I_{EW}$ or $I_{mag}$ from the fitting functions (FFs) of Worthey et al.
(\cite{W94}) and  Worthey \& Ottaviani (\cite{WO97}),
for each elementary bin 
$\Delta \log \frac{L}{L_{\odot}}$ and $\Delta \log T_{eff}$
of a given  isochrone in the HR diagram. 
The residual flux in the central feature $F_R$ 
is recovered from $F_C$ and $I_{EW}$ ($I_{mag}$) by  
inverting Eqs. (\ref{eq1}) and  (\ref{eq2}).
For an SSP of age T and metallicity Z, 
the fluxes in the line and in the continuum, 
$F_{R,SSP}(T,Z)$  and $F_{C,SSP}(T,Z)$,
are then obtained by integrating  $F_C$  and $F_R$ along the isochrone:

\begin{equation}
\label{eq3}
F_{k,SSP}(T,Z)= \int_{M_L}^{M_U} \phi(M) F_k(M,T,Z) dM,
\end{equation}

\noindent where k denotes the flux in the continuum (C) or in the line (R), 
$\phi(M)$ is the Initial Mass Function (IMF) 
defined between $M_L$ and $M_U$, and $F_k(M,T,Z)$ is the flux 
for a star of mass M, age T, and metallicity Z. 
Once the integrated fluxes, 
$F_{R,SSP}$ and $F_{C,SSP}$, are computed, 
the index definition (Eqs. (\ref{eq1}) and  
(\ref{eq2})) is applied to get the integrated SSP index back.
Our SSP models have been computed adopting
 a Salpeter (\cite{Salp55}) IMF slope between 0.15 and 
120 $M_{\odot}$.

\subsection{$\alpha$/Fe-enhanced models}

To compute the SSP models presented  in Sect.~3.1, we have adopted
stellar isochrones based on solar--scaled abundances and FFs 
calibrated on Milky Way stars. For these reasons the standard 
SSPs reflect the solar element abundance pattern. However, there are
several evidences for supersolar $[\alpha/Fe]$ ratios in early-type
galaxies  (Peletier \cite{Pele89};  Worthey et al. \cite{WFG92}; Davies
et al. \cite{DSP93}; Carollo \& Danziger \cite{CD94}; Bender \& Paquet
\cite{BP95}; Fisher et al. \cite{Fish95}; Mehlert et al. \cite{Mehl98};
J\o rgensen \cite{J99}; Kuntschner  \cite{Ku00}; Longhetti et al.
\cite{L00} and others).

The major effect of non--solar abundance patterns is on the stellar
atmospheres. For a given effective temperature, gravity, and total
metallicity, the stellar spectrum, and thus the line-strength indices
that are measured in it, depend on the specific ratios between the
element abundances. 
To account for this effect, we apply a correction to the index derived
from the FFs using tabulated index responses to element abundance
variations. The responses are derived from model atmospheres and
synthetic spectra (Tripicco \& Bell \cite{TB95} (hereafter TB95); Korn
et al. \cite{K05} (hereafter K05), Munari et al. \cite{Mu05} (hereafter
Mu05)). A second, less important effect is on the evolution of the star
and on the stellar opacities. A fully self-consistent $\alpha$-enhanced
SSP model should in principle use $\alpha$/Fe--enhanced stellar
evolutionary tracks (Weiss et al.~\cite{WPM95}; Salaris \&
Weiss~\cite{SW98}; Van den Berg et al.~\cite{Van00}; Salasnich et
al.~\cite{SGWC00}). In the literature, several $\alpha$-enhanced SSP models 
have been presented. In all models a correction is applied to the index 
value to 
account for the impact of the $\alpha$/Fe--enhancement on the stellar
atmosphere (Weiss et al. \cite{WPM95};  Tantalo, Chiosi \& Bressan
\cite{TCB98}; Trager et al. \cite{TFWG00}, Thomas et al. \cite{Thom03}
(hereafter TMB03), Thomas et al. \cite{TMK04}; Tantalo \& Chiosi
\cite{TC04b}; Tantalo et al. \cite{T04}; Korn et al. \cite{K05}). SSPs
adopting $\alpha$/Fe--enhanced stellar evolutionary tracks have been
produced as well (Thomas \& Maraston \cite{TM03}; Tantalo \& Chiosi
\cite{TC04a}), but current enhanced models seem to overestimate the
blueing of the stellar evolutionary tracks. In this section we present
new SSP models that are based on solar--scaled isochrones (Bertelli et
al. \cite{Ber94}) and where only the direct effect of the enhancement
on the stellar atmosphere is taken into account. 
The models are available to the public at http://www.inaoep.mx/$\sim$abressan
or www.stsci.edu/$\sim$annibali/.

\subsection{$\alpha$-enhanced chemical compositions}

Following the procedure adopted by TMB03, we assign elements to three groups:
the {\it enhanced} group ($\alpha$--elements), the {\it depressed} group 
(Fe-peak elements), and the {\it fixed} group (elements left unchanged). 
Taking as a reference model a mixture of total metallicity Z
and solar partitions, a new mixture with the same global metallicity Z, 
but with
supersolar [$\alpha$/Fe] ratio is produced by increasing the abundance of
the $\alpha$-group and decreasing that of the Fe-group in such a way that
the two mass fraction variations balance.
  
If $X_{\alpha,\odot}$ and $X_{Fe,\odot}$ are the total mass fractions of the two
groups for the solar mixture, and   $X_{\alpha}$ and $X_{Fe}$ are the mass
fractions in the new mixture, the degree of enhancement is expressed by the  quantity 

\begin{equation}
\label{eq4}
\left[ \alpha/Fe \right]= \log \frac{X_{\alpha}}{X_{Fe}} -
\log \frac{X_{\alpha,\odot}}{X_{Fe,\odot}}.  
\end{equation}

\noindent The quantities $X_{\alpha,\odot}$ and $X_{Fe,\odot}$ are computed by summing up the
mass fractions of the elements within the $\alpha$ and Fe groups in 
the solar--scaled
mixture, and thus depend on how elements are allocated within the groups.

For an arbitrary mixture, 
the i-th element mass fraction is:

\begin{equation}
\label{eq5}
X_i= f_i \frac{Z}{Z_{\odot}}\Xisol,
\end{equation}

\noindent where $f_i=1$ for solar-scaled abundances, $f_i>1$  
if the element is enhanced with 
respect to the solar-scaled composition, and $f_i<1$ if 
the element is depressed. Once the element allocation within the three
groups and the global enhancement $[\alpha/Fe]$ has been defined, we
derive the quantities $f_i$. Since in our assumption all the 
elements within a group are enhanced/depressed by the same 
amount, it is sufficient to compute  $f_{\alpha}$ and $f_{Fe}$.

Let us call  \zbase \  and \afebase \
the total metallicity and the enhancement of the starting base model,
and  \znew \ and \afenew \ the same quantities in the new mixture.
The enhanced, depressed, and fixed group mass fractions in the new mixture ($n$) are related 
to the  base model ($b$) quantities according to:

\begin{eqnarray}
\label{eq6}
{\rm X_{\alpha,n}} & = & 
{\rm f_{\alpha} \frac{Z_n}{Z_b}  X_{\alpha,b}}, \nonumber \\
{\rm X_{Fe,n}}&=&{\rm f_{Fe} \frac{Z_n}{Z_b}  X_{Fe,b}}, \nonumber \\
{\rm X_{0,n}}&=&{\rm \frac{Z_n}{Z_b}  X_{0,b}} 
\end{eqnarray}

\noindent The quantities $f_{\alpha}$ and $f_{Fe}$ are derived by solving the 
set of equations:

\begin{eqnarray}
\label{eq7}
{\rm Z_n} &= & {\rm X_{\alpha,n}  +  X_{Fe,n} + X_{0,n}} = \nonumber \\ 
& = & {\rm f_{\alpha} \frac{Z_n}{Z_b}  X_{\alpha,b} + f_{Fe}\frac{Z_n}{Z_b} X_{Fe,b}} + \nonumber \\
&& + {\rm \frac{Z_n}{Z_b}  X_{0,b},}  \nonumber \\
{\rm {[\alpha /Fe]}_n - {[\alpha /Fe]}_b} & = & 
{\rm \log \left( \frac{X_{\alpha,n}}{X_{Fe,n}} \right) -  
\log \left(\frac{X_{\alpha,\odot}}{X_{Fe,\odot}} \right)} \nonumber \\
&& - {\rm \log \left(\frac{X_{\alpha,b}}{X_{Fe,b}} \right) +
\log \left(\frac{X_{\alpha,\odot}}{X_{Fe,\odot}} \right)}  \nonumber \\  
& =& {\rm \log \left(\frac{X_{\alpha,n}}{X_{Fe,n}} \right) -
\log \left(\frac{X_{\alpha,b}}{X_{Fe,b}} \right)}  \nonumber \\
&=&{\rm \log \frac{f_{\alpha}}{ f_{Fe}}}
\end{eqnarray}

\noindent The solutions ${\rm f_{\alpha}}$ and ${\rm f_{Fe}}$ to (\ref{eq6}) are:

\begin{eqnarray}
\label{eq8}
{\rm f_{Fe}} & = & 
{\rm \frac{X_{\alpha,b} + X_{Fe,b}}{ X_{\alpha,b} \ 10^{([\alpha /Fe]_{n} - [\alpha /Fe]_{b})} + X_{Fe,b} }}, \nonumber \\
{\rm f_{\alpha}} & = & {\rm f_{Fe}  10^{([\alpha /Fe]_n - [\alpha /Fe]_b)}} 
\end{eqnarray}

\noindent In the case in which the base model corresponds to the solar mixture
(i.e., ${\rm X_{\alpha,b}= X_{\alpha,\odot}}$ and 
${\rm X_{Fe,b}= X_{Fe,\odot}}$), 
the quantities ${\rm f_{\alpha}}$ and ${\rm f_{Fe}}$ are given by:

\begin{eqnarray}
\label{eq9}
{\rm f_{Fe}} & = & {\rm \frac{X_{\alpha,\odot} + X_{Fe,\odot}}{10^{[\alpha /Fe]_{n}}  X_{\alpha,\odot} + X_{Fe,\odot} }}, \nonumber \\
{\rm f_{\alpha}} & = & {\rm f_{Fe}  10^{[\alpha /Fe]_n}}  
\end{eqnarray}

\noindent and the mass fractions in the new mixture are:

\begin{eqnarray}
\label{eq10}
{\rm X_{\alpha,n}} & = & 
{\rm f_{\alpha} \frac{Z_n}{Z_{\odot}} X_{\alpha,\odot}}, 
\nonumber \\ 
{\rm X_{Fe,n}} & = & 
{\rm f_{Fe} \frac{Z_n}{Z_{\odot}}  X_{Fe,\odot}}, \nonumber \\
{\rm X_{0,n}} & = & {\rm \frac{Z_n}{Z_{\odot}}  X_{0,\odot}} 
\end{eqnarray}

\noindent Adopting the solar abundances  
of Grevesse \& Sauval (\cite{GS98}), and assigning the elements 
N, O, Ne, Na, Mg, Si, S, Ca, and Ti to the {\it enhanced} group 
and Cr, Mn, Fe, Co, Ni, Cu, and Zn to the {\it depressed} group,
we get $X_{\alpha,\odot}=1.251 \times 10^{-2}$ and 
$X_{Fe,\odot}=1.4 \times 10^{-3}$.

\subsection{Specific response functions}

Specific response functions allow to isolate the effect on line-strength indices
due to variations of one element at once. The first to quantify this effect were 
TB95 who computed model atmospheres 
and synthetic stellar spectra along a 5--Gyr--old 
isochrone with solar metallicity. 
The model atmospheres were computed for three points along the isochrone 
representative of the evolutionary phases of dwarfs ($T_{eff}=4575$ K, $\log g=4.6$), turn-off (6200 K, 4.1), and giants (4255 K, 1.9). On each synthetic spectra, TB95 measured the absolute values $I_0$ for the original 21 Lick indices. Then they computed new model atmospheres, doubling the abundances of  the dominant elements C, N, O, Mg, Fe, Ca, Na, Si, Cr, and Ti  ($\Delta [X_i]=0.3$) in turn, and derived the index changes $\Delta I$.
The index response for an increase of $+$0.3 dex of the i-th element 
abundance is defined as:

\begin{eqnarray}
\label{eq11}
R_{0.3}(i)&=&\frac{1}{I_{0}} \frac{\partial I}{\partial [X_i]} 0.3=  \frac{\Delta I_{TB95}}{I_{0}}, 
\end{eqnarray}

\noindent where $I_{0}$ and $\Delta I_{TB95}$ are, respectively, the index computed by TB95 for solar composition and the index change measured for a $+$0.3 dex abundance increase.
Recently K05 have extended the work of TB95 to a wide range of metallicities and have included in their analysis the higher
order Balmer lines (H$\gamma$ and H$\delta$) as well.
We have adopted these new responses to compute our SSP models with 
$\alpha$-enhanced composition.

As the index responses will be used in the following to correct 
solar--scaled SSP models based on FFs, 
we have first of all compared the solar--scaled indices 
computed by K05 for their model atmospheres with the FFs of
Worthey et al. (\cite{W94}). We have limited the comparison to metallicities 
$[Z/H] \ge 0$, as at lower metallicities the FFs, 
calibrated on Milky Way stars, 
reflect super solar [$\alpha$/Fe] ratios. We observe that while in some cases there is a good match between FFs and model atmospheres, in other cases large deviations are present (see Fig. 1).

\begin{figure*}
\center{
\resizebox{13.5cm}{!}{ \psfig{figure=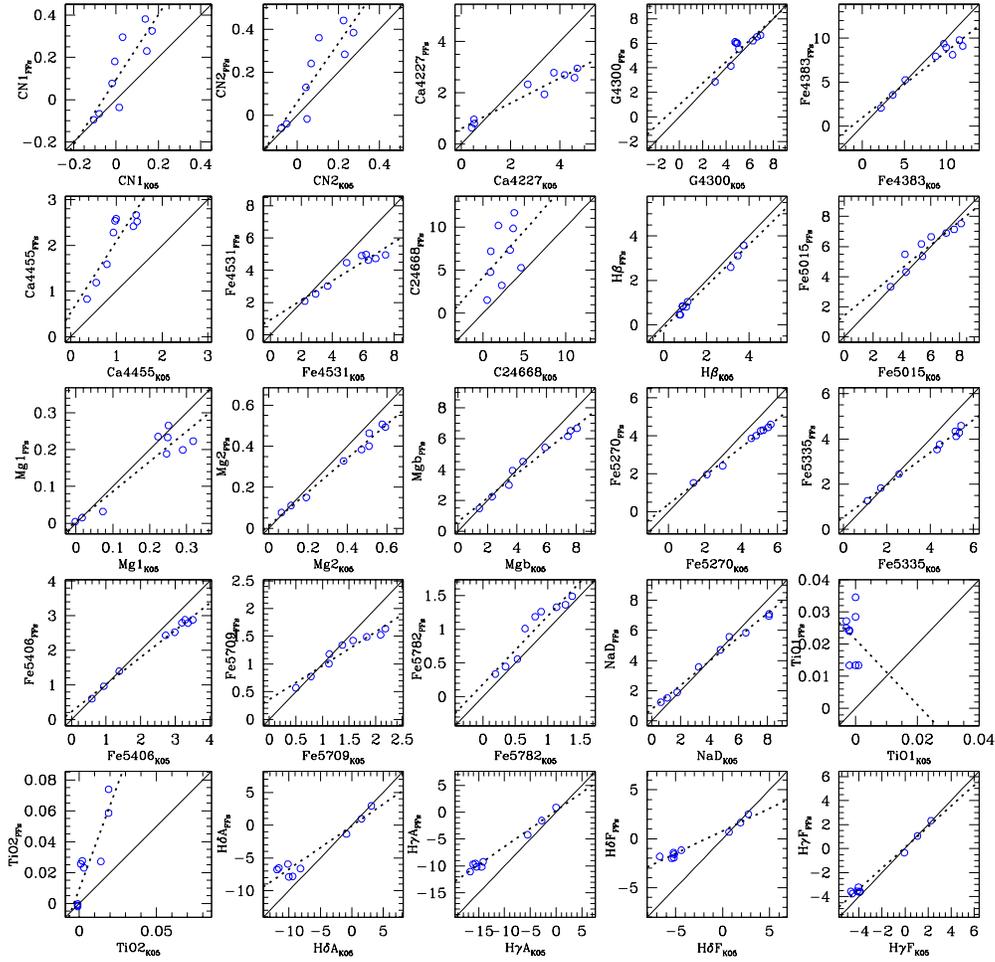,width=13cm,clip=} }}
\caption{Comparison between solar--scaled indices obtained 
by Korn et al. (\cite{K05}) for their model atmosphere with fitting functions.
The comparison is limited to metallicities [Z/H] $\geq$ 0. 
The solid line is the
one--to--one relation, while the dashed line is the performed fit to the
data.}
\label{fig1}
\end{figure*}

Mismatches between theoretical models and observations may be due both to
uncertainties in the input physics of the model and to residual effects in the
calibration of the model into the observed system (Chavez, Malagnini \&
Morossi \cite{CVM96}). We then wonder whether model indices should be
transformed into the Lick system, in analogy with the transformation usually
performed in the literature on measured indices through observation of standard
stars in common with the Lick library (Worthey \& Ottaviani \cite{WO97}). We
derive the linear transformation (${\rm I_{FF}=a \times I_{atmo} + b}$), which
is needed to match the atmospheric indices with the FFs. The coefficients $a$
and $b$ of the transformation are given in Table~2.
In Sect.~3.5 the index correction will be computed adopting the K05
responses revised in the following way:

\begin{table}
\begin{center}
{\bf Table 2}
\begin{tabular}{cccccc}
\noalign{\smallskip}\hline \noalign{\smallskip}
\hline\\
      Index  &   a  & b     &         Index  &  a  & b \\
\hline \\
       CN1  &    1.536      &    0.0939    &     Fe5270    &    0.750   &       0.3796\\
       CN2  &    1.373      &    0.0586    &     Fe5335    &    0.717   &       0.5359\\
    Ca4227  &    0.491      &    0.6089    &     Fe5406    &    0.785   &       0.2249\\
     G4300  &    0.880      &    0.9856    &     Fe5709    &    0.610   &       0.3613\\
    Fe4383  &    0.757      &    0.9480    &     Fe5782    &    1.008   &       0.1879\\
    Ca4455  &    1.564      &    0.5247    &        NaD    &    0.792   &       0.7805\\
    Fe4531  &    0.600      &    0.9362    &       TiO1    &     --     &         --   \\
    C24668  &    1.080      &    4.1530    &       TiO2    &    2.722   &       0.0092\\
    H$\beta$  &  0.936      &   -0.1275    &    H$\delta$  &    0.663   &      -0.1094 \\
    Fe5015  &    0.783      &    1.4040    &    H$\gamma$  &    0.669   &       0.2911 \\
       Mg1  &    0.817      &    0.0040    &    H$\delta$  &    0.461   &       0.8014 \\
       Mg2  &    0.822      &    0.0120    &    H$\gamma$  &    0.864   &       0.0899 \\
       Mgb  &    0.776      &    0.6049    &               &            &            \\ 
\hline \\
\end{tabular}
\end{center}
{Notes: a and b are the coefficients of the linear transformation 
${\rm EW_{Lick}= a \times EW_{atmo} + b}$ adopted to transform the indices derived from the model atmospheres of Korn et al. (\cite{K05}) 
into the Lick system}.
\label{tab2}
\end{table}

\begin{eqnarray}
\label{eq12}
{\rm \frac{\Delta I_{(\alpha, Lick)}}{I_{(\alpha=0, Lick)}}} 
&=& {\rm \frac{ (a \times I_{\alpha} + b) -
(a \times I_0 + b) }{a \times I_0 + b}}= \nonumber \\
&=& {\rm \frac{a \times (I_{\alpha} - I_0)}{a \times I_0 + b}}, 
\end{eqnarray}

\noindent where ${\rm I_0}$ is the index from the solar scaled atmospheres,
${\rm I_{\alpha}}$ is the index from the $\alpha$-enhanced atmospheres,
and ${\rm I_{(\alpha=0, Lick)}}$ and ${\rm \Delta I_{(\alpha, Lick)}}$ are
the solar--scaled index and the index change transformed into
the Lick system, respectively.

An important point to be assessed concerns the metallicity 
dependence of the indices, and more specifically the dependence
of the fractional response $\Delta I/I$.
We recall here that the behavior of the equivalent width of a line
with respect to the abundance of an element is described by the curve
of growth, and characterized by three different regimes: a linear part,
where ${\rm EW \propto N}$, for a small number of absorbers N
(proportional to the metallicity Z); a saturated part,
where ${\rm EW \propto \sqrt{\ln N}}$, in the Doppler 
regime of saturated lines;
and a collisionally broadened regime, described 
by  ${\rm EW \propto \sqrt{N}}$, for large N. 

To assess in which regime the Lick indices are, in a first
attempt we have analyzed the behavior of the K05 indices with global
metallicity. In principle, such analysis would require keeping the
temperature and the gravity fixed and only let the metallicity varying,
while the K05 indices 
were computed for different  (${\rm T_{eff}, \log g}$) pairs at
different metallicities. Nevertheless, ${\rm T_{eff}}$ and ${\rm \log g}$
variations are small within each phase (see Table~2 in K05). 
By analyzing the index behavior, we  observe that the Lick indices are
not linearly related to Z. A viable explanation is that, at the Lick-IDS
system resolution, the contribution to the index absorption strength
from weak lines is negligible, as weak lines are likely to be flattened to
the continuum or blended with stronger lines. In reverse, the 
major contribution to the index equivalent width 
derives from strong saturated absorption lines, 
for which it would be more appropriate to consider the 
flat part or the square root part of the curve of growth.

\begin{figure*}
\center{
\resizebox{13.5cm}{!}{ \psfig{figure=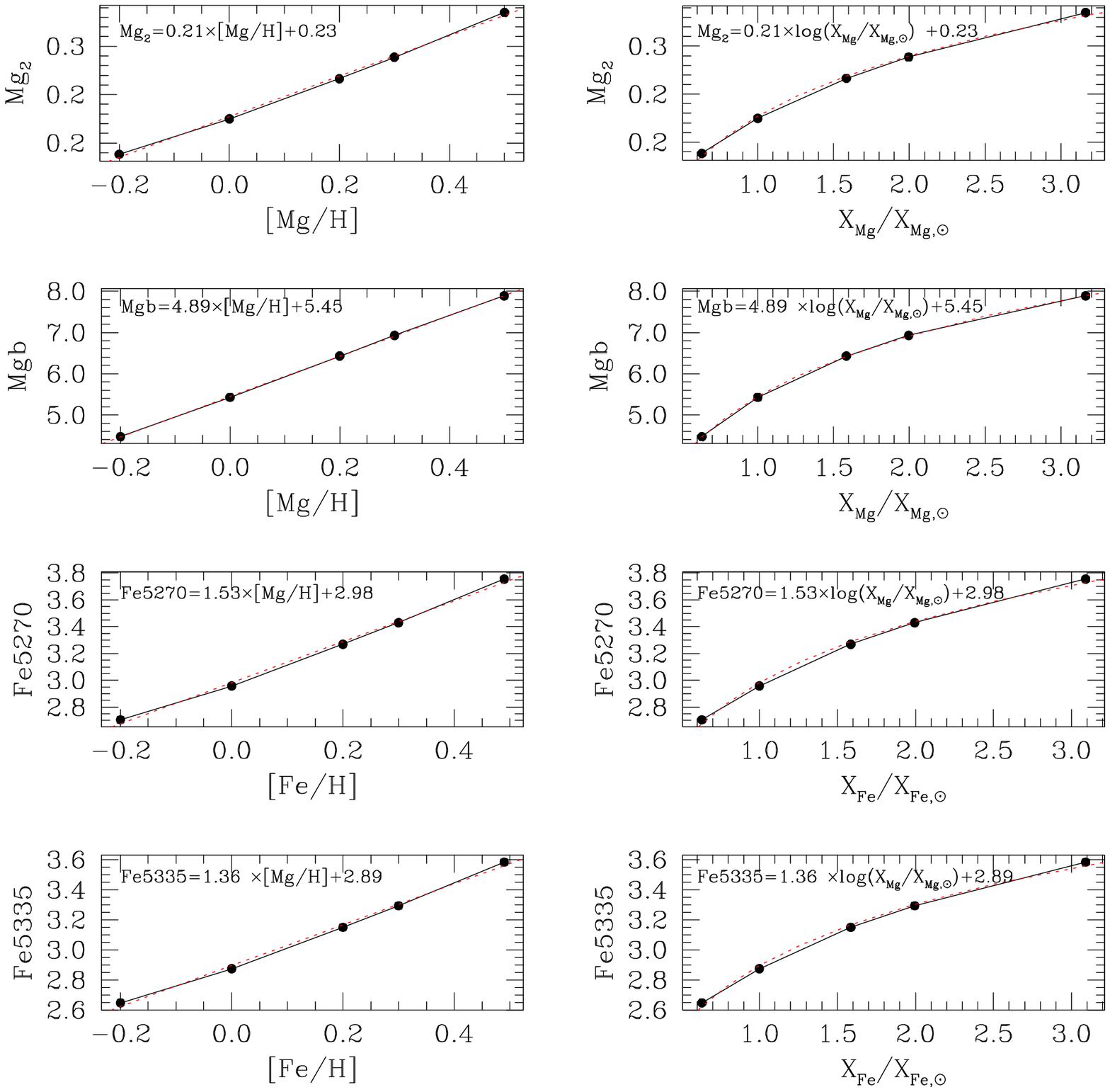,width=13cm,clip=} }}
\caption{Mg$_2$, Mgb, Fe5335, and Fe5270 indices as functions of the element
abundance. In the left panels the index is plotted as a 
function of the logarithmic 
element abundance [X/H], while in the right panels as a function of the
linear abundance ${\rm X/X_{\odot}}$. 
Full circles are the model atmospheres computed with ATLAS12, and the solid
line connects the models, while the dotted line is the performed fit,
which is also given in the box.}
\label{fig2}
\end{figure*}

This evidence is strongly supported by preliminary results
that we have obtained with the last version of the ATLAS code
(ATLAS12, Kurucz \cite{Ku93b}), which is based on the opacity-sampling
(OS) method. Following TB95 and K05, we have computed  
 model atmospheres and synthetic spectra 
for a model with solar abundance ratios and 
for models in which the abundances of the elements  C, N, O, Mg, and Fe
have been changed by different amounts ([X/H]$=-0.2, +0.2, +0.3, +0.5$).
In Figure~\ref{fig2} we plot the index as a function of the logarithmic
(left panels) and linear (right panels) abundance of the index-dominating 
element for the indices (Mg$_2$, Mgb, Fe5335, Fe5270).
For more details we refer to Annibali (\cite{An05}).
We observe that the behavior of the indices as a function of the 
element abundances is very well described by 
a functional form that is linear in [X/H]
(${\rm I=a \times [X/H] + b}$) or, equivalently, logarithmic in X 
(${\rm I=a \times \log X/X_{\odot}+ b}$).

\subsection{The $\Delta I$ correction}

As explained in the previous section,
specific response functions 
quantify the index change due to the deviation 
of one element from the solar composition. 
The total index correction for a mixture in which several 
elements depart from the solar abundance path is derived
through a linear combination of the specific responses,
i.e., assuming that the total effect on the index can be expressed
as the sum of the single effects due to the variation of one
element at once. 
In the literature, different expressions to derive the index
change on the basis of the same specific response functions have been 
proposed. 

Using the TB95 responses, Trager et al. (\cite{TFWG00})  
propose the formula

\begin{eqnarray}
{\rm \frac{\Delta I}{I_0}} &=& {\rm \Big\{ \prod _i [1 + R_{0.3}(X_i)]^{[X_i/H]/0.3} \Big\} -1}
\label{eq13}
\end{eqnarray}

\noindent for the fractional index variation, 
where ${\rm R_{0.3}(X_i)=\frac{1}{I_0} \frac{\partial I}{\partial [X_i]} 0.3}$
is the fractional index change computed by TB95 for 
an increase of the i-th element abundance of 0.3 dex.
The computations of TB95 are limited to the solar metallicity.
Trager et al. (\cite{TFWG00})  propose to extend the  index correction to 
non--solar cases with the reasonable assumption that 
the fractional index change is independent of metallicity.

Tantalo \& Chiosi (\cite{TC04a}) adopt the same functional form to
compute the index variations. TMB03 propose a different approach 
starting from the assumption that the line strength index is 
a linear function of the element abundance. 
TMB03 assume that Lick spectral features are in the linear regime of the
curve of growth. They thus assume that  
${\rm I \propto X_i}$ and consequently ${\rm \ln I \propto [X_i]}$ (note,
however, that the latter implies ${\rm I \propto X_i^{\alpha}}$),  
and then express the index through the Taylor series

\begin{eqnarray}
{\rm \ln I} &=& {\rm \ln I_0 + \sum_{i=1}^n \frac{\partial \ln I}{\partial [X_i]} \Delta [X_i]} 
\label{eq14}
\end{eqnarray}

\noindent However, as discussed in Sect.~3.4, at the Lick resolution 
the major contribution to the index line-strengths arises from strong saturated absorption
lines, and a logarithmic functional form for the index behavior 
with element abundance may be more appropriate than a linear one.

\noindent Then we express the index I through the Taylor expansion:

\begin{eqnarray}
{\rm I} & = & {\rm I_0 + \sum_{i=1}^n  \left[ \frac{\partial I}{\partial [X_i]} \right] _0 \Delta [X_i]}. 
\label{eq15}
\end{eqnarray}

\noindent The fractional index change is

\begin{eqnarray}
{\rm \frac{\Delta I}{I_0}} & = & {\rm \sum_{i=1}^n \frac{1}{I_0} 
\left[ \frac{\partial I}{\partial [X_i]} \right] _0 \Delta [X_i]}. 
\label{eq16}
\end{eqnarray}

The computations of K05 allow us to derive
the fractional index change ${\rm \frac{\Delta I}{I_0}}$ 
at different metallicities.
We assume that we start from a mixture where the total metallicity
is Z and the element abundances are

\begin{eqnarray}
{\rm X_{i,0}} & = & {\rm f_b \frac{Z}{Z_{\odot}} X_{i,\odot}},
\label{eq17}
\end{eqnarray}

\noindent where ${\rm f_b}$ accounts for the [$\alpha$/Fe] bias of the stellar
libraries (see Sect.~3.7), and it is the enhancement/depression 
factor with respect to the solar-scaled abundance for that metallicity
Z. If now we perturb the abundance of the i-th element, the new mass
fraction is given by

\begin{eqnarray}
{\rm X_{i}} &=& {\rm f_i X_{i,b}=f_i f_b \frac{Z}{Z_{\odot}} X_{i,\odot}},
\label{eq18}
\end{eqnarray}

\noindent and the logarithmic abundance change is

\begin{eqnarray}
{\rm \Delta [X_i]} = {\rm \log \frac{X_{i}}{X_{i,0}}= \log f_i}.
\label{eq19}
\end{eqnarray}

\noindent For a given Z, the expression for the fractional index change becomes

\begin{eqnarray}
{\rm \frac{\Delta I}{I_0} |_Z}& = & 
{\rm \sum_{i=1}^n \frac{K_{0.3,Z}}{0.3} \log f_i },
\label{eq20}
\end{eqnarray}

\noindent where  ${\rm K_{0.3,Z}}$ are the new response functions 
${\rm \Delta I / I_0}$ 
computed by K05 
for an increase of the element abundance of +0.3 dex. 

\begin{figure*}
\resizebox{17cm}{!}{ \psfig{figure=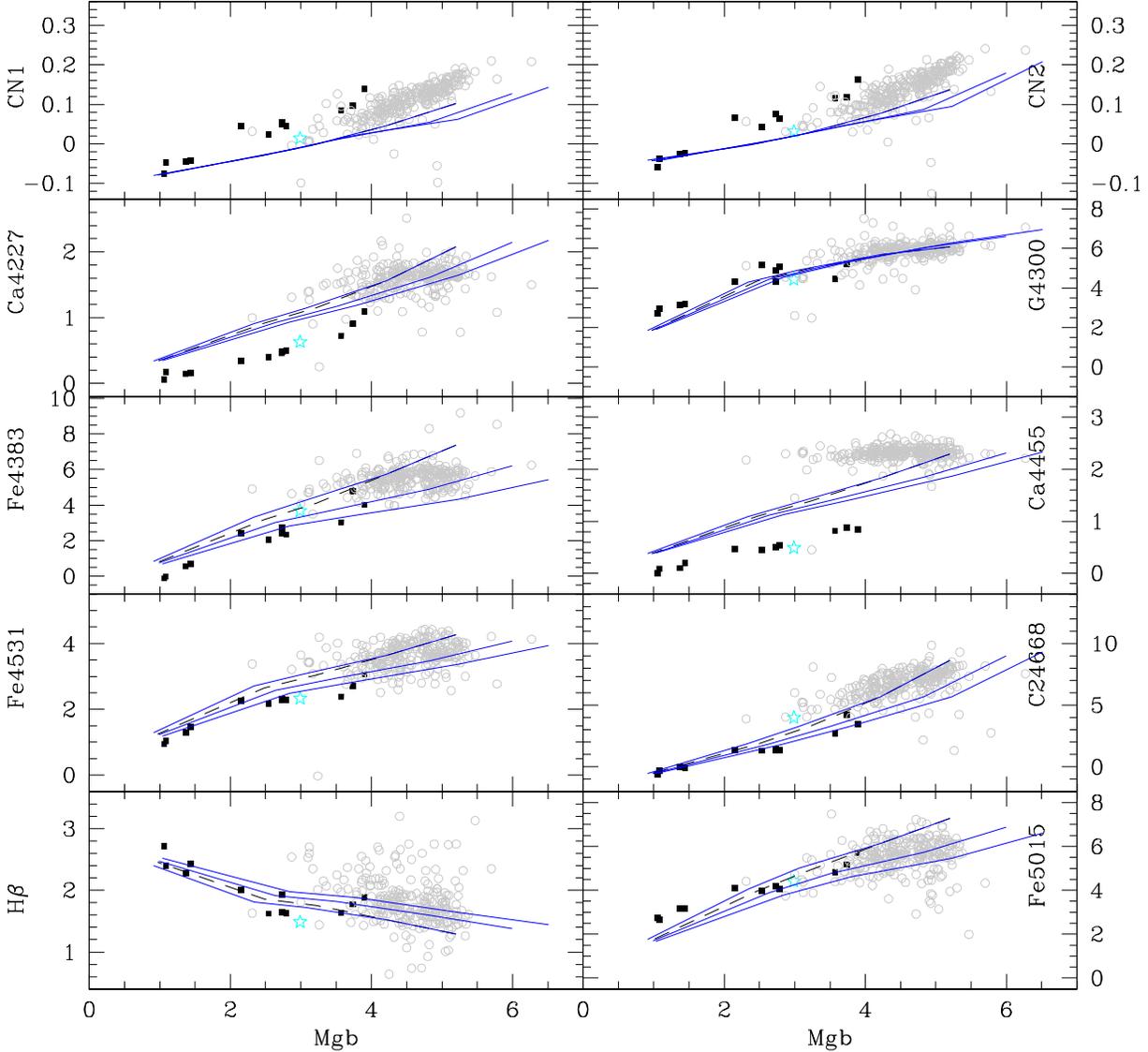,width=13cm,clip=} }
\caption{Mgb index versus the other 24 Lick indices for a fixed age of 
12 Gyr and metallicities in the range 0.0004$\leq$Z$\leq$ 0.05.
The metallicity increases for increasing Mgb values.
The long dashed line is the base model.
The solid line represents the $\alpha$-enhanced models computed for
different [$\alpha$/Fe] ratios (0, 0.3, 0.5), assuming
that the index behavior is logarithmic with element abundance.
Models with the larger Mgb index correspond to larger [$\alpha$/Fe] ratios.
Full squares and open stars are 
the globular clusters data and the integrated bulge light, respectively,
in Puzia et al. (\cite{Puzia2002}). Open circles are our data sample
(Papers~I $+$ II).}
\label{fig3} 
\end{figure*}
\begin{figure*}
\resizebox{17cm}{!}{ \psfig{figure=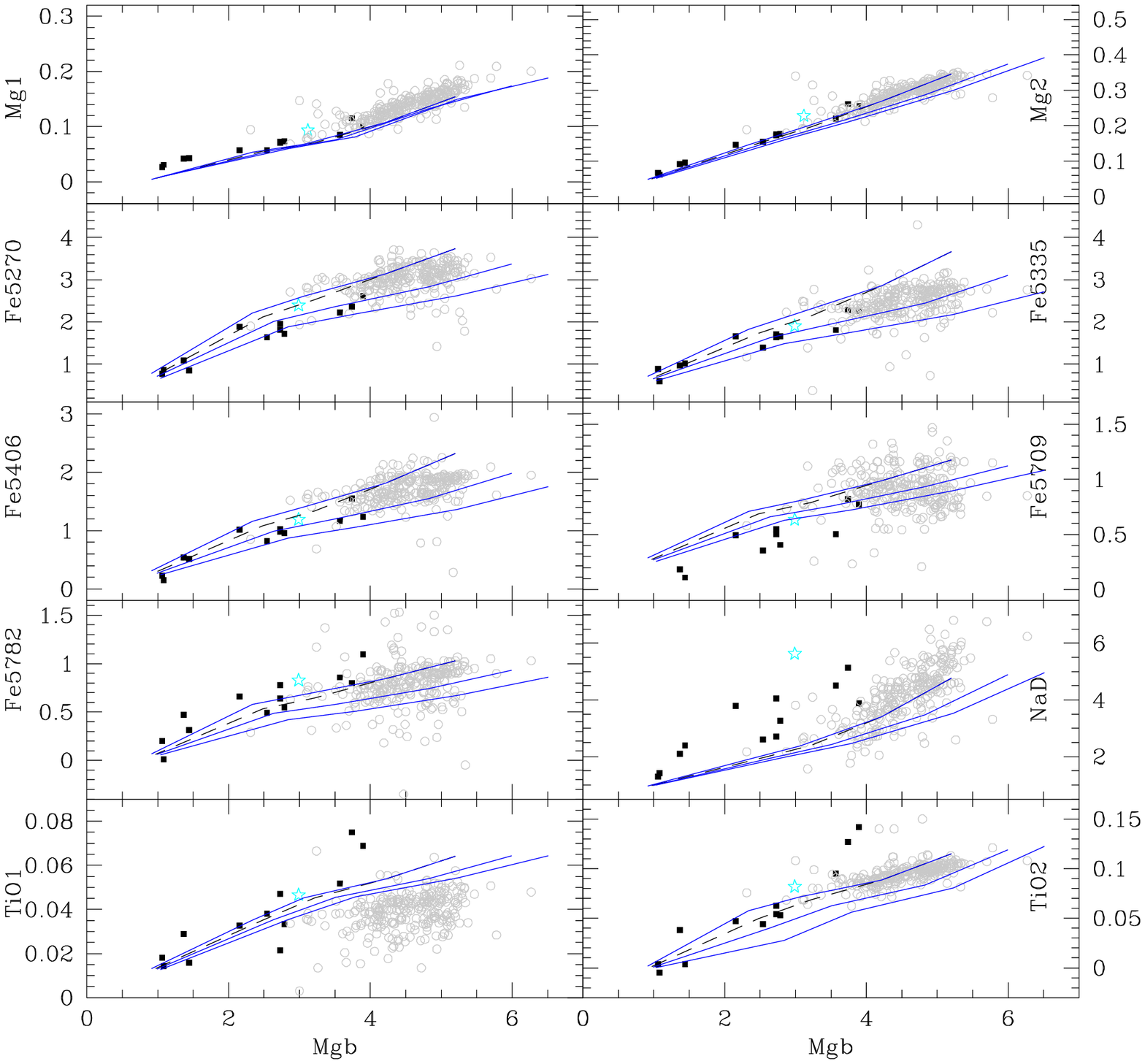,width=13cm,clip=} }
\caption{Same as Fig. 3.}
\label{fig4}
\end{figure*}
\begin{figure*}
\resizebox{17cm}{!}{ \psfig{figure=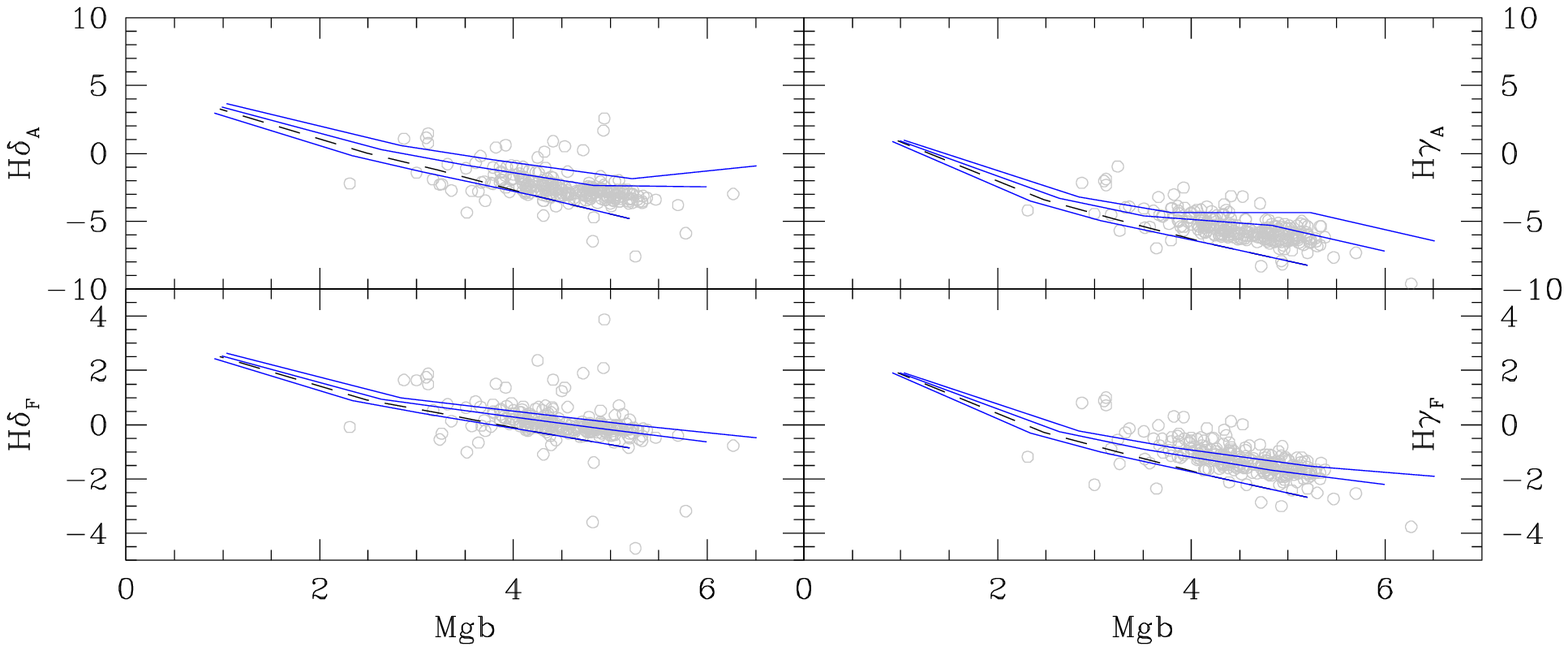,width=13cm,clip=} }
\caption{Same as Fig. 3.}
\label{fig5}
\end{figure*}

\subsection{The final SSP model}

With the functional dependence 
of $\Delta I /I$ on
element abundance described by Eq. (\ref{eq20}), 
the new SSP indices are computed following the
same procedure adopted by TMB03.

We split the basic SSP model into three evolutionary phases, dwarfs (D),
turnoff (TO), and giants (G). Gravity determines the separation between
dwarfs (${\rm \log g > 3.5}$) and giants (${\rm \log g <3.5}$).
Following TMB03, we used the fixed temperature of 5000 K to separate the
turn-off region from the cool dwarfs on the main sequence, independently
of age and metallicity. TMB03 have also demonstrated that the impact of
this choice on the final SSP model is small. The Lick indices of the
base model are computed for each evolutionary phase separately; this is
done by integrating the continua and line fluxes along each phase, and
finally deriving the indices  ${\rm I_D, I_{TO}}$, 
and ${\rm I_G}$ from the  continua  
${\rm F_{C,D}, F_{C,TO}}$, and  ${\rm F_{C,G}}$ and the line fluxes 
${\rm F_{R,D}, F_{R,TO}}$, and ${\rm F_{R,G}}$ of the three phases. 
At this point
the index is corrected for each phase, 
adopting the K05 responses in the appropriate phase and 
computing the ${\rm \Delta I}$ as described in Sect.~3.5.  
Once the new corrected indices 
${\rm I_{D}^*, I_{TO}^*}$, and ${\rm I_{G}^*}$ are computed, the final
$\alpha$-enhanced SSP index is obtained through two different
expressions, depending on if the index is defined as EW or mag.

\noindent{{EW:}}

\begin{eqnarray}
\label{eq21}
{\rm I_{SSP}} &=& {\rm \left[1 -\frac{F_{R,D}+ F_{R,TO}+ F_{R,G}} 
{F_{C,D} +  F_{C,TO} +  F_{C,G}} \right] \Delta \lambda }= \nonumber \\
&=& \left[1  -\frac{\left(1-\frac{I_D^*}{\Delta \lambda} \right) F_{C,D} + 
\left(1-\frac{I_{TO}^*}{\Delta \lambda} \right) F_{C,TO}}{ F_{C,D} +  F_{C,TO} +  F_{C,G}} \right] 
\Delta \lambda + \nonumber \\
&-& \left[ \frac{\left(1-\frac{I_G^*}{\Delta \lambda} \right) F_{C,G}} 
{F_{C,D} +  F_{C,TO} +  F_{C,G}} \right] \Delta \lambda   = \nonumber \\
&=& {\rm \frac{I_D^* \times F_{C,D}+ I_{TO}^* \times F_{C,TO}  + 
I_G^* \times F_{C,G}}{ F_{C,D} +  F_{C,TO} +  F_{C,G}}},
\end{eqnarray}

\noindent {{Magnitude:}} 

\begin{eqnarray}
\label{eq22}
{\rm I_{SSP}} &=& {\rm -2.5 \log  \left(\frac{F_{R,D}+ F_{R,TO}+ F_{R,G}} {F_{C,D} +  F_{C,TO} +  F_{C,G}} \right)} = \nonumber \\
=& 1& -{\rm 2.5 \log \frac{10^{I_D^*} F_{C,D} + 10^{I_{TO}^*} F_{C,TO}+ 10^{I_G^*} F_{C,G} }{ F_{C,D} +  F_{C,TO} +  F_{C,G}}},
\end{eqnarray}

\subsection{The $\alpha$/Fe bias of stellar libraries}

As discussed by TMB03, an additional complication in the construction of
the $\alpha$-enhanced SSP models is due to the fact that the fitting
functions, on which the standard models are based, are constructed from
empirical stellar libraries that reflect the chemical enrichment  
history of the Milky Way. More specifically, metal poor halo stars 
([Fe/H]$<-1$), which formed at early epochs in a short timescale, are
characterized by supersolar [$\alpha$/Fe] ratios ($\sim 0.3$). 
Metal rich disk stars, which formed on longer timescales and
were enriched by Type Ia supernovae, instead  
exhibit an [$\alpha$/Fe] trend that decreases from 0.3 to solar for
increasing metallicity (Edvardsson et al. \cite{Edv93}; Fuhrmann \cite{Fu98}). The
result is that the base models constructed on the 
FFs do not have solar abundance ratios at every metallicity, but reflect
the $\alpha$/Fe bias of the FFs at subsolar metallicity.
In Table~3, we show the [$\alpha$/Fe] values adopted for the bias, which
are obtained by interpolating the values given in Table~3 of TMB03 to
our  metallicity grid. 
For a model of given metallicity and total [$\alpha$/Fe],
the ${\rm [\alpha/Fe]_b}$ of the bias is the starting point
for the computation of the $f_i$ factors according to Eqs.
\ref{eq8} and \ref{eq9}.
Summarizing, we have computed new $\alpha$-enhanced SSP models adopting 
the K05 specific response functions. For comparison with TMB03, we
present the results for models computed for a mixture in which the
enhanced elements are N, O, Ne, Na, Mg, Si, S, Ca, and Ti, while the
depressed ones are Cr, Mn, Fe, Co, Ni, Cu, and Zn. 

\begin{table}
{\bf Table 3. The $\alpha$/Fe bias in the Milky Way}.
\begin{center}
\begin{tabular}{cccccc}
\hline \hline\\
Z   & 0.0004 & 0.004 & 0.008 & 0.02 & 0.05 \\
\\
${\rm [\alpha/Fe]}$  & 0.22 & 0.14 & 0.1 & 0. & 0. \\
\\
\hline
\end{tabular}
\end{center}
\label{tab3}
\end{table}

We plot the Mgb index versus the other 24 
Lick indices for a fixed age of 12 Gyr
and total metallicities in the range 0.0004$\leq$Z$\leq$ 0.05 
in Figs. 3, 4, and 5.
The metallicity increases for increasing Mgb values.
The $\alpha$-enhanced models represented are computed for
different [$\alpha$/Fe] ratios (0, 0.3, 0.5).
Larger [$\alpha$/Fe] ratios  correspond to larger Mgb indices.
In the same figure we have plotted the globular clusters data 
(Puzia et al. \cite{Puzia2002}), the integrated bulge light
(Puzia et al. \cite{Puzia2002}), and the 65 elliptical 
galaxies of the sample of Papers~I $+$ II.
The plot is analogous to that presented in Fig.~2 of TMB03
and Fig.~3 of K05.

\subsection{Calibration of an enhancement-independent index}

We have checked if  the [MgFe] index, proposed by G93 
and revised by TMB03, is still a tracer of total 
metallicity independently of the  $\alpha$-enhancement. G93 defines the
index

\begin{eqnarray}
{\rm [MgFe]}&=&{\rm \sqrt{Mgb <Fe>}} 
\label{eq23}
\end{eqnarray}

\noindent with

\begin{eqnarray}
{\rm <Fe>}&=&{\rm \frac{1}{2} (Fe5270 + Fe5335)}, 
\label{eq24}
\end{eqnarray}

\noindent while TMB03 propose the new index ${\rm [MgFe]^{'}}$, 
whose definition is still given by Eq. (\ref{eq23}),
although the ${\rm <Fe>}$ index is replaced by

\begin{eqnarray}
{\rm <Fe>}&=& {\rm  (0.72 \cdot Fe5270 + 0.28 \cdot Fe5335)}.
\label{eq25}
\end{eqnarray}

\begin{figure}
\resizebox{8.5cm}{!}{ \psfig{figure=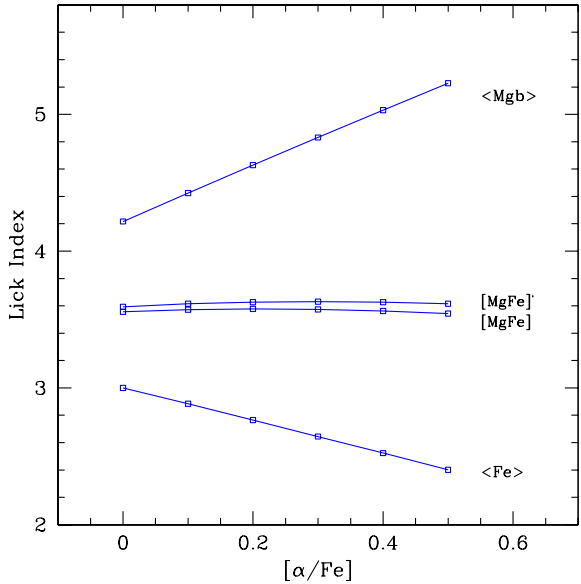,width=7cm,clip=} }
\caption{Lick indices as a function of 
[$\alpha$/Fe] ratio at fixed solar 
metallicity. We plot the indices Mgb, 
${\rm <Fe>=\frac{1}{2} (Fe5270 + Fe5335)}$,
[MgFe], and [MgFe]$^{'}$ (see text for details) for a 12--Gyr--old SSP
of our new $\alpha$-enhanced models.} 
\label{fig6}
\end{figure}

In Fig. \ref{fig6} we plot the indices Mgb, ${\rm <Fe>}$, 
[MgFe], and [MgFe]$^{'}$
as a function of [$\alpha$/Fe] for our new SSP models.
In all the models, the Mgb index increases with the $\alpha$-enhancement, 
while ${\rm <Fe>}$  decreases. On the other hand, the [MgFe] and [MgFe]$^{'}$
indices remain essentially constant.
More specifically, the fractional variations of [MgFe] and  [MgFe]$^{'}$
amount to $\sim -0.004$ and $\sim 0.006$, respectively, passing from 
[$\alpha$/Fe]$=0$ to [$\alpha$/Fe]$=0.5$.

\subsection{Comparison with TMB03 SSPs}

We have compared our new SSP models with those of TMB03. 
Since both models are based on the fitting functions of 
Worthey et al. (\cite{W94}) and Worthey \& Ottaviani (\cite{WO97}), 
the possible differences between the models for solar--scaled 
compositions must be due to 
the different evolutionary tracks adopted.
For $\alpha$-enhanced compositions, an additional difference 
is the new index dependence 
on element abundance introduced in our models (see Sect. 3.5).

\begin{figure}
\resizebox{9.cm}{!}{ \psfig{figure=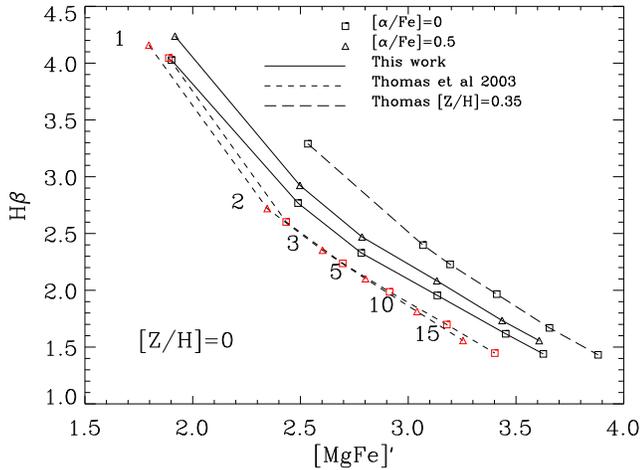,width=9cm,angle=90,clip=}}
\caption{Comparison of our SSP models 
with the models of TMB03 for
solar total metallicity, ages in the range
1--15 Gyr, and [$\alpha$/Fe]$=$0, 0.5. 
In the H$\beta$ vs. ${\rm [MgFe]^{'}}$ plane, our models 
are denoted with the solid line, the TMB03 models with 
the short-dash line. The long-dash line is for the TMB03
models at [Z/H]$=$0.35. We observe that our solar metallicity models
are between the [Z/H]$=$0 and  [Z/H]$=$0.35 TMB03 models.}
\label{fig18}
\end{figure}

In Fig.~\ref{fig18} we plot the H$\beta$ and [MgFe]$^{'}$ indices predicted by 
our models and those of TMB03 for solar total metallicity, ages in the 
range 1--15 Gyr, and [$\alpha$/Fe]$=$0, 0.5, plus the TMB03 models 
for [Z/H]$=$0.35 and [$\alpha$/Fe]$=$0.
In particular, we observe that our solar metallicity models
are between the [Z/H]$=$0 and  [Z/H]$=$0.35 TMB03 models. Thus we expect 
our models to yield lower metallicities for the same line--strength 
index data.
The comparison is also showed in Table~4.
We compare our models with those of TMB03 and Tantalo \& Chiosi 
(\cite{TC04a}, (TC04) for solar metallicity, 
[$\alpha$/Fe]$=$0, and ages=5, 10 Gyr, for some commonly used indices.
We observe that there is a significant difference in the metallic indices,
in particular the Mgb. Our models predict Mgb indices that are quite
larger than TMB03, and this results in higher [MgFe]$^{'}$.
The TC04 models are instead between our models and the TMB03 ones.
Bluer metallic indices are probably the results of hotter stellar tracks.
In particular it is well known that the Girardi et al. (\cite{G02}) 
tracks used in the TC04 models predict warmer red giant branches 
(see, e.g., Bruzual \& Charlot \cite{BC03}).

\begin{table}
{\bf Table 4. Comparison with SSPs in the literature 
}
\begin{center}
\begin{tabular}{ccccccc}
\hline \hline\\
    &  H$\beta$  & Mg$_2$ & Mgb & Fe5270 & Fe5335 & [MgFe]$^{'}$ \\
\hline\\
5 Gyr & & & & & &\\
\hline\\
US     & 1.95 &   0.23 &  3.60  &    2.86 &    2.59 & 3.17 \\ 
TMB03  & 1.99 &   0.21 &  3.18  &    2.83 &    2.52 & 2.95 \\ 
TC04   & 1.90 &   0.23 &  3.30  &    2.90 &    2.63 & 3.06 \\   
\hline\\
10 Gyr & & & & & &\\
\hline\\
US     & 1.62 &  0.26 &   4.08  &   3.06 &    2.79 & 3.49 \\ 
TMB03  & 1.70 &  0.23 &   3.58  &   2.98 &    2.66 & 3.22 \\
TC04   & 1.50 &  0.27 &   3.76  &   3.12 &    2.85 & 3.38 \\ 
\\
\hline
\end{tabular}
\end{center}
\label{tab4}
\end{table}

To perform a more direct and quantitative comparison, we have 
considered the (H$\beta$, Mgb, and $<$Fe$>$) 
indices for some selected TMB03 models
of a given age, [Z/H], and [$\alpha$/Fe], 
and have fitted those indices with our 
models. This allows us to directly compare the (age, [Z/H], 
and [$\alpha$/Fe])  
solutions provided by our models and those of TMB03 
for the same index triplet.
Our test shows that there are significant differences in the derived stellar 
population parameters.
The major discrepancy concerns metallicity:
our models provide metallicities that are systematically lower 
(up to 0.2 dex) 
than those predicted by the TMB03 models.
On the other hand, our derived  ages tend to be older 
(10\% higher at 10 Gyr).  
The [$\alpha$/Fe] ratios are instead in good agreement. We only observe that
for high [$\alpha$/Fe] ratios ($\sim 0.5$), we tend to predict values 
that are larger by $\sim$ 20 \%.
The same metallicity-age  
discrepancy has also been discussed by Denicolo' et al. 
(\cite{Den05b})
when comparing the TMB03 models with those of Worthey (\cite{W94}): 
fitting the line strength index data of the galaxy NGC~3379 with both 
the set of models, they derived a higher metallicity ($\Delta [Fe/H]=0.08$)
and a younger age ($\Delta$ age =-0.5 Gyr) with the TMB03 SSPs than with the
Worthey (\cite{W94}) ones.

\section{Derivation of stellar population parameters}

\subsection{The line--strength index diagnostic}

The use of absorption-line indices has proven to be a powerful tool in the derivation of ages and metallicities of unresolved stellar populations (e.g., Worthey \cite{W94}).
To this purpose it is important to identify those absorption features, among the Lick indices, that possess a marked sensitivity either to abundance or age, that are 
not strongly affected by uncertainties in the measurement process,
and, finally, that can be described by fairly robust models.
Among the Balmer lines, which, measuring the presence of warm A-type stars,
can be used as age indicators, the H$\beta$ index turns out relatively
insensitive to abundance ratio variations, but it may be contaminated by
nebular emission. In reverse, higher order Balmer lines, less affected than
H$\beta$ by emission infilling (in particular the H$\delta$ less than the
H$\gamma$), are not independent from abundance ratio variations, as shown in
Sect.~3.

Line--strength indices, like the Mg and Fe indices, derived from metallic lines are mainly sensitive to element abundances, but they also depend on  age.
In particular, we have shown in Sect.~3.8 that the [MgFe] and [MgFe]$^{'}$ 
indices are good tracers of total metallicity and are almost 
completely independent of $\alpha$/Fe ratio variations.
Thus, a three--dimensional space defined by a Balmer absorption line,
the [MgFe] (or [MgFe]$^{'}$), and a metallic index sensitive 
to $\alpha$-enhancement 
(like Mgb) demonstrates itself to be a suitable diagnostic tool for
the derivation of ages, metallicities, and [$\alpha$/Fe] ratios.

\begin{figure*}
\resizebox{17cm}{!}{ \psfig{figure=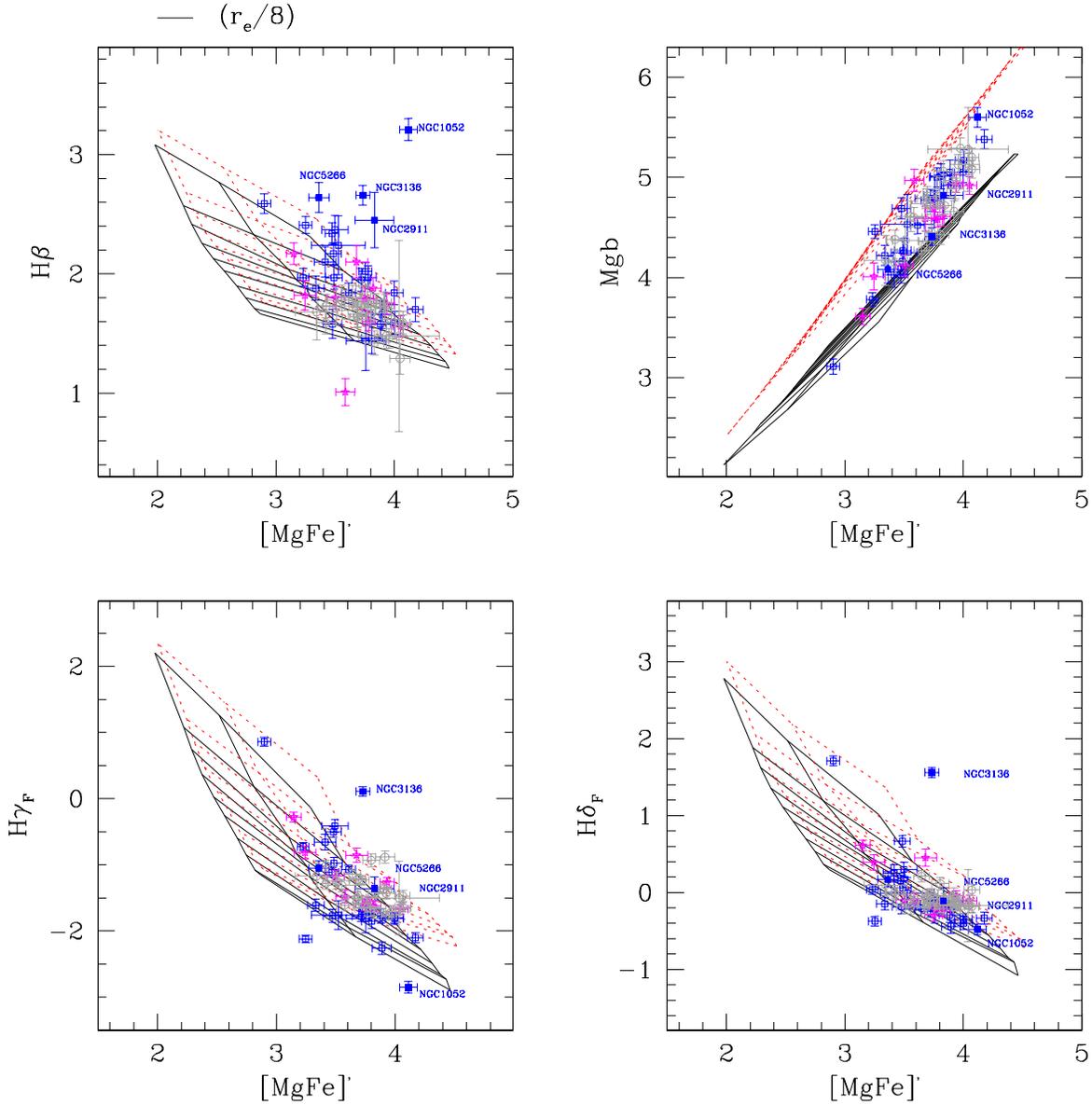,width=13cm,clip=} }
\caption{ 
Lick line--strength indices ($r_e/8$ aperture) of 
our sample compared with SSP models with metallicities 
Z$=0.008, 0.02$, and $0.05$, ages in the range (2--15) Gyr, and 
[$\alpha$/Fe]=0 (solid line) and 0.4 (dotted line). We have plotted 
${\rm H\beta - [MgFe]^{'}}$ (top left panel), 
${\rm Mgb -  [MgFe]^{'}}$ (top right),
${\rm H\gamma_{F}~-  [MgFe]^{'}}$ (bottom left), and 
${\rm H\delta_{F}~-  [MgFe]^{'}}$ (bottom right) planes.
The metallicity increases for increasing  ${\rm [MgFe]^{'}}$ values,
while the age increases for decreasing Balmer index.
Vertical lines in the Balmer index vs.  ${\rm [MgFe]^{'}}$ plane are models
of constant metallicity, while horizontal lines are models
of constant age. The data have been plotted with different symbols
according to their emission properties: open circles for 
[OIII$\lambda$5007] emission detected
under 1$\sigma$, triangles for emission between 1 and 2 $\sigma$,
and squares $>2 \sigma$.
The galaxies IC~5063, NGC~6875, and NGC~2962 have been excluded from the original sample of 65 galaxies. 
}
\label{fig7} 
\end{figure*}

In Fig.~\ref{fig7}, we plotted four projections
of the spaces defined by (H$\beta$, [MgFe]$^{'}$, Mgb), 
(H$\gamma_{F}$, [MgFe]$^{'}$, Mgb), and (H$\delta_{F}$, [MgFe]$^{'}$, Mgb).
In the planes  Mgb -- [MgFe]$^{'}$, H$\beta$ -- [MgFe]$^{'}$, 
H$\gamma_{F}$ -- [MgFe]$^{'}$, and H$\delta_{F}$ -- [MgFe]$^{'}$,  
we superposed the SSP models on the line--strength indices 
derived for our sample within $r_e/8$. 
Galaxies with different detection levels of the [OIII]$\lambda$5007 
emission line are plotted using different symbols.
In this plot we show only 62 out of the initial 
sample of 65 galaxies. We have excluded: 
the Seyfert galaxy IC~5063 because of the presence of 
very strong emission lines, which prevented us from deriving reliable 
absorption line indices; NGC~6875, which lacks a measure of 
velocity dispersion (consequently the indices have not been corrected for this effect), and NGC~2962, which clearly shows calibration problems in the higher order  
Balmer indices and presents significant emission that can affect the H$\beta$.
These three galaxies will not be included in our results.
In the Balmer indices vs. [MgFe]$^{'}$ planes,  
models of constant age (almost horizontal lines)
and of constant metallicity (almost vertical lines) are fairly well separated 
as a result of the markedly different sensitivity of the 
Balmer index to age and of the [MgFe]$^{'}$ index to metallicity
(see Worthey \cite{W94}). 
The effect of $\alpha$-enhancement is that of slightly strengthening
the H$\beta$ feature. Though small, the variation at old ages corresponds
to a few Giga years. On the contrary, [MgFe]$^{'}$ is almost unaffected, due to
the opposite response of Mg and Fe indices to $\alpha$-enhancement. 
The H$\beta$ vs. [MgFe]$^{'}$ plane, where
the enhanced SSPs are almost superposed on the standard ones,
is in principle well suited to directly reading off 
ages and total metallicities almost independently of abundance ratio effects.
ùIn the H$\beta$ vs. [MgFe]$^{'}$ plane, but also in the other planes
defined by high order Balmer lines, the data  
cluster in the lower right region of the diagram, between solar and twice solar metallicity models, and at old ages. A tail of objects extends 
at higher H$\beta$ values and, as already noticed by 
Longhetti et al. (\cite{L00}), does not follow a line of constant metallicity.
A large scatter in age is observed, and galaxies with equivalent--SSP ages 
of few Gyrs are present.

There are, in particular, few galaxies that fall outside the models
at very high H$\beta$ values, even extrapolating the models at metallicities higher than Z$=0.05$: NGC~1052, NGC~3136, NGC~5266, and NGC~2911.
We have labeled these objects in all the four panels to better
understand the origin of the discrepancy. In particular we want to know 
if the high H$\beta$ values are caused by 
an overestimation of the emission correction or are rather due to 
intrinsic properties.
We observe that NGC~3136 has high values for all the three Balmer indices.
We note in particular that while the H$\beta$  has been 
corrected for emission,
H$\gamma_{F}$ and H$\delta_{F}$ have not been. Thus the plotted 
H$\gamma_{F}$ and H$\delta_{F}$ are lower limits, and by applying the (small) 
emission correction to them, we would get even higher values.
Given the high quality and the high S/N of the spectrum also, 
we conclude that the observed strong Balmer lines for  NGC~3136 
are a real effect. 
As noted by Bressan et al. (1996) and Longhetti et al. (2000),
a galaxy in which a recent burst of star formation is 
superimposed on an old stellar population can fall above 
the H$\beta$ - [MgFe] region occupied by the SSPs.
The reason is that if in the secondary burst the gas metallicity 
was higher than the metallicity of the old population, 
the galaxy moves vertically toward the top of the  
H$\beta$ - [MgFe] plane. Thus, it is likely that the position of 
NGC~3136 in the H$\beta$ - [MgFe]$^{'}$ plane is due to the occurrence 
of secondary bursts of star formation.

The situation seems to be different for NGC~1052, NGC~5266, and NGC~2911.
They all show high H$\beta$ values,
but H$\gamma_{F}$ and H$\delta_{F}$ fall within the models 
(NGC~5266 and NGC~2911) or are too low (H$\gamma_{F}$ for NGC~1052).
The observed inconsistency among the Balmer lines seems to be due to
an overestimation of the H$\beta$ emission, with consequent overcorrection 
of the index.
Given the accurate subtraction of the continuum around H$\alpha$
performed to determine the emission (see Papers~I and II), the only
viable explanation for the high H$\beta$ emission is 
the presence of significant dust extinction: in this case the 
FH$\alpha$/FH$\beta$ ratio is higher than 2.86, which is the value
adopted throughout our analysis.
Indeed HST observations have revealed
dust absorption features in the center of NGC~1052 (van Dokkum \& Franx
\cite{Vand05}; Gabel et al. \cite{Gab02}); 
NGC~2911 is a disk dominated S0 with an important dust component
(Michard \& Marchal \cite{MM94}), and NGC~5266 is an HI-rich 
elliptical galaxy with a dust ring along the minor axis 
(Varnas et al. \cite{Var87}; Morganti et al. \cite{Mor97}).

The index errors displayed in Fig. 8 have been obtained 
adding a Poissonian fluctuation to each spectrum and generating a set
of 1000 Monte Carlo modifications (see Papers I and II). The error in the 
H$\beta$ index does not include the systematic uncertainties in the emission
correction due to the choice of a particular template for the 
continuum subtraction and of a FH$\alpha$/FH$\beta$ ratio. 
Thus, we warn the reader that the H$\beta$ error bars plotted in Fig. 8 
may be underestimated.

In the  Mgb -- [MgFe]$^{'}$ plane, models of constant age and metallicity are 
almost degenerate, while the  effect of the enhancement is well separated. 
We observe in particular that the bulk of data is confined 
between [$\alpha$/Fe]$=0$ and [$\alpha$/Fe]$=0.4$.
Furthermore, there is a trend  for galaxies with
larger [MgFe]$^{'}$ values to lie closer to the enhanced models than
to the standard ones, suggesting a correlation between metallicity
and $\alpha$/Fe enhancement.
The three--dimensional spaces described in this section
will be used in our following analysis to derive
ages, metallicities, and [$\alpha$/Fe] ratios for our 
sample of galaxies. For this purpose we have devised a simple but robust
algorithm, which we describe in the next section.

\begin{figure*}
\resizebox{17cm}{!}{ \psfig{figure=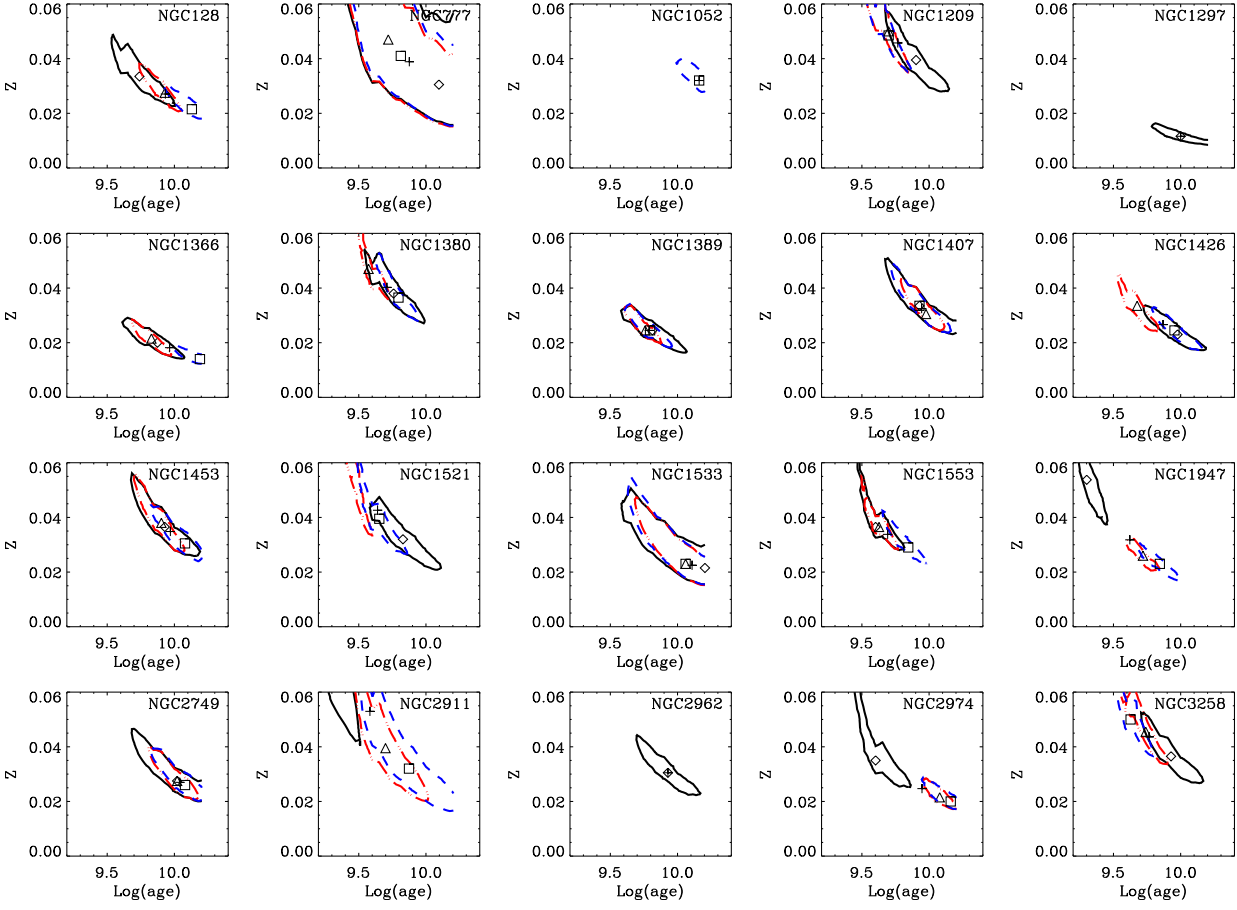,width=13cm,angle=90,clip=} }
\caption{ Example of contour levels enclosing solutions that are within
one $\sigma$ from the observed value. Black solid lines
are for (H$\beta$, $<$Fe$>$, Mgb) triplets, 
red dotted lines for the (H$\gamma_{F}$, $<$Fe$>$, Mgb) triplets,
and blue dashed lines for the  (H$\delta_{F}$, $<$Fe$>$, Mgb) triplets.
Diamonds, triangles, and squares refer to the most probable 
solution for each triplet, 
while crosses indicate the average of the logarithmic values of the solutions.
}
\label{fig8}
\end{figure*}

\subsection{The probability distribution of the physical parameters}
\label{method}

A SSP model of given age, metallicity, and $\alpha$/Fe ratio
corresponds to a point in a multi-dimensional space defined by 
the line--strength indices.
If we exclude ages older than $\sim$ 17 Gyr, where a blueing of the indices
may be introduced by the presence of Horizontal Branch 
(HB) stars even at high metallicity,
there is a unique correspondence between the index multiplet 
and a point in the age, metallicity, and [$\alpha$/Fe] 
parameter space. 
In principle, one could chose three
arbitrary indices and, by comparison with SSP models,
derive the (age, Z, [$\alpha$/Fe])  solution.
Nevertheless, as discussed in the previous section,
some combinations of indices among the Lick set are more
useful than others to disentangle the age-metallicity degeneracy, and 
thus turn out to be particularly suitable for deriving the stellar population parameters.
More specifically, for our analysis we select the three dimensional spaces
defined by (H$\beta$, $<$Fe$>$, Mgb), 
(H$\gamma_{F}$, $<$Fe$>$, Mgb), and 
(H$\delta_{F}$, $<$Fe$>$, Mgb).
Once the index 3D space has been defined, the 
stellar population parameters are derived according to the following 
procedure.

Let us call $(x_0, y_0, z_0)$ the observed point in the index space and 
$(\sigma_{x_0}, \sigma_{y_0}, \sigma_{z_0})$ the
associated error. 
For each point (t, Z, $\alpha$) in the (age, metallicity, $\alpha$/Fe enhancement) 
space (G space), we can define the probability density that
the generic SSP model $(x_{(t,Z,\alpha)}, 
y_{(t,Z,\alpha)}, z_{(t,Z,\alpha)})$ is the solution
to the observed line--strength indices.
In particular, if the errors are Gaussian, the 
probability density assumes the expression:

\begin{eqnarray}
P_{t,Z,\alpha}&=& \frac{1}{\sqrt{(2 \pi)^3} \sigma_{x_0}  \sigma_{y_0}  \sigma_{z_0}} 
\exp \left[ -\frac{1}{2} \left( \frac{x-x_0}{\sigma_x} \right)^2 \right.   \nonumber \\
 &-&\left. \frac{1}{2} \left( 
\frac{y-y_0}{\sigma_y}\right)^2  -\frac{1}{2} \left(\frac{z-z_0}{\sigma_z}\right)^2 \right].
\label{eq6.1}
\end{eqnarray}

\noindent Equation ({\ref{eq6.1}}) allows us to construct a probability density 
map $P_{t,Z,\alpha}$
defined on the three dimensional grid ($t,Z, \alpha$).

In Fig.~\ref{fig8}, we show the contour levels in the ($t, Z$) plane, enclosing the solutions that are within one $\sigma$ (in the three indices) from the observed values for some galaxies of our sample, as an example. Each panel refers to a different galaxy, and the three curves refer to the solutions obtained with the different triplets adopted, i.e., H$\beta$, $<$Fe$>$, Mgb; H$\gamma_{F}$, $<$Fe$>$, Mgb; and H$\delta_{F}$, $<$Fe$>$, Mgb. The most probable solution in each index triplet is obtained by finding the maximum of the $P_{t,Z,\alpha}$ function.

Figure~\ref{fig8} shows that measurement errors combine with the
relative index sensitivity to age and metallicity variations 
(the age metallicity-degeneracy). Levels of constant probability define a narrow elongated 
region of anti-correlation between age and metallicity solutions.
Some authors have shown this effect by using Montecarlo simulations 
to generate a mock catalog of synthetic galaxies (e.g., Thomas et al.~\cite{Thom05}). 
Notice that the age-metallicity degeneracy introduces
an asymmetry in the errors of the solution, as a Gaussian error in the 
measured indices translates into a non-Gaussian error in the solution. 
We have used the three triplets to minimize spurious 
observational effects in the age indicator H$\beta$, which may be affected
by emission filling. For each galaxy we combined only those index 
triplets that provide 
more reliable results, depending on the strength of the present emission (see below).

For most of the sources, the 
contour levels of the three index spaces possess 
a clear region of intersection. This indicates
that models predict consistent results. However, there are
galaxies for which the three G spaces do not 
have common solutions (within one-$\sigma$, if we consider two-$\sigma$, then
there are always common solutions, but the uncertainty is large).
In particular, for some galaxies, H$\beta$ provides solutions that are 
really detached from those provided by the other two age indicators.
As discussed in Sect.~4.1, discrepancies between the H$\beta$
and the higher order Balmer lines could be due to problems in the derivation
of the emission correction.

For each 3D index space, the solution is computed adopting a subspace ($G^*$) defined by  $P_{t,Z,\alpha} > f \times P_{max}$, and performing an average on the ($G^*$) values  weighted for the probability density:

\begin{eqnarray}
t_{\mu}= \frac{\int\!\!\!\int\!\!\!\int_{G^*} t \  P(t,Z,\alpha) \ dt \ dZ \ d\alpha}{\int\!\!\!\int\!\!\!\int_{G^*}  P(t,Z,\alpha) \ dt \ dZ \ d\alpha} \nonumber \\
Z_{\mu}= \frac{\int\!\!\!\int\!\!\!\int_{G^*} Z \  P(t,Z,\alpha) \ dt \ dZ \ d\alpha}{\int\!\!\!\int\!\!\!\int_{G^*}  P(t,Z,\alpha) \ dt \ dZ \ d\alpha} \\
\alpha_{\mu}= \frac{\int\!\!\!\int\!\!\!\int_{G^*} \alpha \  P(t,Z,\alpha) \ dt \ dZ \ d\alpha}{\int\!\!\!\int\!\!\!\int_{G^*}  P(t,Z,\alpha) \ dt \ dZ \ d\alpha}. \nonumber
\label{partial_sol}
\end{eqnarray}

\noindent More specifically, we consider the subspace of solutions obtained with f=0.9.
The uncertainties on the solutions are computed considering the whole G space:

\begin{eqnarray}
{\sigma_t}^2 &=& \frac{\int\!\!\!\int\!\!\!\int_{G} (t-t_{\mu})^2 \  P(t,Z,\alpha) \ dt \ dZ \ d\alpha}{\int\!\!\!\int\!\!\!\int_{G}  P(t,Z,\alpha) \ dt \ dZ \ d\alpha} \nonumber \\
{\sigma_Z}^2 & =& \frac{\int\!\!\!\int\!\!\!\int_{G} (Z-Z_{\mu})^2 \  P(t,Z,\alpha) \ dt \ dZ \ d\alpha}{\int\!\!\!\int\!\!\!\int_{G}  P(t,Z,\alpha) \ dt \ dZ \ d\alpha} \\
{\sigma_{\alpha}}^2 &=& \frac{\int\!\!\!\int\!\!\!\int_{G} (\alpha-\alpha_{\mu})^2 \  P(t,Z,\alpha) \ dt \ dZ \ d\alpha}{\int\!\!\!\int\!\!\!\int_{G}  P(t,Z,\alpha) \ dt \ dZ \ d\alpha}. \nonumber 
\label{partial_err}
\end{eqnarray}

\noindent The analysis, repeated for the triplets 
(H$\beta$, $<$Fe$>$, Mgb), 
(H$\gamma_{F}$, $<$Fe$>$, Mgb), and 
(H$\delta_{F}$, $<$Fe$>$, Mgb), provides three different partial solutions,
$t_{\mu,i}, Z_{\mu,i}$, and $\alpha_{\mu,i}$ 
for i=1, 3, with the corresponding uncertainties.

At this point we compare the results provided by the different index triplets and 
check for consistency within the errors.
We observe that for those galaxies with low or absent emission 
($|EW_{em}(H\beta)| \le 0.2$), there is a good agreement among the stellar 
population parameters derived with the use of the three different Balmer indices.
For larger emissions, the results obtained with the H$\beta$ index 
deviate from the H$\gamma$ and  H$\delta$ predictions however, with a tendency toward 
younger ages. Finally, for strong emissions ($|EW_{em}(H\beta)| > 0.6$),
the infilling of the 
H$\gamma$ starts to become significant and the index, 
which has not been corrected for this effect,
systematically provides very old ages. In this last case, the H$\delta$ index, 
that presents the lowest emission contamination, 
is perhaps the most reliable one. 
Thus we combine the information from the different index triplets,
performing a weighted mean over those N indices which, according to 
the galaxy emission properties, provide a reliable solution as explained 
above:

\begin{equation}
t_{f} = \frac{\displaystyle \sum_{i=1}^N t_{\mu,i} 
           W_{t,i}}{\displaystyle \sum_{i=1}^N W_{t,i}}; 
Z_{f} = \frac{\displaystyle \sum_{i=1}^N Z_{\mu,i} 
           W_{Z,i}}{\displaystyle \sum_{i=1}^N W_{Z,i}}; 
\alpha_{f} = \frac{\displaystyle \sum_{i=1}^N \alpha_{\mu,i} 
           W_{\alpha,i}}{\displaystyle \sum_{i=1}^N W_{\alpha,i}}, 
\label{final_sol}
\end{equation} 

\noindent where the weights are derived from the uncertainties on the parameters:

\begin{equation}
W_{t,i}=\frac{1}{\sigma_{t,i}^2}; 
W_{Z,i}=\frac{1}{\sigma_{Z,i}^2}; 
W_{\alpha,i}=\frac{1}{\sigma_{\alpha,i}^2}.  
\label{eq6.2}
\end{equation} 

\begin{figure}
\centerline{
\resizebox{8cm}{!}{ \psfig{figure=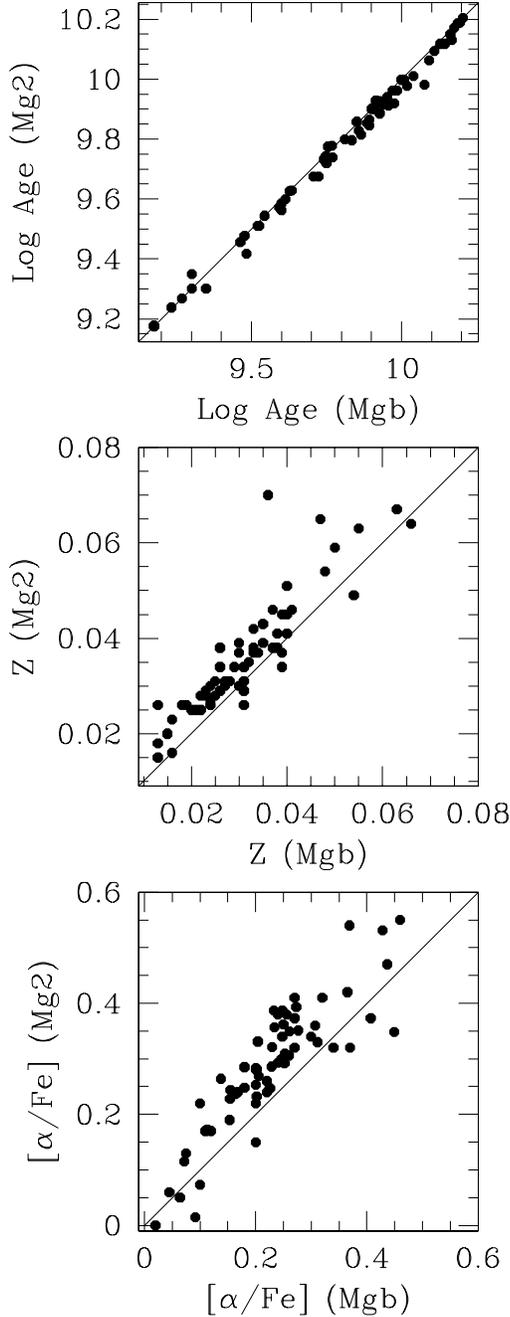,width=5cm} } }
\caption{Comparison of age, metallicity, and $\alpha$-enhancement
values derived using Mg2 instead of Mgb index.}
\label{mgbmg2}
\end{figure}

\section{Stellar populations in early-type galaxies}

Through the procedure described in Sect.~4.2, we have derived stellar population parameters 
(age, metallicity, and [$\alpha$/Fe] ratio) for the apertures and
gradients of the 62 galaxies plotted in Fig.~8.
At the base of our analysis
is the assumption that early-type galaxies contain almost coeval stellar populations
and may be well described by only one SSP. This is the most direct method to derive hints about 
the correlations between stellar population parameters and galaxy properties, like mass and 
environment. In the following analysis, we will focus both on the correlation between central 
galaxy populations and velocity dispersion/environment 
and on the analysis of the radial gradients.

\subsection{Ages, metallicities, and $\alpha$/Fe in the central region}

We provide  ages, metallicities, and [$\alpha$/Fe] ratios derived from 
the three G-spaces (H$\beta$, $<$Fe$>$, Mgb), 
(H$\gamma_{F}$, $<$Fe$>$, Mgb), and (H$\delta_{F}$, $<$Fe$>$, Mgb) for 
the apertures of the sample in Table~5 to 7.
Column (1) gives the galaxy identification name, Col. (2) the 
aperture  number (see Papers~I and II for the aperture radii),
and Cols. (3), (4), and (5) provide the age, the metallicity 
mass fraction Z ($Z=0.018$ is the solar value) and the [$\alpha$/Fe] ratio respectively.
The complete tables are given in electronic form.
In Table~8 we provide the final combined solutions for a  $r_e/8$ aperture, 
obtained through Eqs. (26) and (27).
Columns (1) and (2) give the galaxy identification name and the 
central velocity dispersion respectively. Columns (3), (4), and (5) provide the age, the metallicity 
mass fraction Z ($Z=0.018$ is the solar value), and the [$\alpha$/Fe] ratio, respectively.

Before continuing the discussion, we remark that our results
are fairly independent of the choice of either magnesium index.
Figure \ref{mgbmg2} compares the results obtained
using the Mg2 index instead of the Mgb index. 
We notice a very good consistency in the derived ages,
while the metallicity and $\alpha$--enhancement would be only slightly
higher adopting Mg2 instead of Mgb. However,
the size of the latter differences will not affect our conclusions.

\setcounter{table}{7}

\begin{figure*}
\centerline{
\resizebox{19cm}{!}
{\psfig{figure=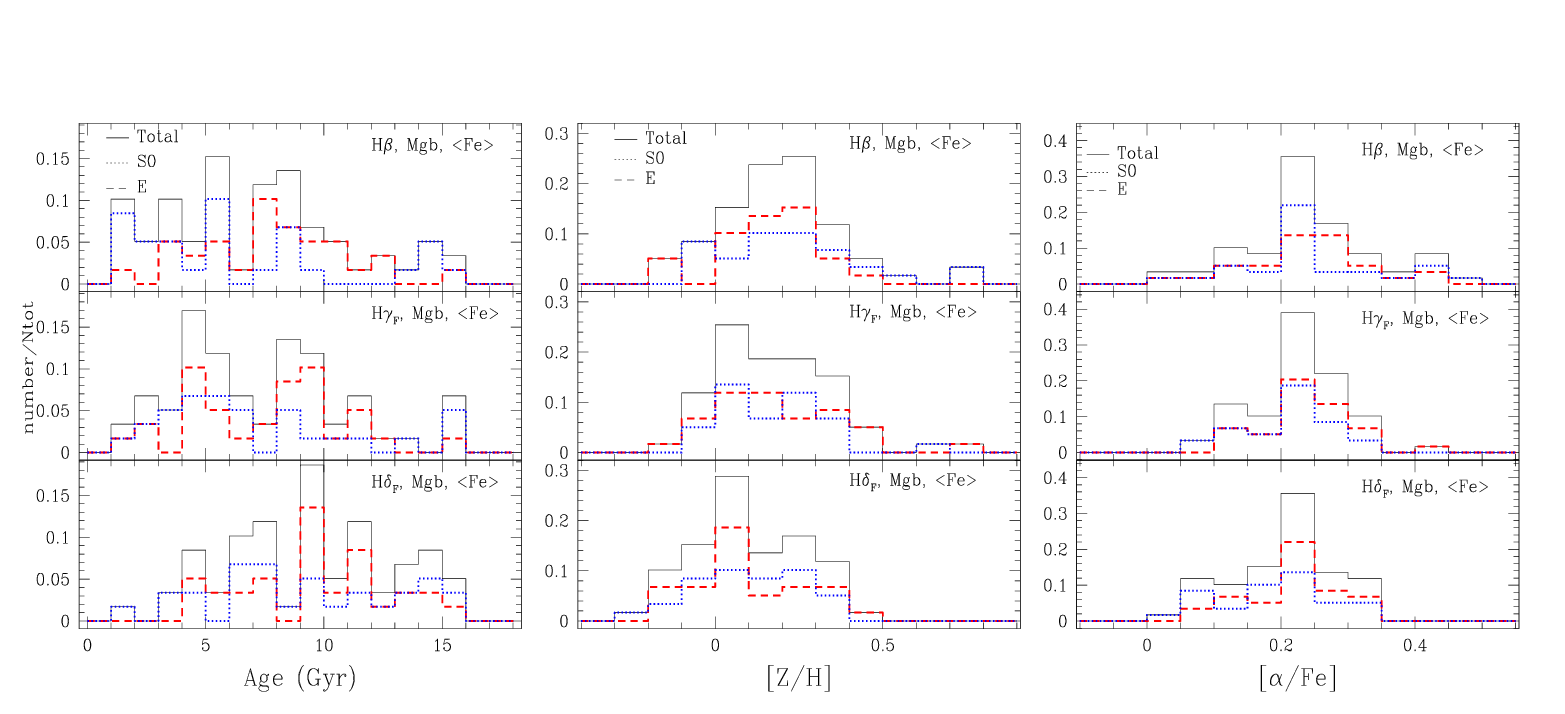,width=17cm}}}
\caption{Distribution of ages (left panel), 
metallicities (central), and [$\alpha$/Fe]
ratios (right) measured at $r_e/8$, derived from  the (H$\beta$, $<$Fe$>$, Mgb), (H$\gamma_{F}$, $<$Fe$>$, Mgb), and (H$\delta_{F}$, $<$Fe$>$, Mgb)
index spaces. In the plot we have included only the galaxies for which the solution is derived within 1~$\sigma$ from the observed indices. The solid line refers to the total sample, while the dotted and dashed lines are for S0s and Es, 
respectively.
For each distribution, the used index triplet is provided in the
label inside the box.} 
\label{fig9}
\end{figure*}

The distributions of age, metallicity, and enhancement within $r_e/8$  are shown in 
Figure~\ref{fig9}. In each panel, we present 
the results obtained from the three different G spaces.
We plot both the distributions for the total sample and the partial distributions for
lenticular and elliptical galaxies. 
For all three index triplets,
the age distribution shows a large spread (1 Gyr$<$ age$<$ 15 Gyr).
From the H$\beta$ and H$\gamma$ indices, a bimodality,
with a group of old objects (ages $\geq$ 7 Gyr), mainly composed by 
ellipticals, and a tail of young objects, mainly lenticulars, appears.
We also observe that H$\beta$ tends to provide 
a larger number of very young galaxies than H$\gamma$ and H$\delta$ . 
This effect is likely the result of the high
uncertainty in the applied emission correction. 
However, we stress again that the method we adopted to combine 
the results from the different indices (see Sect. 4.2) allows us to 
clean our sample from spurious effect and obtain unbiased results for 
the final ages, metallicities, and [$\alpha$/Fe] ratios. 
The average ages for the whole sample, E and S0 obtained by properly combining the information from the three index triplets, are 8, 8.7, and 6.3 Gyr, respectively.

The metallicity distribution shows a broad peak, located between 
0$<$[Z/H]$<$0.3.
There is a good agreement among the values derived from the
analysis of the different index triplets, with the H$\delta$ distribution 
being peaked at slightly lower metallicities, which is an evident effect of 
the age-metallicity degeneracy.
The average metallicities for the whole sample, E and S0, 
are 0.21, 0.22, and 0.19, respectively.
Finally, the [$\alpha$/Fe] ratio presents a maximum at $\sim$ 0.22 
with a narrower distribution than those of age and metallicity.
The average ratios for the whole sample, E and S0, 
are 0.21, 0.23, and 0.17, respectively.

\begin{table*}
\begin{center}
{\bf Table 8. Average ages, metallicities, and $\alpha$/Fe ratios ($r_e/8$ aperture)}
\begin{tabular}{ccccc}
\hline
\hline
Ident. & $\sigma_c$  & Age  & Z & [$\alpha$/Fe] \\ 
       & km~s$^{-1}$  & Gyr  &   &  \\
\hline\\  
NGC~128  &                 183       &    9.7     $\pm$       1.7   &      0.024   $\pm$      0.004     &    0.16    $\pm$     0.03      \\   
NGC~777  &                 317       &    5.4     $\pm$       2.1   &      0.045   $\pm$      0.020     &    0.28    $\pm$     0.10      \\       
NGC~1052 &                 215       &   14.5     $\pm$       4.2   &      0.032   $\pm$      0.007     &    0.34    $\pm$     0.05       \\
NGC~1209 &                 240       &    4.8     $\pm$       0.9   &      0.051   $\pm$      0.012     &    0.14    $\pm$     0.02       \\      
NGC~1297 &                 115       &   15.5     $\pm$       1.2   &      0.012   $\pm$      0.001     &    0.29    $\pm$     0.04       \\
NGC~1366 &                 120       &    5.9     $\pm$       1.    &      0.024   $\pm$      0.004     &    0.08    $\pm$     0.03       \\      
NGC~1380 &                 240       &    4.4     $\pm$       0.7   &      0.038   $\pm$      0.006     &    0.24    $\pm$     0.02        \\     
NGC~1389 &                 139       &    4.5     $\pm$       0.6   &      0.032   $\pm$      0.005     &    0.08    $\pm$     0.02        \\     
NGC~1407 &                 286       &    8.8     $\pm$       1.5   &      0.033   $\pm$      0.005     &    0.32    $\pm$     0.03       \\      
NGC~1426 &                 162       &    9.0     $\pm$       2.5   &      0.024   $\pm$      0.005     &    0.07    $\pm$     0.05       \\      
NGC~1453 &                 289       &    9.4     $\pm$       2.1   &      0.033   $\pm$      0.007     &    0.22    $\pm$     0.03        \\     
NGC~1521 &                 235       &    3.2     $\pm$       0.4   &      0.037   $\pm$      0.006     &    0.09    $\pm$     0.02        \\     
NGC~1533 &                 174       &   11.9     $\pm$       6.9   &      0.023   $\pm$      0.020     &    0.21    $\pm$     0.10       \\      
NGC~1553 &                 180       &    4.8     $\pm$       0.7   &      0.031   $\pm$      0.004     &    0.10    $\pm$     0.02        \\     
NGC~1947 &                 142       &    5.9     $\pm$       0.8   &      0.023   $\pm$      0.003     &    0.05    $\pm$     0.02        \\     
NGC~2749 &                 248       &   10.8     $\pm$       2.3   &      0.027   $\pm$      0.006     &    0.25    $\pm$     0.04        \\     
NGC~2911 &                 235       &    5.7     $\pm$       2.0   &      0.034   $\pm$      0.019     &    0.25    $\pm$     0.10         \\     
NGC~2974 &                 220       &   13.9     $\pm$       3.6   &      0.021   $\pm$      0.005     &    0.23    $\pm$     0.06        \\     
NGC~3136 &                 230       &    1.5     $\pm$       0.1   &      0.089   $\pm$      0.004     &    0.36    $\pm$     0.02        \\
NGC~3258 &                 271       &    4.5     $\pm$       0.8   &      0.047   $\pm$      0.013     &    0.21    $\pm$     0.03         \\    
NGC~3268 &                 227       &    9.8     $\pm$       1.7   &      0.023   $\pm$      0.004     &    0.34    $\pm$     0.04         \\
NGC~3489 &                 129       &    1.7     $\pm$       0.1   &      0.034   $\pm$      0.004     &    0.05    $\pm$     0.02         \\    
NGC~3557 &                 265       &    5.8     $\pm$       0.8   &      0.034   $\pm$      0.004     &    0.17    $\pm$     0.02         \\    
NGC~3607 &                 220       &    3.1     $\pm$       0.5   &      0.047   $\pm$      0.012     &    0.24    $\pm$     0.03         \\    
NGC~3818 &                 191       &    8.8     $\pm$       1.2   &      0.024   $\pm$      0.003     &    0.25    $\pm$     0.03         \\    
NGC~3962 &                 225       &   10.0     $\pm$       1.2   &      0.024   $\pm$      0.003     &    0.22    $\pm$     0.03         \\    
NGC~4374 &                 282       &    9.8     $\pm$       3.4   &      0.025   $\pm$      0.010     &    0.24    $\pm$     0.08         \\    
NGC~4552 &                 264       &    6.0     $\pm$       1.4   &      0.043   $\pm$      0.012     &    0.21    $\pm$     0.03         \\     
NGC~4636 &                 209       &   13.5     $\pm$       3.6   &      0.023   $\pm$      0.006     &    0.29    $\pm$     0.06         \\    
NGC~4696 &                 254       &   16.0     $\pm$       4.5   &      0.014   $\pm$      0.004     &    0.30    $\pm$     0.10         \\    
NGC~4697 &                 174       &   10.0     $\pm$       1.4   &      0.016   $\pm$      0.002     &    0.14    $\pm$     0.04          \\   
NGC~5011 &                 249       &    7.2     $\pm$       1.9   &      0.025   $\pm$      0.008     &    0.25    $\pm$     0.06          \\   
NGC~5044 &                 239       &   14.2     $\pm$       10.   &      0.015   $\pm$      0.022     &    0.34    $\pm$     0.17           \\  
NGC~5077 &                 260       &   15.0     $\pm$       4.6   &      0.024   $\pm$      0.007     &    0.18    $\pm$     0.06           \\   
NGC~5090 &                 269       &   10.0     $\pm$       1.7   &      0.028   $\pm$      0.005     &    0.26    $\pm$     0.04          \\   
NGC~5193 &                 209       &    6.8     $\pm$       1.1   &      0.018   $\pm$      0.002     &    0.26    $\pm$     0.04         \\   
NGC~5266 &                 199       &    7.4     $\pm$       1.4   &      0.019   $\pm$      0.003     &    0.15    $\pm$     0.05          \\   
NGC~5328 &                 303       &   12.4     $\pm$       3.7   &      0.027   $\pm$      0.006     &    0.15    $\pm$     0.05          \\    
NGC~5363 &                 199       &   12.1     $\pm$       2.3   &      0.020   $\pm$      0.004     &    0.16    $\pm$     0.05          \\   
NGC~5638 &                 168       &    9.1     $\pm$       2.3   &      0.024   $\pm$      0.008     &    0.24    $\pm$     0.05          \\   
NGC~5812 &                 200       &    8.5     $\pm$       2.1   &      0.027   $\pm$      0.008     &    0.22    $\pm$     0.05          \\   
NGC~5813 &                 239       &   11.7     $\pm$       1.6   &      0.018   $\pm$      0.002     &    0.26    $\pm$     0.04         \\    
NGC~5831 &                 164       &    8.8     $\pm$       3.5   &      0.016   $\pm$      0.011     &    0.21    $\pm$     0.09          \\   
NGC~5846 &                 250       &    8.4     $\pm$       1.3   &      0.033   $\pm$      0.005     &    0.25    $\pm$     0.03          \\   
NGC~5898 &                 220       &    7.7     $\pm$       1.3   &      0.030   $\pm$      0.004     &    0.10    $\pm$     0.03           \\  
NGC~6721 &                 262       &    5.0     $\pm$       0.8   &      0.040   $\pm$      0.007     &    0.24    $\pm$     0.02           \\  
NGC~6758 &                 242       &   16.0     $\pm$       2.5   &      0.016   $\pm$      0.002     &    0.32    $\pm$     0.05          \\   
NGC~6776 &                 242       &    2.7     $\pm$       0.5   &      0.033   $\pm$      0.010     &    0.21    $\pm$     0.05          \\   
NGC~6868 &                 277       &    9.2     $\pm$       1.8   &      0.033   $\pm$      0.006     &    0.19    $\pm$     0.03          \\   
NGC~6876 &                 230       &    9.8     $\pm$       1.6   &      0.023   $\pm$      0.003     &    0.26    $\pm$     0.03         \\   
NGC~6958 &                 223       &    3.0     $\pm$       0.3   &      0.038   $\pm$      0.006     &    0.20    $\pm$     0.03         \\   
NGC~7007 &                 125       &    3.4     $\pm$       0.6   &      0.031   $\pm$      0.010     &    0.15    $\pm$     0.05         \\    
NGC~7079 &                 155       &    6.7     $\pm$       1.1   &      0.016   $\pm$      0.003     &    0.21    $\pm$     0.05         \\   
NGC~7097 &                 224       &   10.5     $\pm$       2.4   &      0.024   $\pm$      0.005     &    0.30    $\pm$     0.05         \\   
NGC7135  &                 231       &    2.2     $\pm$       0.4   &      0.047   $\pm$      0.010     &    0.46    $\pm$     0.04        \\
NGC~7192  &                257       &    5.7     $\pm$       2.0   &      0.039   $\pm$      0.015     &    0.09    $\pm$     0.05          \\  
NGC~7332  &                136       &    3.7     $\pm$       0.4   &      0.019   $\pm$      0.002     &    0.10    $\pm$     0.03        \\    
NGC~7377  &                145       &    4.8     $\pm$       0.6   &      0.020   $\pm$      0.002     &    0.10    $\pm$     0.03        \\

\noalign{\smallskip}\hline\hline\noalign{\smallskip}\\[-2mm]
\end{tabular}
\end{center}
\label{tab8}
\end{table*}

\begin{table*}!
\begin{center}
{\bf Table 8 Continue. Average ages, metallicities, and $\alpha$/Fe ratios ($r_e/8$ aperture)}
\begin{tabular}{ccccc}
\hline
\hline
Ident. & $\sigma_c$  & Age  & Z & [$\alpha$/Fe] \\ 
       & km~s$^{-1}$  & Gyr  &   &  \\
\hline\\  
IC1459   &                 311       &    8.0     $\pm$       2.2   &      0.042   $\pm$      0.009     &    0.25    $\pm$     0.04        \\    
IC2006   &                 122       &    8.1     $\pm$       0.9   &      0.026    $\pm$     0.003     &    0.12     $\pm$    0.02        \\     
IC3370   &                 202       &    5.6     $\pm$       0.9   &      0.022   $\pm$      0.004     &    0.17    $\pm$     0.04       \\     
IC4296   &                 340       &    5.2     $\pm$       1.0   &      0.044   $\pm$      0.008     &    0.25    $\pm$     0.02       \\

\noalign{\smallskip}\hline\hline\noalign{\smallskip}\\[-2mm]
\end{tabular}
\end{center}
{Notes: the values are obtained combining the H$\beta$, H$\gamma$, H$\delta$, Mgb,  and $<$Fe$>$ indices as described in Sect. 4.2}. 
\label{tab4}
\end{table*}

\subsubsection{Correlation with central velocity dispersion}

\begin{figure*}
\center{
\psfig{figure=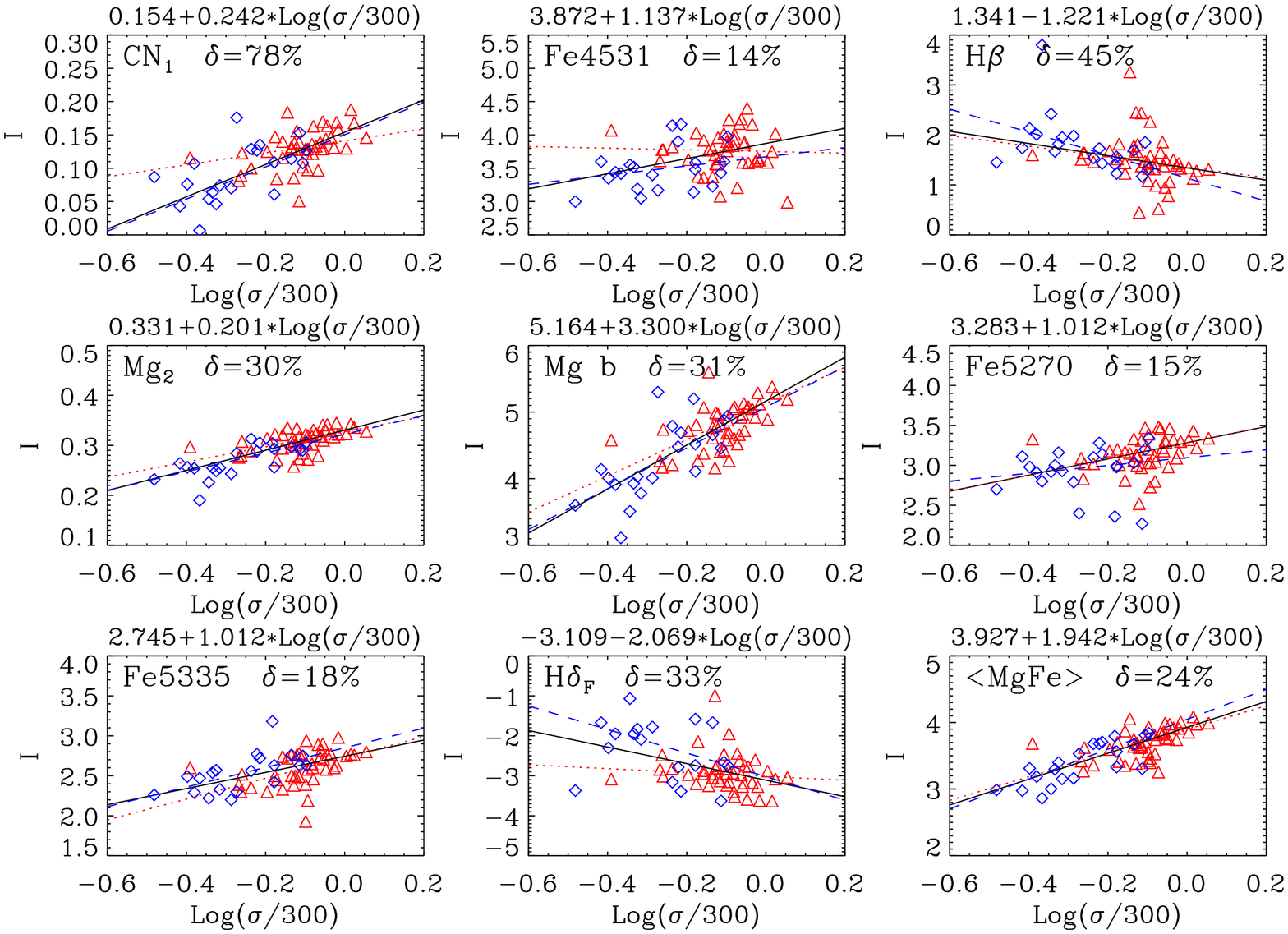,width=0.95\textwidth,angle=90} }
\caption{Selected Lick indices,  measured at $r<r_e/8$, 
for the total sample of 65 galaxies as a 
function of $\log(\sigma_c/300)$, where $\sigma_c$ is the
central velocity dispersion. Triangles and diamonds indicate E and S0 
galaxies, respectively. The dashed,
dotted, and solid lines mark the linear fit obtained for S0 galaxies, 
E galaxies, and the total sample, respectively. For each index, the linear fit to the total sample is labeled above
each panel.}
\label{fig10}
\end{figure*}

Figure~\ref{fig10} shows selected Lick indices measured
in the central aperture ($r<r_e/8$)
for the 65 galaxies of our sample 
as a function of central velocity dispersion.
We observe that the separation between the two classes
(those galaxies classified as E according to RC3 and those classified as S0)
is quite well marked in
velocity dispersion, with very few E galaxies populating the region below 
$\sigma_c \sim$ 180 km/s. The fit to the total sample
with coefficients is given in the upper label.
We also show the percentage index variation along the regression when 
the velocity dispersion changes from $\sigma_c =$ 300 km/s
to  $\sigma_c =$ 100 km/s.
Separate fits are made for the S0 and E galaxies.

It is clear that metallic indices 
show a well--established positive trend with 
$\sigma_c$ (with a significantly shallower variation for the  Fe indices). 
On the contrary, Balmer indices decrease with velocity dispersion.
The observed behavior suggests metallicities or ages increasing
with galaxy mass, but for the quantitative results we refer to the 
following comparison with SSP models.

\begin{figure*}
\center{
\resizebox{13.5cm}{!}{ \psfig{figure=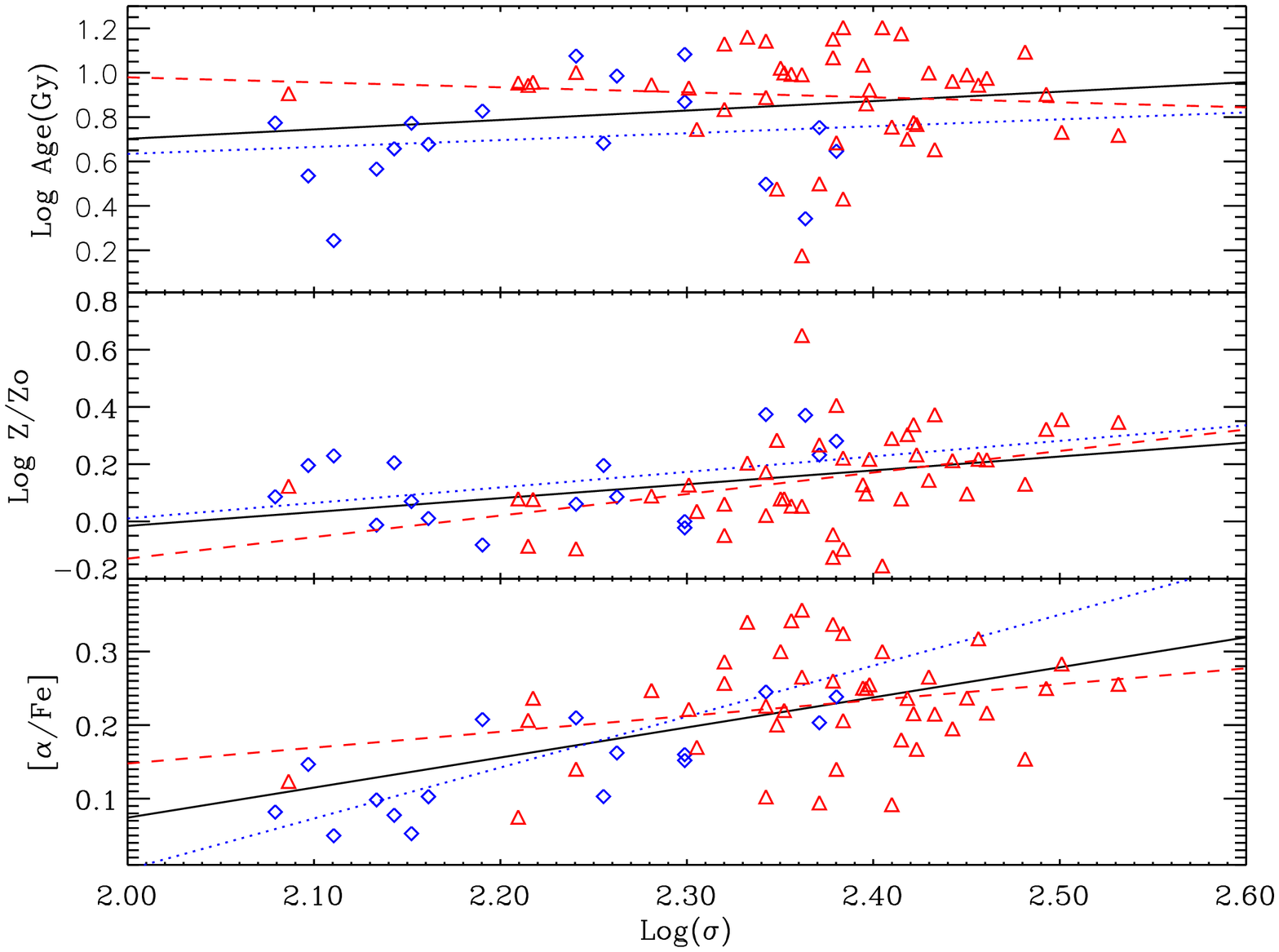,width=13cm,angle=90,clip=} }}
\caption{Ages, metallicities, and [$\alpha$/Fe] ratios, measured at $r_e/8$,
vs.  the central velocity dispersion, $\log \sigma_c$.
The plotted stellar population parameters are derived from the combined
analysis of the (H$\beta$, $<$Fe$>$, Mgb), 
(H$\gamma_{F}$, $<$Fe$>$, Mgb), and (H$\delta_{F}$, $<$Fe$>$, Mgb)
index spaces (see text for details).
Triangles and diamonds denote, respectively, E and S0 galaxies.
The solid line is the linear fit performed to all the galaxies,
while dashed and dotted lines are the best fit to Es and S0s subsamples.}
\label{fig11}
\end{figure*}

To seek for possible correlation of the derived  stellar population
parameters with central velocity dispersion/galaxy mass, 
we plot the final solutions derived within
a $r_e/8$ aperture (see Table~8) as function of $\sigma_c$ 
in Fig.~\ref{fig11}. 
In the top, central, and bottom panels, 
we plot ages, metallicities, and [$\alpha$/Fe] ratios, respectively.
Figure~\ref{fig11} shows a trend of increasing age 
with central velocity 
dispersion. Considering the separate fits for ellipticals and 
lenticulars, we obtain that the trend is positive for lenticulars 
and negative for ellipticals.
However, a Spearman rank order test provides no proof for the 
existence of a correlation
between age and $\sigma_c\simeq$ for both the whole sample and the 
E and S0 subsamples.
The separate fits suggest that on average lenticulars are younger 
than ellipticals. The former, however, shows a large scatter. 
There is also a group
of young E galaxies that deviate significantly from the 
main trend, at $\sigma_c\simeq$250 km~s$^{-1}$. 
The middle panel shows that there is a significant positive trend of metallicity with  $\sigma_c$, with no significant difference in the slope 
between the two morphological types. 
However, lenticular galaxies appear on
average more metal rich than ellipticals at the same velocity dispersion.
A Spearman test provides p-levels of significance\footnote{The p level of significance represents the probability that the relation found between the variables is not true.} of 0.0001, 0.0002, and 0.07
for the whole sample, E and S0, respectively, confirming that there is strong
correlation between the parameters.
The same result is found  for the $\alpha$-enhancement, 
which again increases
with central velocity dispersion. For the whole sample, the Spearman 
test results in a p level of 0.0008.
The absence of a clear trend of age with $\sigma_c$, and the presence 
of strong correlations for  metallicities and  $\alpha$-enhancement 
suggest that 
the galaxy gravitational potential must mainly affect 
the chemical enrichment process of the galaxy.

\subsubsection{Correlation with  environment}

\begin{figure*}
\center{
\resizebox{13.5cm}{!}{ \psfig{figure=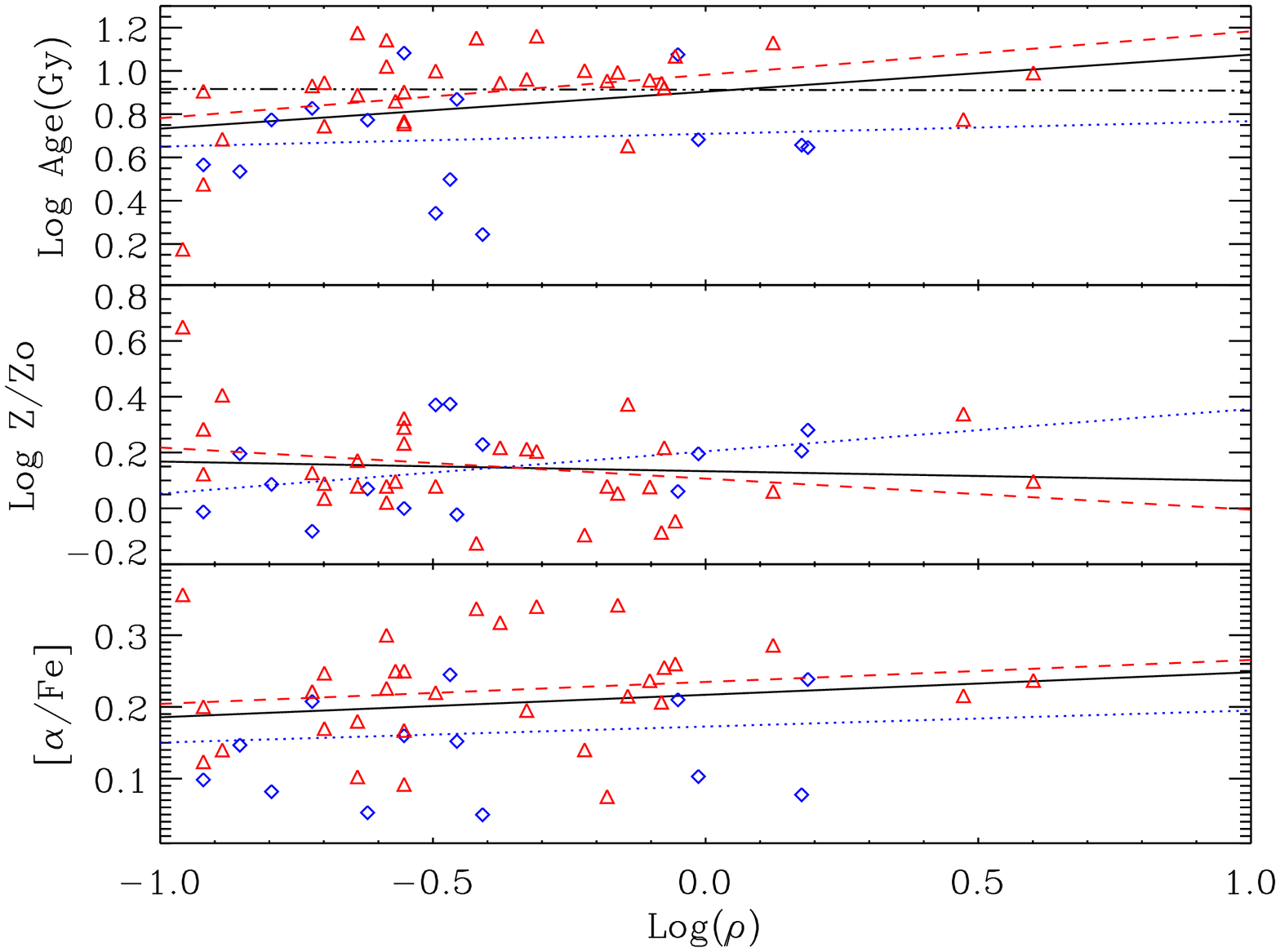,width=13cm,angle=90,clip=} }}
\caption{Ages, metallicities, and [$\alpha$/Fe] ratios, measured at $r_e/8$,
vs.  the density of the environment, $\log(\rho)$,  
in galaxies~Mpc$^{-3}$ (Tully \cite{Tu88}, see Table~1).
Triangles and diamonds denote, respectively, E and S0 galaxies.
The solid line is the linear fit performed to all the galaxies.
Dashed and dotted lines are the best fit to E and S0 subsamples.
The dot-dashed line is the fit for the galaxies older than 4 Gyr.
}
\label{fig12}
\end{figure*}

A measure of the richness of the environment, $\rho_{xyz}$, 
surrounding each galaxy in galaxies~Mpc$^{-3}$,
is reported by Tully (1988) for 47 galaxies (73\% of the whole sample)
(see Table~1).
Even though our sample is prevalently made of field galaxies and
we are far from the densities found in rich galaxy clusters,
we may investigate how (eventually) the environment affects the 
trend of the physical parameters with velocity dispersion.
In Fig.~\ref{fig12}, we plot the derived 
age, metallicity, and enhancement values within an $r_e/8$ aperture 
as a function of the richness parameter  $\rho_{xyz}$.
Analogously to Fig.~\ref{fig11}, E and S0 galaxies are denoted, 
and a linear fit is performed 
for all the galaxies and for the E and S0 
subsamples.
The upper panel indicates a positive trend of age with the 
richness parameter (the Spearman test provides a 85 \% probability
for a correlation to exist).
As for what concerns metallicity and $\alpha$-enhancement, there are
no clear relations with  $\rho$.
The fitted relations to the data suggest that galaxies in lower density environments are slightly more metal rich and less enhanced in $\alpha$ --elements than galaxies in  richer environments. 
However, a Spearman test does not provide any evidence for the
existence of significant correlations.

The relation of age with environment is particularly interesting, in view of the fact
that we have not found any trend with velocity dispersion.
It is surprising that, in spite of containing more than 
40\% of the sample with determined $\rho_{xyz}$,
the region above log($\rho_{xyz}$) $\geq$-0.4
does not contain galaxies younger than 4 Gyr.
If we consider only galaxies with ages older than 4 Gyr, 
the resulting fit is flat, indicating that
the age--environment relation is actually due to 
the presence of very young objects in the poorer environments.
In fact, in such environments we have the contemporary presence of
objects with both high and low velocity dispersions. 
It is also clear, after comparison with Fig.~\ref{fig11}, that the clump of young E
galaxies at $\sigma_c\simeq$250 km~s$^{-1}$ is due to objects
located in a very low density environment.

\begin{figure*}
\center{
\resizebox{13.5cm}{!}{ \psfig{figure=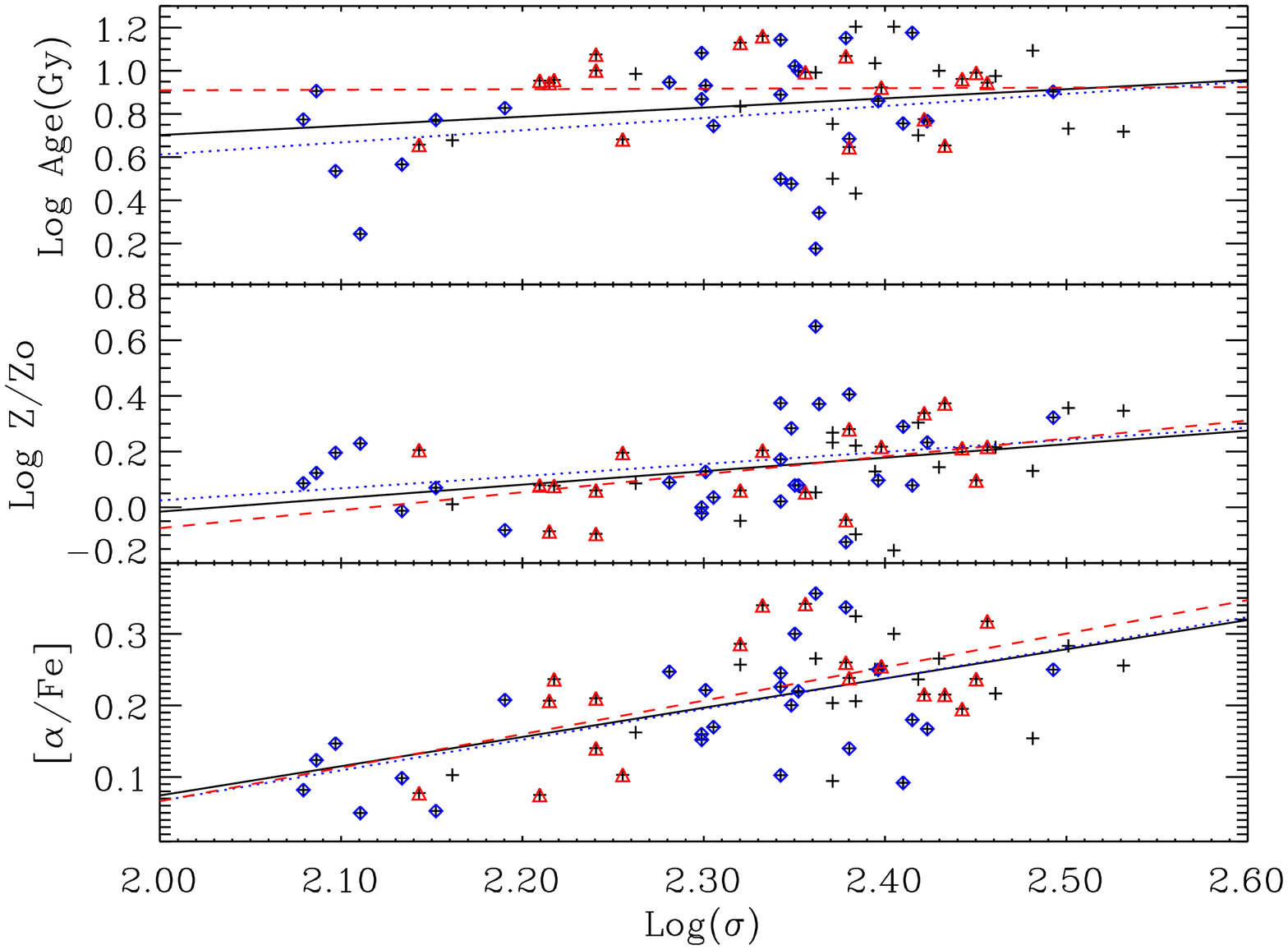,width=13cm,angle=90,clip=} }}
\caption{Ages, metallicities, and [$\alpha$/Fe] ratios, measured at $r_e/8$,
vs. the central velocity dispersion, $\log \sigma_c$. In this figure triangles
denote galaxies with a relatively higher richness parameter, log($\rho$)$\geq$-
0.4. Diamonds denote the other objects, while crosses denote the total sample.
The solid line is the linear fit performed to all the galaxies. The dashed and
the dotted lines refer to galaxies in high and low density environments, respectively.}
\label{fig13}
\end{figure*}

Figure~\ref{fig12} also suggests that the relation with velocity dispersion
may be "disturbed" by the presence of such 
objects populating very low density environments.
In Fig.~\ref{fig13}, we have redrawn 
the fit with velocity dispersion separating objects
according to their richness parameter, instead of morphological type.
A linear fit is performed for all the galaxies, while
fits to the galaxies in high and low density environments 
are also performed separately.
We notice that at values log($\rho_{xyz}$) $\approx$ -0.4,
we may identify small associations of galaxies,
such as pairs and poor galaxy groups,
in the Tully~(\cite{Tu88}) catalogue. 
On the upper side, above log($\rho_{xyz}$) $\approx$ 0.0 typical of
the outskirts of Virgo cluster, we 
count only four objects,
and we stress again that our {\sl rich environment}
is not representative of rich clusters.
Again there is no clear trend of age with velocity dispersion, 
but now galaxies in relatively richer environments are more tightly clustered
around the average relation.
The following correlations are derived for age, metallicity, and $\alpha$/Fe enhancement,
respectively:

{\small
\begin{eqnarray}
{\rm Log(age/Gyr)} &=& {\rm -0.15 + 0.43 \times  log(\sigma_c) \qquad   (Total: r=0.2)}\nonumber \\
{\rm  Log(age/Gyr)}&=& {\rm -0.51 + 0.56 \times  log(\sigma_c) \qquad   (LDE: r=0.24) } \nonumber \\
{\rm Log(age/Gyr)}&=&  {\rm  0.86 + 0.03 \times  log(\sigma_c) \qquad   (HDE: r=0.01)}  \nonumber \\
\label{fitage}
\end{eqnarray} 

\begin{eqnarray}
{\rm  Log(Z/Z_\odot)}&=& {\rm -0.99 + 0.48 \times  log(\sigma_c) \qquad (Total: r=0.34)} \nonumber \\
{\rm  Log(Z/Z_\odot)}&=& {\rm -0.85 + 0.44 \times  log(\sigma_c) \qquad (LDE: r=0.3) }   \nonumber \\
{\rm  Log(Z/Z_\odot)}&=& {\rm -1.36 + 0.64 \times  log(\sigma_c) \qquad (HDE: r=0.47)}    \nonumber \\
\label{fitzet}
\end{eqnarray} 

\begin{eqnarray}
{\rm [\alpha/Fe]} &=& {\rm -0.74 + 0.41 \times  log(\sigma_c) \qquad (Total: r=0.53)}  \nonumber \\
{\rm [\alpha/Fe]} &=& {\rm -0.79 + 0.43 \times  log(\sigma_c) \qquad (LDE: r=0.52) }    \nonumber \\
{\rm [\alpha/Fe]} &=& {\rm -0.87 + 0.47 \times  log(\sigma_c) \qquad (HDE: r=0.58)},     \nonumber \\
\label{fitafe}
\end{eqnarray} 
}

\noindent where LDE and HDE indicate galaxies populating low and high density 
regions, respectively, and $r$ is the linear correlation coefficient.

At 200~km~s$^{-1}$, the average age of galaxies inhabiting HDE is
of about 8.5 Gyr, while those in LDE is $\simeq$6 Gyr.
As far as the relation between metallicity and $\sigma_c$ is concerned,
we notice that there is almost no dependence on the environment.
The different relations are almost superimposed, with the only 
noticeable difference being that the galaxies in HDE are less dispersed
than those in LDE. The same is valid for the enhancement.

Summarizing, combining the information of Fig.~\ref{fig13} 
with the behavior of the stellar population parameters with environment,
we argue that the metal enrichment is essentially dominated by the
gravitational binding of the parent galaxy. In reverse, 
the way in which the populations are assembled (the growth
of baryons within galaxies) seems modulated by the environment,
with galaxies in richer environments being on average older than galaxies in
less rich environments. Galaxies in more rich environments show a lack of very 
young members. Since environment seems to have no effect on the
metallicity-$\sigma_c$ and [$\alpha$/Fe]--$\sigma_c$ relations,
we argue that very young members in the lowest density regions
are more likely due to rejuvenation episodes than to 
more prolonged star formations.

\subsection{Stellar population gradients}

Gradients within galaxies can provide further insight 
on the process  of formation and evolution of early-type 
galaxies. Our sample shows the presence of clear
gradients in the observed line--strength indices. 
Gradients have been extracted from a linear fit to
the index values measured in the four radial regions
$0 \le r \le r_e/16$, $r_e/16 \le r \le r_e/8$, 
$r_e/8 \le r \le r_e/4$, and
$r_e/4 \le r \le r_e/2$. 

\begin{figure*}
\center{
\psfig{figure=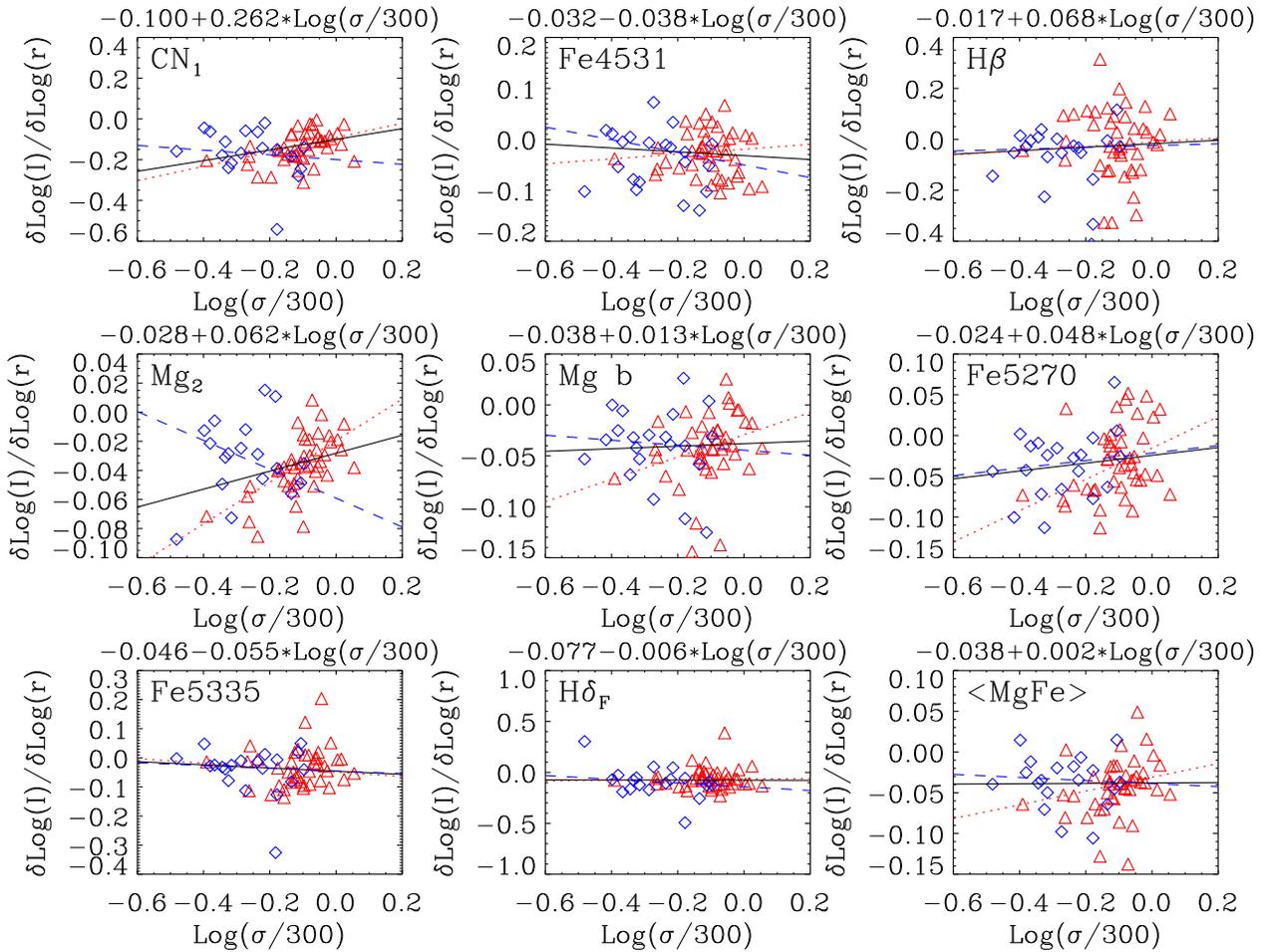,width=0.95\textwidth,angle=90} }
\caption{Gradients of Lick indices computed as $\delta I/\delta log(r)$
for the total sample of 65 galaxies as 
a function of $\log(\sigma_c/300)$, where $\sigma_c$ is the central velocity 
dispersion. Triangles and diamonds denote, respectively, E and S0 galaxies.
The dashed, dotted, and solid lines mark the linear fit obtained for
S0 galaxies, E galaxies, and the total sample, respectively.
For each index, the linear relation that fits the total sample is labeled above
each panel.}
\label{fig14}
\end{figure*}

In Fig.~\ref{fig14} we show as an example
the observed index gradients as a function of $\sigma_c$
for some selected 
indices.
Analogously to Fig.~\ref{fig11}, we have used different 
symbols for S0 and E galaxies, and different lines to denote the fits to
S0 and E separately, and the fit to the total sample. 
The figure, besides showing the presence of eventual
gradients of indices, highlights whether a correlation between 
the gradient and the gravitational potential of the 
galaxy exists.
From Fig.~\ref{fig14}, we notice that in general metallic indices
decrease from the center of the galaxy outwards, while Balmer indices (i.e., H$\beta$, H$\delta$a) increase.
Such a behavior suggests that either metallicity or age
decreases from the central regions of the galaxy to the periphery.
As far as the modulation of the gradients with velocity dispersion
is concerned, we notice that larger gradients tend to correspond to lower $\sigma_c$ values 
(at least for the metallic indices), suggesting that stellar population gradients tend to flatten with increasing galaxy mass.
However, these are only qualitative conclusions, and 
the quantitative derivation of the stellar population gradients  
requires the comparison of the observed indices with 
SSP models.
Age, metallicity, and [$\alpha$/Fe] values derived for 
the gradients ($0<r<r_e/16$ ,$r_e/16<r<r_e/8$, 
$r_e/8<r<r_e/4$, $r_e/4<r<r_e/2$)  are provided for the
three G-spaces (H$\beta$, $<$Fe$>$, Mgb), 
(H$\gamma_{F}$, $<$Fe$>$, Mgb), and (H$\delta_{F}$, $<$Fe$>$, Mgb)
and for the final solution (Eq. \ref{final_sol})
in Tables~9 to 12.
The complete tables are given in electronic form.
For a more detailed presentation of the results with complete figures for 
the stellar population gradients, we refer to Annibali (\cite{An05}).
Gradients are expressed as
$\delta\log($Age/Gyr)/$\delta\log$(r/r$_e$),
$\delta\log$(Z)/$\delta\log$(r/r$_e$), and   
$\delta$[$\alpha$/Fe]/$\delta\log$(r/r$_e$), respectively.

\begin{figure*}
\center{
\resizebox{17cm}{!}{ \psfig{figure=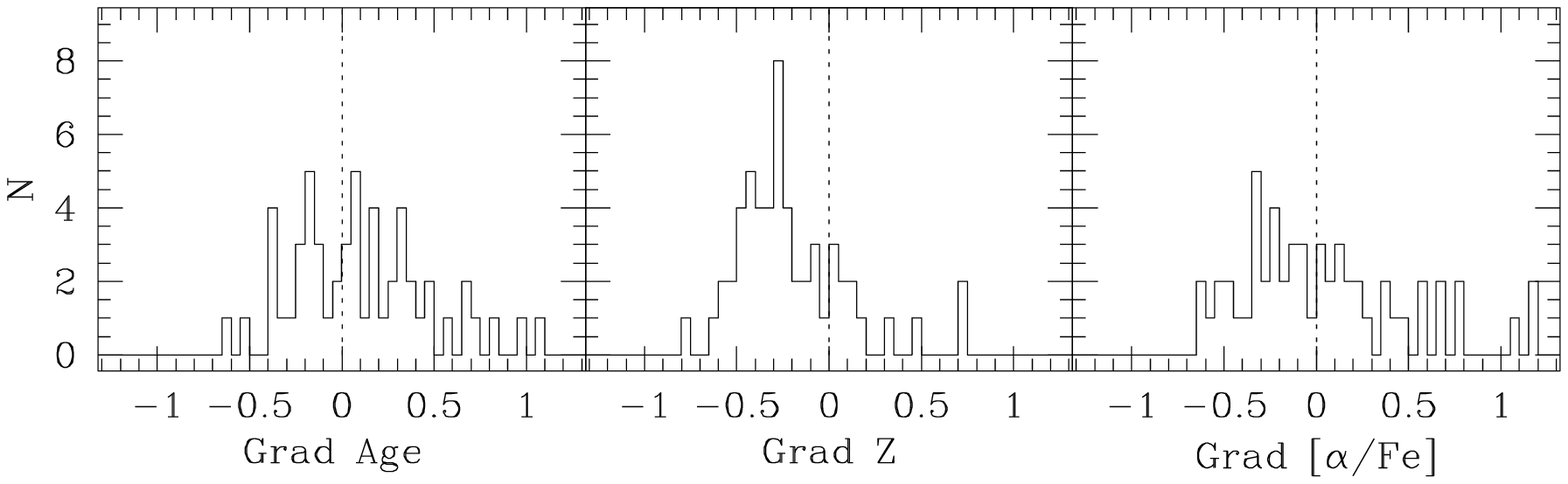,width=13cm,clip=}
}}
\caption{
Distributions of the age, metallicity, and [$\alpha$/Fe] gradients
computed as $\delta log(Age/Gyr)/\delta log(r/r_e)$, 
$\delta\log$(Z)/$\delta\log$(r/r$_e$), and 
$\delta$[$\alpha$/Fe]/$\delta\log$(r/r$_e$), respectively.
The dotted vertical line separates negative from positive gradient values.}
\label{grad}
\end{figure*}

We plot the distributions of age, metallicity, and 
$\alpha$-enhancement gradients in Fig.~\ref{grad} .  
The age and [$\alpha$/Fe] gradients present a large dispersion.
The age gradients are symmetrically distributed around zero, while
the [$\alpha$/Fe] gradients show a low peak at negative values.
The metallicity gradient distribution is the only one that presents a 
clear peak at negative values. We derive an average metallicity 
gradient of -0.21.

\begin{figure*}
\center{
\resizebox{13.5cm}{!}{ \psfig{figure=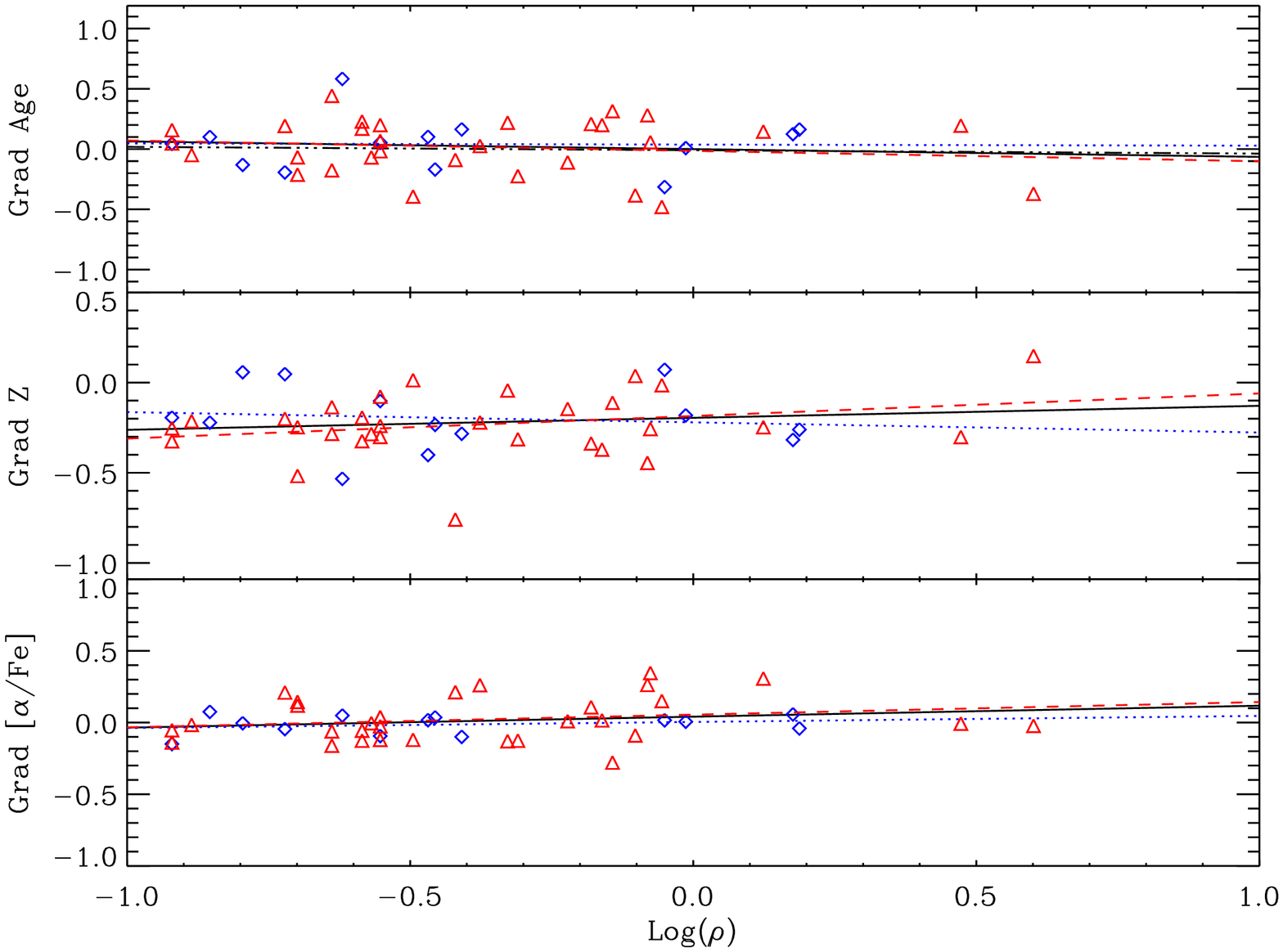,width=13cm,angle=90,clip=}
}}
\caption{Age, metallicity, and [$\alpha$/Fe] gradients
against the density parameter, $\log(\rho)$.
Triangles and diamonds represent, respectively, E and S0 galaxies.
The solid line is the linear best fit performed to all the galaxies.
Dashed and dotted lines are the best fit to the E and S0 subsamples.
The dot-dashed line best fits all galaxies older than 4 Gyr.}
\label{fig15}
\end{figure*}

\begin{figure*}
\center{
\resizebox{13.5cm}{!}{ \psfig{figure=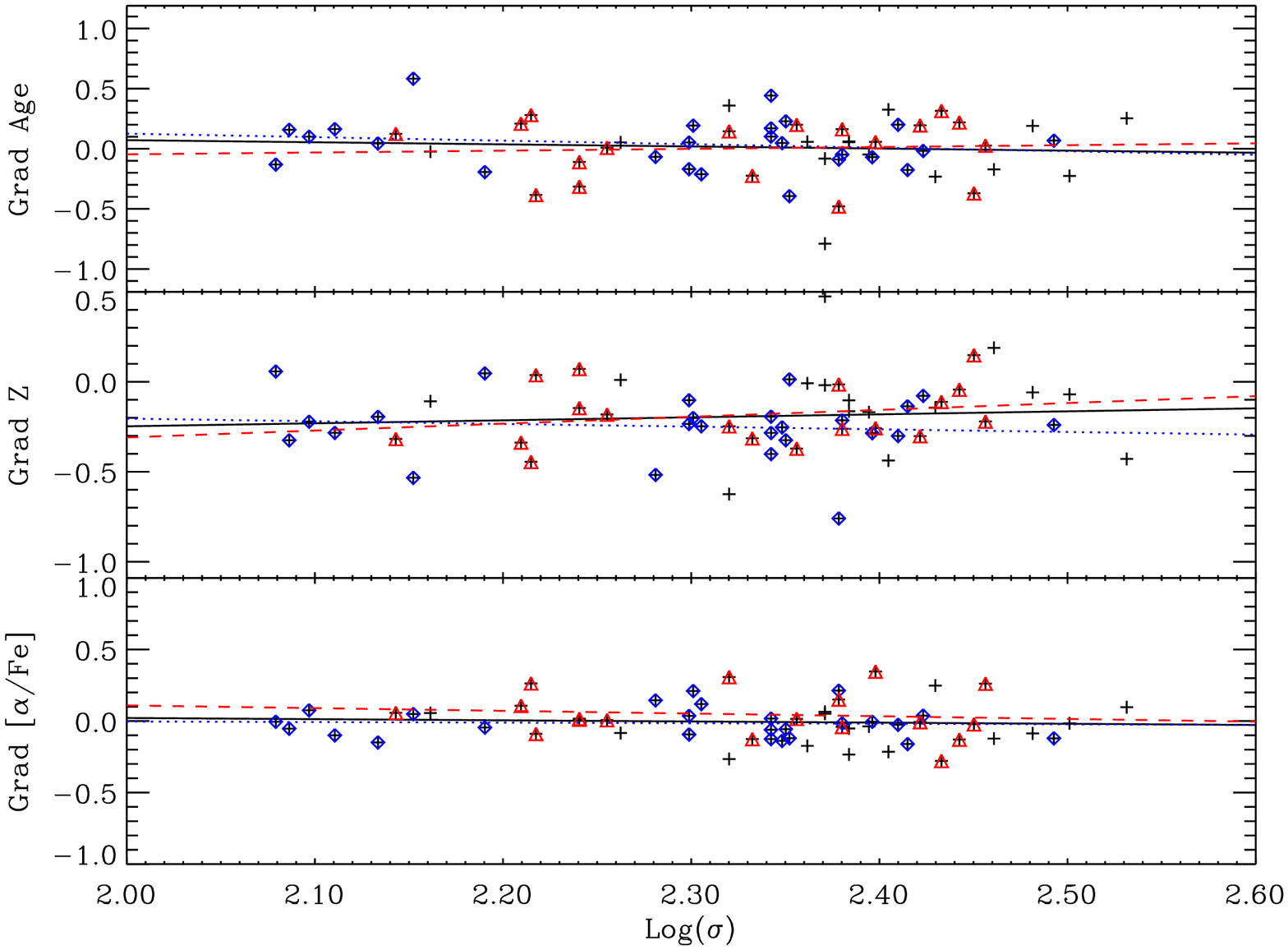,width=13cm,angle=90,clip=}
}}
\caption{Age, metallicity, and [$\alpha$/Fe] gradients vs. the central velocity dispersion $\log \sigma_c$.
The plotted stellar population parameters are derived from the combined
analysis of the (H$\beta$, $<$Fe$>$, Mgb), 
(H$\gamma_{F}$, $<$Fe$>$, Mgb), and (H$\delta_{F}$, $<$Fe$>$, Mgb) index spaces
(see text for details). Triangles are galaxies with relatively high density
parameter  (log($\rho$)$\geq$-0.4), and diamonds the other objects, 
while crosses
mark the whole sample. The solid line is the linear best fit performed to all
the galaxies. The dashed  and dotted lines are the best fits to the galaxies
located in relatively  high and low density  environments, respectively.}
\label{fig16}
\end{figure*}

As was done for the stellar population parameters 
derived within a $r_e/8$ aperture, 
we investigate possible trends of the gradients with the richness parameter  
$\rho$ (Fig.~\ref{fig15}) and the  central velocity dispersion  
(Fig.~\ref{fig16}).
In Fig.~\ref{fig15} E and S0 galaxies are denoted. 
A linear fit is performed for all the galaxies, and fits are also performed  
for the E and S0 subsamples.
Like in Figure~\ref{fig12}, a fit to all the galaxies older than 4 Gyr
is performed. 
We notice that for the gradients there is not the 
dichotomy, revealed  by Fig.~\ref{fig12}, between objects of
relatively high and relatively low richness parameters.
However, there are hints of the existence of trends in the gradients with the
richness parameter. The most evident trends are revealed by the chemical
enrichment path. The metallicity gradient is at values around -0.25 (in good
agreement with, e.g., Davies et al. \cite{DSP93}), but seems to vanish at increasing
richness. The tendency seems even more clear if only ellipticals are considered. 
The
gradient of $\alpha$--enhancement is slightly negative at very low density and
increases with the richness parameter. 
The dispersion is, however, larger than in
the case of the metallicity gradient, and, in spite of the trend suggested by
the linear fits, according to a Spearman rank order test, 
there is no significant
correlation between the gradients and the  density parameter $\rho$.

In Fig.~\ref{fig16} we consider the relation between gradients
and central velocity dispersion once, as in Fig.~\ref{fig13},
we have separated galaxies according to their richness parameter.
There is no significant trend of the age gradient
with velocity dispersion  for the whole sample.
Again, the most important trends are those of metallicity and $\alpha$/Fe--
enhancement.
For the whole sample, the metallicity gradient 
shows an increasing trend with velocity dispersion
(larger negative gradients for less massive systems);
a Spearman test provides a significance level of 0.07 (93\% probability
that a correlation exist).
The correlation is very strong if only E galaxies are considered
($p=0.009$).
As for the [$\alpha$/Fe] gradient, it is slightly positive and constant
for HDE galaxies ($\delta$[$\alpha$/Fe]/$\delta\log$(r/r$_e$)$\sim$0.05).
LDE galaxies have preferentially negative gradients,
with all galaxies above 200~km/s lying below the relation
defined by HDE galaxies.
There are essentially no differences if we considered the sample
of E and S0 galaxies separately (the plot
is not shown here).

\section{Discussion}

How and when early-type galaxies formed is still a matter of debate. 
Two alternative formation scenarios 
have been proposed in the past. In the monolithic one, 
the spheroidal component forms by the gravitational collapse 
of a gas cloud with considerable energy dissipation
(Larson \cite{Lar74b}; Arimoto \& Yoshii \cite{AY87}).
As a result of the rapidity of this collapse,
the bulk of stars in ellipticals should be relatively old.
Unfortunately, this scenario is not supported by
detailed hydrodynamical calculations:
any angular momentum initially present in the 
overdensity region prevents the rapid collapse of the gas.
Small disky entities are formed first, and only  
after subsequent major merging events may they give rise to the
spheroidal component of the galaxies
(Toomre \cite{T77}; Kauffmann et al. \cite{KWG93}; Baugh, Cole \& Frenk \cite{Bau96};
Steinmetz \& Navarro \cite{SN02}). The star formation
process is delayed in this  so--called hierarchical
scenario, which then predicts
the presence of significant intermediate-age stellar populations.

The study of absorption line indices in local 
early--type galaxies has proven to be one of the most
powerful diagnostics to constrain star formation history
and to trace star evolution over time. Besides "direct" age estimates,
indirect evidences of the duration of the star formation
process can be analyzed through the study of
the chemical enrichment pattern, and more specifically of
the $\alpha$/Fe enhancement.
While the $\alpha$-elements O, Na, Mg, Si, S, Ar, Ca, and Ti
(particles that are built up with $\alpha$-particle nuclei) plus the
elements N and Ne are derived mainly by Type II supernova 
explosions of massive progenitor stars, a substantial fraction 
of the Fe-peak elements Fe and Cr comes from the delayed
exploding Type Ia supernovae (Nomoto et al. \cite{N84}; 
Woosley \& Weaver \cite{WW95}; Thielemann et al. \cite{T96}).
Hence, the $\alpha$/Fe ratio quantifies the relative importance
of Type II and Type Ia supernovae 
(Greggio \& Renzini \cite{GR83}; Matteucci \& Greggio \cite{MG86};
Pagel \& Tautvaisiene \cite{PT95}; Thomas et al. \cite{Thom98}), 
and therefore carries information about the  
timescale over which star formation occurs.
Thus, the $\alpha$/Fe ratio can be considered
to be an additional robust measure of the duration
of the star formation process.
Finally, a great deal of information on the formation process
comes nowadays from simulations that are able to predict
the radial distribution of the stellar populations 
within the galaxy (Larson \cite{Lar74a}; Larson \cite{Lar75}; 
Carlberg \cite{Car84}; Bekki \& Shioya \cite{BS99}; Chiosi \& Carraro \cite{CC02}: 
Kawata \& Gibson \cite{KG03}; Kobayashy \cite{Kob04}).
In the following we will summarize all the information
we have deduced from the analysis of our sample,
and see what scenario is more suited
for early--type galaxies in low density environment.

\subsection{The scaling relations for field early-type galaxies}

An important result of our analysis is the derivation of correlations  
between the stellar population parameters and the galaxy potential 
well as a function of environment for our sample of galaxies.
Other studies in the literature have presented scaling relations for 
early--type galaxies on the basis of different samples and/or stellar population models,
among which the most recent are those of
Thomas et al. (\cite{Thom05}, T05), Denicol{\'o} et al. 
(\cite{Den05b}, D05), Gallazzi et al. (\cite{Gal05}, G05) and Clemens et al. 
(\cite{Cl06}, C06).

T05 have assembled three different data sets: G93, 
Beuing et al. (\cite{Beu02})  and 
Mehlert et al. (\cite{Mehl00}, \cite{Mehl03}). 
Their sample contains objects located in both
"high" and "low" {\sl surface} density environments.
It is worth noticing that H$\beta$ emission corrections have been applied
by G93 and by Mehlert et al. (\cite{Mehl00}, \cite{Mehl03})
but not by Beuing et al. (\cite{Beu02}).
T05 generated mock catalogues
starting from the velocity dispersion distribution of the sample and 
assuming suitable scaling relations between the SSP parameters
and the velocity dispersion (allowing for dispersion and for observational errors). The scaling relations are then modified until the best 
"by-eye" fit to the indices (H$\beta$, Mgb, and $<$Fe$>$) is obtained.

D05 have derived ages, metallicities, and [$\alpha$/Fe] ratios  at r$_e/8$ 
for a sample of 83 early-type galaxies essentially in groups, the field, or
isolated objects, which consist of 52 elliptical galaxies and 
31 bulges  of S0s or early-type spirals.
Their results are based on the comparison of the
measured H$\beta$, Mgb, Mg2, $<$Fe$>$ and [MgFe] indices with the TMB03 
models. The  H$\beta$ values used in their analysis were corrected
for emission contamination through the H$\alpha$ line (see also Sect.~2
of this paper).
 
G05 presented stellar metallicities and ages  
for a magnitude limited sample
of $\sim$ 26,000 early-type galaxies drawn from the Sloan Digital Sky Survey
Data Release Two (SDSS DR2). Their analysis rests on the recent
population synthesis models of Bruzual \& Charlot (\cite{BC03}) and is based 
on the simultaneous fit
of five spectral absorption features 
(among which are H$\beta$ and ${ \rm H\delta_A + H\gamma_A}$), 
which depend only weakly on the
$\alpha$/Fe element abundance ratio.
Before comparison with models,
spectra were corrected for nebular emission lines.
In their analysis [$\alpha$/Fe] ratios have not been derived.

C06 have used the SSPs presented in this work
to analyze a volume limited subsample of about 4000 early--type 
galaxies in the 
Third  Release of SDSS catalogue. By considering 11 Lick indices at once,
C06 have derived mean age, metallicity, 
and enhancement variations as a function of the dispersion velocity and 
environment.

When comparing our results with the above literature we should keep 
in mind that:
{\mybold  (a)} the simple stellar population models adopted are different
(in particular TMB03 models are based on different stellar evolutionary tracks and on a different index dependence with element abundance with respect to our new models);
{\mybold  (b)} different line--strength indices are used in the analysis:
in particular, while T05 and D05 adopt a single age indicator, 
the H$\beta$ index, G05, and this
work are based on the use of multiple age
indicators (H$\beta$, H$\delta$, and H$\gamma$), which allow us to minimize 
uncertainties due to emission contamination. Finally, C06 use a simultaneous
least--square fit to 11 index derivatives that takes 
their age/metallicity/enhancement (and also [C/Fe]) sensitivity into account;
{\mybold  (c)} Our sample is biased not only towards early-type 
galaxies in LDE, but also towards
galaxies whose spectra show emission lines,
since we selected them to find ISM traces.
T05 divide their sample into low and high density classes, 
but in both works the classification is performed on the basis of a 
subjective estimate of galaxy density 
(while we use density parameter, which is a properly measured volume density).
The D05 sample contains galaxies biased toward low density 
environment and 
G05 and C06 samples are drawn from the SDSS, which  
contains galaxies belonging to 
a wide range of environments, but predominantly composed of galaxies in 
low density regions.
Finally, D05 and G05 provide only qualitative trends 
of the derived stellar population 
parameters, while C06 provide
differential variations as a function of velocity dispersion.
In the following we will perform a detailed comparison of 
the results obtained in this work with those obtained
in the above quoted papers. However, for the sake of conciseness,
we will limit the graphical comparison of the scaling relations 
to the work of T05, while we will only quote the results obtained by
the other authors.

\begin{figure}
\center{
\resizebox{8cm}{!}{ \psfig{figure=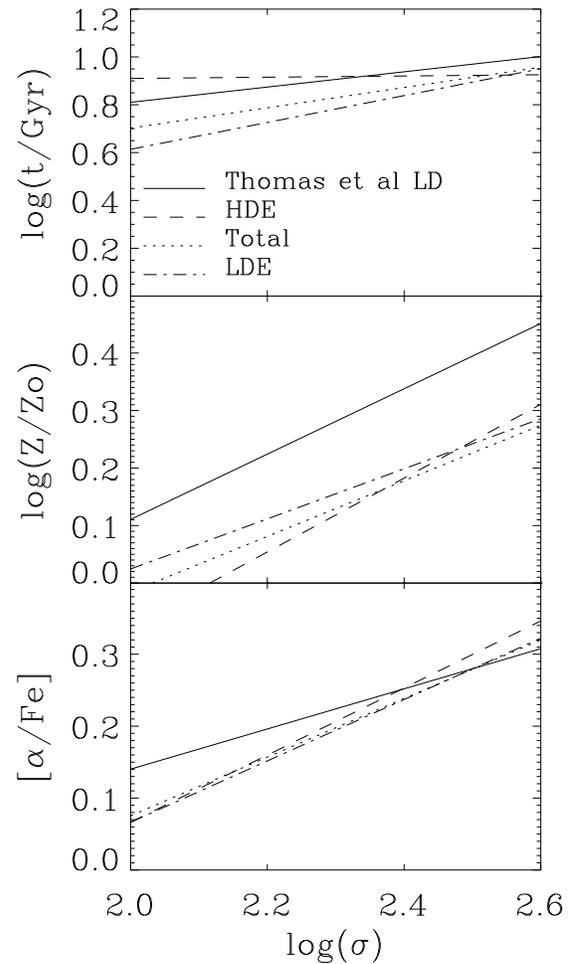,clip=}
}}
\caption{Scaling relations obtained in this work compared with those
of Thomas et al. (\cite{Thom05}) for their low density environment galaxies.
Solid lines refer to Thomas et al. (\cite{Thom05}), while dashed, dot-dashed,
and dotted lines refer to our HDE and LDE galaxies
and to the total sample.}
\label{fig17}
\end{figure}

\subsubsection{The age scaling relation}

In the top panel of Fig.~\ref{fig17},
we compare the age scaling relations obtained by T05 
for the galaxies in low density environments with our results
for the total sample, for HDE galaxies and for LDE galaxies,
respectively.
It is worth noticing that the fit of T05 refers to a group of
"passively" evolving galaxies (in the low density environment subsample)
selected in the H$\beta$-[MgFe]$^{'}$ diagram, while
we have selected, in our field sample, a subsample of "passively" evolving 
galaxies by considering a density threshold limit above which
there is no evidence of rejuvenation episodes (see Sect.~5.1.2).
Thus, when comparing to T05, we consider their scaling relations 
for the galaxies in a low density environment 
and our results for the HDE.
The most noticeable difference between the age--$\sigma_c$ relation
obtained in this work and that of T05 is
in the slope.
T05 find increasing ages
with velocity dispersion (slope $\sim$ 0.32),
while we derive a flat relation (slope of $\sim$ 0.03).
However, we notice that the linear correlation coefficient 
is poor in both cases (Eq. (\ref{fitage})),
indicating that the spread of the data around the relation is large,
and, given the limited number of objects, the error on the coefficients
is also large.
We also notice that if we consider the whole HDE + LDE sample, 
where young galaxies are present, we get a positive trend with a 
slope of 0.43. However, as already discussed in Sect. 5.1.1, a Spearman 
rank order test does not provide significant correlation between age and 
velocity dispersion.

Our results are more consistent
with those of G05 and C06. G05 do not find any significant variation of age 
with stellar mass (in their early--type sub-sample), except for a population
of young metal--poor galaxies in the lowest mass bins. C06
find a significant flattening of the age above $\sigma\sim$ 170km/s.
D05 estimate an average age of 5.8 Gyr for Es and 3 Gyr for S0s,  
with hints of a trend for massive galaxies to be older.

\subsubsection{The metallicity scaling relation}

The relation between metallicity and velocity dispersion is shown in the 
central
right panel of Fig.~\ref{fig17}. The slope we have derived ($\sim$ 0.64) 
falls between the one derived by T05 ($\sim$0.57) and
that derived by C06 (0.76). However, there is a
significant zero point offset between T05 and our relation, 
with the metallicity derived by T05
being about 0.15 dex larger than ours. 
At 100 km~s$^{-1}$, T05 find an average metallicity for galaxies in low density environment that is 26\% larger than solar. Instead, we find that at the same velocity dispersion
the average metallicity of old field galaxies should be $\sim$80\% of the solar one.
Our derived relation is in good agreement with metallicity estimates
of stars of our own Bulge: the average metallicity of the Bulge is found to
be [Fe/H]$\simeq$-0.2 (Ramirez  et al. \cite{Ram00}; Mc William \& Rich \cite{MR94})
and, with its velocity dispersion of $\sigma_c\simeq$104 km~s$^{-1}$ (Blum, {\cite{B95}), 
it fall just slightly below our relation.
The zero point offset is most likely caused by the different evolutionary 
tracks adopted as fully discussed in Sect. 3.

Apart from this offset, the estimate of the slope of the relation 
seems very robust, because it is quite independent from the
adopted sample and from the analysis process.
The linear correlation coefficient and the Spearman test indicate that the  relation between metallicity and $\sigma_c$ is fairly tight. If we add the additional evidence that
the slope seems to be almost independent from the environment
(Eq. (\ref{fitzet}) and T05), we may conclude that the metallicity-$\sigma_c$
relation is perhaps witnessing one of the tightest links between stellar populations and structural parameters.

A metallicity increase with galaxy mass is found by 
D05 and G05 as well. However, the metallicities derived by D05 are  
larger when compared to our work and that of G05, likely because of the use
of TMB03 models:
at $\sigma \sim 300 km/s$ 
D05 derive metallicities above 0.3 dex, while
the G05 metallicities are peaked  
around ${\rm \log Z/Z_{\odot} \sim 0.15}$ and
our relation gives ${\rm \log Z/Z_{\odot} \sim 0.22}$.
We find only a slight dependence of the  metallicity-$\sigma$ relation on the 
richness parameter, in agreement with T05, while C06 find none, in their much larger sample.

\subsubsection{The $\alpha$-enhancement scaling relation}

Our $\alpha$-enhancement scaling relation is compared
with the one obtained by T05 in the 
lower left panel of  Fig.~\ref{fig17}.
Our derived slope (0.47) falls between that found by
T05 (0.28) and that of C06 (0.74) and agrees well
with that derived by D05 (0.59). Our [$\alpha$/Fe] ratio 
coincides with that of T05 at
$\sigma_c$=250 km~s$^{-1}$, while at 
$\sigma_c$=100 km~s$^{-1}$ 
it is $\sim$ 0.07 dex lower.
The $\alpha$-enhancement scaling relation
is almost independent of the richness parameter, in very good agreement
with C06 and T05}.

As discussed in Sect.~3.9, there are no significant differences in the 
[$\alpha$/Fe] ratios predicted by TMB03 and our models. Moreover,
since C06 have also used the models presented here,
we may exclude that the different slopes found by T05, C06, and us
are due to the models. The only noticeable differences are 
in the data samples and in the adopted methods. In particular, the
use of different index sets coupled with the different sensitivity
of the indices to the enhancement may explain the 
significant variation of the slope found by different authors.

\subsection{Gradients and galaxy formation}

Besides global stellar population parameters and their
relation with galaxy structural parameters,
population gradients within galaxies give one of the most
strongest constraints on  galaxy formation.
Spectrophotometric and chemical properties 
at various locations within a galaxy are in fact closely related
to the process of galaxy formation and evolution.
Numerical simulations of dissipative collapse of galaxies 
including star formation definitely predict strong
radial gradients in chemical enrichment 
(Larson \cite{Lar74a}, \cite{Lar75}; Carlberg \cite{Car84}).
The predicted metallicity gradients 
are as steep as 
$\Delta \log$ Z/ $\Delta \log$ $\sim$ -0.35 
(Larson \cite{Lar74a}), $\sim$ -1.0 (Larson \cite{Lar75}), 
and $\sim$ -0.5 (Carlberg {\cite{Car84}). 
A further prediction of monolithic models 
is a steepening of the  metallicity
gradient with the potential well of the galaxy, 
which increases from about zero (in low mass galaxies)
to $\sim$ -0.5 in the most massive ones
(Carlberg \cite{Car84}; Bekki \& Shioya \cite{BS99}; Chiosi
\& Carraro \cite{CC02}; Kawata \& Gibson \cite{KG03}; Kobayashi \cite{Kob04}).

On the contrary, the dissipationless collapse models predict no gradients
in chemical enrichment (Gott \cite{Gott73}, \cite{Gott75}) and 
the occurrence of major mergers predicts a significant dilution
(White \cite{W80}; Bekki \& Shioya \cite{BS99}).
More recently, Kobayashi (\cite{Kob04}) 
has simulated the formation and chemodynamical
evolution of elliptical galaxies including radiative cooling,
star formation, feedback from Type II and Ia supernovae and stellar winds,
and chemical enrichment. Galaxies are supposed to form through the
successive merging of galaxies with various masses, which
varies between a major merger at one extreme and a 
monolithic collapse of a slow-rotating gas cloud at the other
extreme.  They predict an average metallicity gradient
of $\Delta \log$ Z/ $\Delta \log$ $\sim$ $-0.3 \pm 0.2$ and
no correlation between metallicity gradients and mass.
The variety of gradients stems from the difference in
the merging histories; in line with previous findings,
galaxies that form monotonically
have steeper gradients, while galaxies that undergo major mergers 
have shallower gradients. 

If metallicity gradients are more directly related
to the degree at which star formation 
has reprocessed the gas inside the galaxy, 
[$\alpha$/Fe] gradients give very strong constraints
about the {\it{duration}} of the chemical enrichment process 
at different radii. 
According to Pipino \& Matteucci (\cite{PM04}), an outside-in formation
scenario would predict [Mg/Fe] ratios increasing with galaxy radius
(and slightly older ages in the external regions 
compared to the central ones).
On the other hand, an inside-out formation, as suggested to
explain the abundance gradients in the Milky Way
(e.g., Matteucci \& Francois \cite{MF89}), would produce a decrease of the
[Mg/Fe] ratio with radius, since the outermost regions
would evolve slower than the inner ones.

From an observational aspect,
evidence of metallicity gradients comes from the increase of 
line--strength indices (Carollo et al. \cite{CDB93};
Davies et al. \cite{DSP93}; Saglia et al. \cite{SMGBZ00}; Trager et al. \cite{TFWG00};
Mehlert et al. \cite{Mehl03}; Wu et al. \cite{Wu05}) and the reddening of the
colors (e.g., Peletier \cite{P90}) towards the center of 
early-type galaxies.
However, there is not yet a clear picture of 
the radial gradient behavior with respect to 
velocity dispersion.
Some authors have shown that
elliptical galaxies with larger values of the
central Mg$_2$ index tend to have steeper  Mg$_2$
gradients (Gorgas, Efstathiou \& Aragon-Salamanca \cite{GES90};
Carollo, Danziger \& Buson \cite{CDB93}; Gonzalez \& Gorgas \cite{GG95}).
On the contrary,
Kobayashi \& Arimoto (\cite{KA99})  did not find
any correlation between gradients
and physical properties in the 80 early--type galaxies of their study.
Along the same line are the studies of Proctor \& Sansom (\cite{PS02})
and Mehlert et al. (\cite{Mehl03}).
More recently, Forbes et al. (\cite{F05}) have re-analyzed the data of 
Coma cluster ellipticals (Mehlert et al. \cite{Mehl03}; Sanchez-Blasquez \cite{San04})
and have found evidence for
the presence of stronger metallicity gradients in
more massive ellipticals.
The only significant gradient shown by our study is that of the metallicity,
while age and $\alpha$ enhancement seem to be flat within R$_e$/2,
though with a significant dispersion 
(Figs. \ref{fig15} and \ref{fig16}).
The average metallicity gradient is
$\Delta\log Z/\Delta\log(r/r_e)\sim -0.21$.
There is also evidence for a trend of the Z gradient with central
velocity dispersion, in the sense that more massive galaxies tend to have
shallower gradients (Fig.~\ref{fig16}).}
We thus do not find evidence for 
a steepening of the metallicity gradient
toward larger masses, even if we consider only
galaxies in the relatively richer environment.

The flat distribution of $\alpha$ elements, on average, 
in spite of a well--established metallicity gradient, is a little intriguing.
The latter suggests that stellar populations are not mixed within
r$_e$/2 and that metal enrichment was more efficient in the central region than
in the outskirts of the galaxies. An outside-in formation scenario predicts
the existence of gradients in both metallicity and $\alpha$ enhancement,
while merging would smear out
stellar populations and dilute any gradient. 
Our observations indicate that the star formation proceeded on 
typical lifetimes not significantly different across r$_e$/2, 
but evidently with a larger efficiency in the center.
It is easy to show, however, that the observed flat gradient in 
$\alpha$ enhancement is within the prediction of galaxy models.
Using GALSYNTH\footnote{GALSYNTH is a web based interface 
for the chemo-spectrophotometric code GRASIL at  
http://web.pd.astro.it/galsynth/index.php},
we have run a chemical evolution model with parameters suited for a typical
massive early--type galaxy (briefly, high SFR for a fraction of a Gyr and
passive evolution thereafter)
and we have obtained that the run of [Mg/Fe] between [Z/H]=-0.5 and
[Z/H]=0.5, where the majority of stars are formed, is well approximated by
\begin{equation}
        [Mg/Fe] = 0.46 - 0.13 \times [Z/H],  
\end{equation}
Thus the expected gradient in [Mg/Fe] corresponding to 
$\Delta\log Z/\Delta\log(r/r_e)\sim -0.21$ is 
$\Delta~[Mg/Fe]/\Delta\log(r/r_e)\sim 0.03$ in very good agreement
with what was observed.
To find full support for an outside-in formation scenario,
more external regions should be observed and analyzed.

\subsection{Rejuvenation}

The lack of environmental effects on the metallicity--$\sigma_c$
and [$\alpha$/Fe]--$\sigma_c$ relations suggests that 
{\sl young} ages are not the result of 
a more prolonged star formation in the low density environment.
Moreover, NGC 3136, NGC3607, NGC7135, and NGC 6776 are massive galaxies,
metal rich and $\alpha$-enhanced, for which simple stellar population fits
provide ages of only a few Gyrs.
If recently formed and on the short timescale suggested by their
$\alpha$-enhancement, these objects would appear like the very powerful ultra
luminous infrared galaxies (ULIRGs) observed in the local universe. 
However, it is easy to show that their luminosity would exceed by at 
least one order of magnitude that of the local powerful ULIRGs. 
In fact, recent determinations of star formation timescales in these 
latter objects, by different methods, indicate
that the gas content cannot sustain the observed luminosity for more than a few
tens of Myr (e.g., Vega et al.~\cite{Ve05}; Prouton et al.~\cite{Pr04}). This
implies SFR in excess of 10$^4$ M$\odot/yr$ for the masses  that correspond
to the velocity dispersions of these galaxies and, consequently, bolometric
luminosities that by far
exceed the typical ones in the local ULIRGs. 

All these considerations, together with the
presence of morphology/kinematics signatures 
of recent interaction episodes 
in all the four quoted galaxies
(see Appendix in Papers I and II), strongly support the
idea that the young ages obtained are due
to recent rejuvenation episodes of star formation and do not correspond to
their epoch of formation.
If so, what is the mass involved in the rejuvenation episode?
It is difficult to answer this question because the
fading of the indices (H$\beta$ for example)
in a composite population depends on the fraction 
of the young component and the epoch of the event.
We can only proceed with a statistical argument, though we are aware that
the number of galaxies is not large.
Looking to the data of Fig.~\ref{fig12}, we determine that
about 15\% of all the objects look rejuvenated.
The threshold limit for rejuvenation is taken from the age limit of the 
higher density objects.
Let us suppose that, since the average epoch of formation, 
{\sl all} the galaxies
underwent a single rejuvenation event with a
probability that follows the halo merging rate, 
namely $\sim$(1+z)$^{3.2}$ (Le F{\` e}vre et al. \cite{Lef00}).
The average epoch of formation is taken from
the line in Fig. \ref{fig12} indicating an average age 
of about 8 Gyr.

Assuming the above merging rate law, and a flat $\Omega_m$=0.3,
$\Omega_\Lambda$=0.7, and H$_0$ = 70
km~sec$^{-1}$~Mpc$^{-1}$  Universe, we determine that 15\% of the total
merging events since a look back time of 8 Gyr occur within the 
last 2.2 Gyr. 
This time must correspond to the time  during which
the H$\beta$ index of a combined  (old plus young) population 
remains above the selected threshold for rejuvenation and, of course,
it will depend on the relative mass fractions of the old and young 
populations.
Thus we have to seek for what fractional 
mass of the young component a combined population 
(8 Gyr plus the young component) 
maintains its H$\beta$ index
above the threshold value, for a duration of about 2.2 Gyr.

After selecting a threshold value of H$\beta\sim$2.0 
(from the data in Fig. \ref{fig12}), 
we find that the mass fraction of such a young population 
is $\sim$ 25\% of the total mass. 
We conclude that the data of Fig.~\ref{fig12} are compatible with
{\sl all} galaxies having undergone a merger event since their formation, 
where not more than 25\% of their total mass has been converted into stars.
For a larger mass fraction, the fraction of rejuvenated 
objects we would see today would be larger.
In other words, 75\% of the total mass is definitely old,
even if the galaxies suffered a major merger in their life.

\section{Summary and conclusions}

We have analyzed the stellar populations of the 
sample of 65 early--type galaxies mainly located in the field, 
presented in Papers~I and II.
Our sample is biased towards early-type
galaxies selected on the basis of ISM traces, in particular optical
emission lines, tracing the warm ISM component. On the other hand,
considering the incidence of galaxies showing
a warm ISM in a randomly selected sample
of early-type galaxies quoted by Falcon-Barroso et al.
(2006), i.e., 75 \%, (83\% in the field), we are confident
that our sample is not very dissimilar from those with which
we compare our results.

Stellar population parameters age, metallicity, and [$\alpha$/Fe] ratio
have been derived for all the apertures and gradients of the sample 
by the comparison of the Lick indices with
new SSP models, presented in this work,  which account for the presence
of non-solar element abundance patterns.
New SSP models have been developed for a wide range of ages
($10^{9}-16 \times 10^{9}$ yr), metallicities (Z=0.0004, 0.004, 0.008, 0.02,
0.05), and [$\alpha$/Fe] ratios (0-0.8). The SSPs are based on the Padova
stellar evolution tracks and accompanying isochrones (Bertelli et al.
\cite{Ber94}; Bressan et al. \cite{Bre94}), which have been previously
calibrated on color-magnitude diagrams 
of several star clusters in the Galaxy and in the Magellanic Clouds.

The standard composition line-strength indices are computed on the basis
of the fitting functions of Worthey et al. (\cite{W94}) and Worthey \&
Ottaviani (\cite{WO97}). The enhanced SSPs are computed by   
applying an index correction, which we derive
following the main guidelines provided by previous works in the
literature (TMB03, Tantalo \& Chiosi \cite{TC04a}), 
although with revised index dependence on the element abundance,
to the standard SSPs.
More specifically, we find that a logarithmic dependence of the index with
element abundance provides a more satisfactory description 
of the index behavior.
This evidence is also supported by preliminary results obtained 
running new model
atmospheres and synthetic spectra with the code ATLAS12, that consistently 
computes both interior and emergent spectra models with arbitrary abundances.
The correction for the effects of $\alpha$-enhancement is computed on the
basis of the most recent specific index responses computed by
Korn et al. (\cite{K05}) for different metallicities.

The new $\alpha$-enhanced SSPs have been implemented 
within an algorithm devised to derive stellar population parameters 
from the observed line--strength indices. 
The method, which rests on the probability density function, provides, 
together with the most probable solution in the (age, Z, [$\alpha$/Fe]) 
space, the solutions along the age-metallicity degeneracy that are within 
1~$\sigma$ error from the observed index values.

Summarizing our results:

\begin{itemize}

\item The derived age distribution shows a wide spread,
with SSP-equivalent ages from a few Gyr to 15 Gyr and 
with lenticular galaxies being generally younger than ellipticals
(the average ages for the whole sample, E and S0 are 8, 8.7, and 6.3 Gyr,
respectively); 
the metallicity distribution shows a broad peak at $0<[Z/H]<0.3$; finally, 
the [$\alpha$/Fe] ratio definitely presents a peak at $\sim$0.22, 
with a narrower distribution than for age and metallicity.

\end{itemize}

We also seek possible correlation of the stellar population 
parameters with both central velocity dispersion and local galaxy density.

\begin{itemize}

\item Concerning the relation with the galaxy potential well, we do not find 
 the clear signature of a global trend of age with velocity dispersion; 
on the other hand, significant positive trends are derived for metallicity and $\alpha$/Fe enhancement.

\item The robust metallicity-$\sigma_c$ relation testifies that chemical
enrichment is more efficient in more massive galaxies and,  within a universal
initial mass function, the relation between the $\alpha$/Fe enhancement and
the velocity dispersion indicates that the overall duration of the star
formation process is shorter within deeper potential wells. These two
relations do not depend on galaxy morphological type, unlike the
age-$\sigma_c$ relation, indicating that the galaxy gravitational potential is
the main driver of the chemical enrichment process of the galaxy.

\end{itemize}

We analyzed the relations of the stellar population parameters with 
the local galaxy density $\rho_{xyz}$ and we obtained the following
results:

\begin{itemize}

\item Concerning the age--$\rho_{xyz}$ relation, we derive a clear dichotomy:
while very young objects (from 1 Gyr to 4 Gyr, some of which have high
$\sigma_c$) are found  in very low density environments ($\rho_{xyz} \le
0.4$), none of the galaxies in high density regions (40 \% of the sample with
measured density) is younger than 4-5 Gyr. 
No trend of the chemical enrichment path is found
with environmental conditions. 
The lack of environmental effect on the 
($\alpha$-enhancement)--$\sigma_c$  relation indicates that in very low
environments rejuvenation episodes, rather than more prolonged star formation,
are frequent.
\end{itemize}

We analyzed the stellar population gradients
within the single galaxies:

\begin{itemize}

\item We definitely derive negative metallicity gradients (metallicity
decreases from the central regions outwards) of the order of
$\Delta \log Z/\Delta \log (r/r_e) \sim -0.21$, consistently with literature
values. On average the $\alpha$-enhancement remains quite flat
within r$_e$/2, but this is consistent with the observed metallicity gradient.
The only strong correlation is that of the metallicity gradient 
with velocity dispersion: larger negative gradients are derived for less massive systems and become shallower at increasing galaxy mass.

\end{itemize}

From the above results, we try to sketch a possible scenario
for galaxy formation. The potential well of the galaxy seems to be the  main 
driver of the chemical path during its formation,
with more massive galaxies exhibiting the more efficient chemical enrichment 
and shorter star formation timescale.
Furthermore, galaxies in denser environments 
are on average older than galaxies in less dense environments.
This is in agreement with recent models of galaxy formation that combine
feedback from QSO and SN, and cooling of the gas 
(Granato et al. \cite{G01}, \cite{G04}). 

Galaxies in less dense environments also show 
signatures of very recent rejuvenation episodes. 
By comparing the number of ``young'' objects with
the total number of galaxies, and by means of simple two-SSP
component models, we estimate that in these rejuvenation episodes 
(like major mergers), not more than 25\% of the galaxy mass could be formed.
Thus we estimate that, even if spheroidal galaxies are  affected
by major mergers during their evolution, about 75\% of the galaxy mass 
is formed during the initial epoch of formation.

\begin{acknowledgements} 
The authors would like to thank the referee, Dr. H. Kuntschner, whose comments 
and suggestions significantly contributed to improve the paper.
We acknowledge P. Bonifacio and F. Castelli for having
kindly provided us with the ATLS12 code and for their contribution in the
computation of model atmospheres. RR and WWZ acknowledge  the partial support of
the Austrian and Italian Foreign Offices in the framework of the science and
technology bilateral collaboration (project number 25/2004). WWZ acknowledges the
partial support of the Austrian Science Fund (project P14783) and of the
Bundesministerium f\"ur Bildung, Wissenschaft und Kultur. AB acknowledges
warm hospitality by INAOE (Mex).
\end{acknowledgements}

\end{document}